%% file: main.tex
\documentclass{article}

\usepackage{arxiv}

\usepackage[utf8]{inputenc} 
\usepackage[T1]{fontenc}    
\usepackage{hyperref}       
\usepackage{url}            
\usepackage{booktabs}       
\usepackage{amsfonts}       
\usepackage{nicefrac}       
\usepackage{microtype}      
\usepackage{lipsum}
\usepackage[inline]{enumitem}
\usepackage{graphicx}
\graphicspath{ {./images/} }

\usepackage[numbers]{natbib}
\usepackage{yfonts}
\usepackage{subfigure}
\usepackage{multirow}
\usepackage{tcolorbox}

\newcommand{\myexample}[2]{
    \begin{tcolorbox}[colback=black!5!white,colframe=black,title={#1}]
        #2
    \end{tcolorbox}
}

\newcommand{\furl}[1]{\footnote{\url{http://#1}}}

\title{The Four Dimensions of Social Network Analysis:\\ An Overview of Research Methods, Applications, \\and Software Tools\footnote{This paper is currently under evaluation in Information Fusion journal.}}

\author{
  David Camacho\\
  \texttt{david.camacho@upm.es}\\
  Departamento de Sistemas Inform\'{a}ticos\\
  Universidad Polit\'{e}cnica de Madrid, Spain
  \and
  \'{A}ngel Panizo-LLedot\\
  \texttt{angel.panizo@urjc.es}\\
  Computer Science Department\\
  Universidad Rey Juan Carlos, Spain
  \and
  Gema Bello-Orgaz\\
  \texttt{gema.borgaz@upm.es}\\
  Departamento de Sistemas Inform\'{a}ticos\\
  Universidad Polit\'{e}cnica de Madrid, Spain
  \and
  Antonio Gonzalez-Pardo\\
  \texttt{antonio.gpardo@urjc.es}\\
  Computer Science Department\\
  Universidad Rey Juan Carlos, Spain
  \and
  Erik Cambria\\
  \texttt{cambria@ntu.edu.sg}\\
  School of Computer Science and Engineering\\
  Nanyang Technological University, Singapore
}

\begin{document}
\maketitle
\begin{abstract}
Social network based applications have experienced exponential growth in recent years. One of the reasons for this rise is that this application domain offers a particularly fertile place to test and develop the most advanced computational techniques to extract valuable information from the Web. The main contribution of this work is three-fold: \begin{enumerate*}[label=(\arabic*)] \item we provide an up-to-date literature review of the state of the art on social network analysis (SNA);\item we propose a set of new metrics based on four essential features (or \textit{dimensions}) in SNA; \item finally, we provide a quantitative analysis of a set of popular SNA tools and frameworks\end{enumerate*}. We have also performed a scientometric study to detect the most active research areas and application domains in this area. This work proposes the definition of four different dimensions, namely \textit{Pattern \& Knowledge discovery}, \textit{Information Fusion \& Integration}, \textit{Scalability}, and \textit{Visualization}, which are used to define a set of new metrics (termed \textit{degrees}) in order to evaluate the different software tools and frameworks of SNA (a set of 20 SNA-software tools are analyzed and ranked following previous metrics). These dimensions, together with the defined degrees, allow evaluating and measure the maturity of social network technologies, looking for both a quantitative assessment of them, as to shed light to the challenges and future trends in this active area. 
\end{abstract}

\keywords{Social Network Analysis \and Social Media Mining \and Social Data Visualization \and Data Science \and Big Data}


\input{sections/introduction.tex}
\input{sections/scientometric.tex}
\input{sections/sna-research.tex}
\input{sections/sna-applicationDomain.tex} 
\input{sections/4-dimensions.tex}

\input{sections/sna-tools.tex}
\input{sections/challenges.tex}

\bibliographystyle{unsrt}  
\bibliography{biblioSocialErik}

\end{document}

%% file: sections/introduction.tex

\section{Introduction}
\label{sec:introduction}

Currently, online social networks (OSNs) are seen as an essential element for interpersonal relationships in a large part of the world. OSNs allow the elimination of physical and cultural barriers through the globalization of the technology. For this reason, OSNs have billions of active users around the world. OSNs can be defined as a social structure made up of people, or entities, connected by some type of relationship or common interest (professional relationship, friendship, kinship, etc.). From~\cite{boyd2007social}, an OSN can be defined as: ``a service that allow individuals to \begin{enumerate*}[label=(\arabic*)] \item define a public (or semi-public) profile within an application or specific domain (friendship, professional, common interests, etc.), \item manage a list of other users with whom the individual (or entity) will share a connection, and \item view and traverse their list of connections and those made by others within the social site\end{enumerate*}''. Although the origins correspond mainly to different areas from Social Sciences, the term is attributed to the British anthropologists Alfred Radcliffe-Brown~\cite{radcliffe1930social, radcliffe1940social} and John Barnes~\cite{barnes1954class}. 

OSNs were created in the late 90s, when Randy Conrads founded \textit{Classmates} in 1995 and Andrew Weinreich launched, in 1997, \textit{SixDegrees.com}. The first mathematical and statistical tool used for studying networks was developed in sociology~\cite{scott1988social}. However, from the last decades, and thanks to the increasing capacity of computing systems in both computational power and information storage, new disciplines have contributed to the area of OSNs, from Sciences and Engineering (such as Computer Science, Mathematics, Physics, or Biology) to Social Sciences (such as Sociology, Anthropology, Psychology, Linguistics, or Advertising).

The fast growth of OSN sites has led to an enormous interest in the analysis of this type of networks (the interconnections that originate, their structure, the evolution of the network, the information flow and dissemination, or the patterns that can be extracted from them, among many others). The easy access to this type of information, the availability of vast amounts of data, the simple and straightforward codification in form of graph-based representation, as well as the direct application of any practical results drawn from them, has made OSNs one of the hot research areas in several disciplines as Data Mining~\cite{aggarwal2011introduction}, Big Data~\cite{wu2013data}, Machine Learning~\cite{wang2015link}, Information Visualization~\cite{borgatti2018analyzing}, or Complex systems~\cite{pastor2015epidemic}, among many others. This growth has been exponential in terms of the number of people connected to the network, (currently, there are around $3.484$ billion of active social media users, up $9\%$ year-on-year, connected to different OSNs~\cite{Chaffey2019}), as well as, in the number of different social-based applications, tools and frameworks available to analyze this social data. However, the current growth of social media has caused serious problems for traditional data analysis algorithms and methods (such as data mining, statistics or machine learning)~\cite{Guille-SIGMOD-13, tang2008arnetminer}. The just mentioned areas have to face the challenge of designing and implementing new methods able to work efficiently with the huge amount of data generated in the OSN.

This exciting area generates thousands of papers per year, hundred of different algorithms, tools, and frameworks, to tackle the challenges and open issues related to OSNs. Therefore, when anyone tries to develop a comprehensive analysis of the state of the art in such a complex and multidisciplinary area, it is necessary to establish a formal method to allow the analysis of the immense amount of information available. To do that, it has been carried out a scientometric analysis (Section~\ref{sec:scientometric}), which has led us to organize the research process of this paper. This process can be summarized as follows:

\begin{enumerate}
  \item We have performed a scientometric study over the papers published in the last $5$ years to extract both, the most active research areas and the most relevant application domains in SNA. The research areas selected have been: graph theory and network analytics, community detection algorithms, information diffusion models, user profiling, topic extraction, and finally sentiment analysis area. The selected application domains have been: Healthcare, Marketing, Tourism \& Hospitality, and Cybersecurity. This analysis has allowed us also to find a set of emerging areas, and we have selected and studied the areas of Politics, detection of fake news, and Multimedia in OSNs. These emerging areas have been selected due to their current research activity level, and their high potential impact in the next years.
  
  \item Once the SNA application domains and fundamental research areas have been selected, we have analyzed in detail the most relevant research works published in the last years in the corresponding research fields and application domains.
  
   \item Due to the complexity of the field, and the increasing amount of available technologies for OSN, we have selected, studied, and analyzed an extensive set of SNA software tools and frameworks.
  
  \item We have defined a set of metrics, which allow any researcher to assess any SNA framework, tool or algorithm. These metrics are based on some key aspects, or features, that are relevant for any algorithm belonging to SNA research field (such as knowledge discovery, information fusion or visualization among others).
  
  \item Finally, we have assessed some relevant frameworks and tools available on the Internet to perform SNA tasks by using the metrics previously mentioned. This assessment allows us to rate the degree of maturity of SNA technologies, so any engineer or researcher can better understand the strengths and weaknesses of the different tools and frameworks currently available.
\end{enumerate}
 
 The main contributions of this paper can be summarized as follows:
 
 \begin{enumerate}
  \item A detailed review of the current state of the art of a set of highly relevant research works, grouped in different categories depending on the fundamental research area addressed, and the application domain. The different research areas, and application domains, have been selected according to a scientometric analysis, which takes into account the keyword (\#hashtag) ``\emph{\#social network analysis (SNA)}", used in a set of relevant papers published in the last $5$ years, along with some seminal papers from these areas.
  
  \item The definition of four new ``SNA-\textit{Dimensions}". These dimensions are inspired by the popular V-models~\citep{laney20013d} used in the Big Data area, and its goal is to measure the capacity of the different frameworks and tools to perform SNA tasks. These dimensions will be used to define a set of metrics (that we named \textit{degrees}), that will allow any researcher to identify the technology maturity or technology readiness level, and the main challenges and trends, under the area of SNA. The dimensions defined will be directly related to \textit{Pattern \& Knowledge discovery, Information Fusion \& Integration, Scalability}, and \textit{Visualization} research topics. These dimensions try to answer the following Research Questions:
  \begin{enumerate}
    
    \item[RQ1)] \textbf{{\Large W}hat can I discover?} (\textit{Pattern \& Knowledge discovery})~\cite{Guille-SIGMOD-13, tang2008arnetminer}: This is the most classical and studied characteristic in OSNs. From the past century different methods and models from Social Sciences, and more recently from Science and Engineering, have been used to discover new knowledge and patterns (in form of network structure and topology, statistical correlations, information flow and diffusion, etc.) in networks. We will define this feature as a dimension related to the capacity of algorithms, methods and techniques to gather knowledge (usually complex and non-trivial patterns) from OSNs. 
    
    \item[RQ2)] \textbf{{\Large W}hat is the limit?} (\textit{Scalability}):~\cite{pujol2011little, wakita2007finding}: Due to the fast growth of OSNs, this is a critical dimension based on the capacity of algorithms, methods and frameworks to work with large amounts of data in an adequate time (when the timing would be necessary for a particular application, as could happen in Marketing or Recommender Systems in OSNs). As occur with \textit{Volume} dimension in Big Data~\citep{bello2016social}, this dimension joins both computational time and volume of data to obtain a qualitative measure related to the capacity to manage large networks.
     
     \item[RQ3)] \textbf{{\Large W}hat kind of data can I integrate?} (\textit{Information Fusion \& Integration})~\cite{poria2016fusing, beach2010fusing}: This dimension will be related to the capacity of fusing different kinds of sources (text, video, images, audio). The currently available OSNs provide data in different formats, which allow defining different types of networks (as Multilayer SNA), which can be later integrated to generate new knowledge. 
     
    \item[RQ4)] \textbf{{\Large W}hat can I show?} (\textit{Visualization})~\cite{freeman2000visualizing, brandes2004analysis}: Visualization in OSN is one of the most powerful tools used in this area to analyze the knowledge and patterns hidden in these networks. Due to the huge amount of available data, and the complex information (and patterns), which can be stored in a network, an adequate visualization of this information is always a challenging task for any OSN analyst. For this reason, the capability to visualize, filter and represent adequately the information stored in a network will be one of the dimensions selected.
  \end{enumerate}

  \item The definition of a new \textit{global Capability metric}, named \textgoth{C}$_{SNA}$, based on the just mentioned metrics that can be used to rank the capabilities of the SNA technology, framework or tool analyzed.
  
  \item The assessment of $20$ of the most relevant frameworks and tools to perform SNA tasks using these different dimensions and the global Capability metric defined. This analysis not only shows what are the main strengths and weaknesses of the different frameworks, but it also provides a useful guide for those beginners or senior researchers in the area of SNA. 
\end{enumerate}

\begin{figure}[h!]
\centering
\includegraphics[width=1\linewidth]{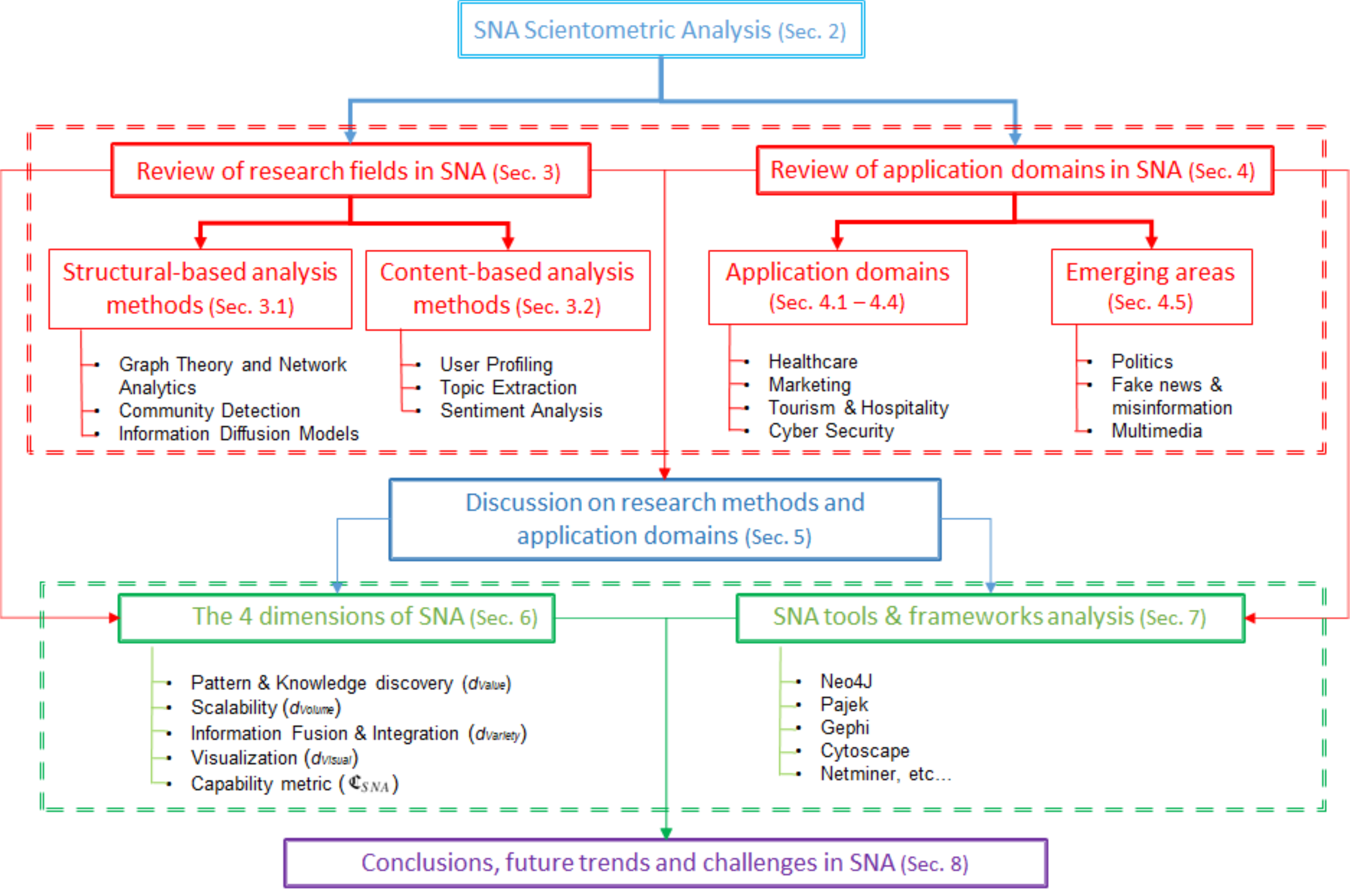} 
\caption{Structure of the paper and the main contents of different sections. Red boxes show the sections related to the review on both the current state of the art in SNA research, and their application domains, whereas green boxes show the contributions on the four dimensions defined, and the software tools and frameworks assessment carried out.}
\label{fig:structurepaper}
\end{figure}

To help readers through the contents of this paper, Fig.~\ref{fig:structurepaper} shows the overall organization of this work. Using this figure, readers can easily understand the main contents of each section and go directly to those contents that can be more relevant for their future work.

 This paper is structured as follows: Section~\ref{sec:scientometric} contains the scientometric analysis performed to highlight the different research areas with high impact in SNA research field. Section~\ref{sec:sna-research} provides an analysis of the different concepts and research works that belongs to the SNA research field from the Computer Science point-of-view. In this sense, there are two approaches based on what kind of information is used to perform the analysis: the structure of the network, or the content of the messages published. The different application domains in SNA, as well as the emerging areas, are analyzed in Section~\ref{sec:applications}. Section~\ref{sec:soadiscusion} provides a discussion and some conclusions from the state of the art analyzed in sections~\ref{sec:sna-research} and~\ref{sec:applications}. Section~\ref{sec:4dimensions} contains the definition of the four dimensions and their related metrics (or degrees), which are used to compare the different SNA tools in Section~\ref{sec:sna-tools}. Finally, some challenges, future trends in this area, and the main conclusions of this work are given in Section~\ref{sec:challenges}.

%% file: sections/scientometric.tex

\section{SNA Scientometric Analysis}
\label{sec:scientometric}

Although bibliometrics emerged to support the daily work of librarians, nowadays it is used to evaluate the scientific achievements of people and institutions~\cite{ball2017introduction}. The main goal of bibliometrics is the quantification of scientific production, measuring the performance of institutions and people. One of the contributions of this paper is an updated review of the current state of the art in the area of SNA. Nevertheless, the number of research papers in this domain increases every day and it is near to impossible to cover in detail all the application domains in which SNA methods, algorithms, tools and frameworks are applied. This section presents a scientometric analysis of the research papers published in SNA, which has been carried out to detect and select those relevant areas that will be analyzed. 

This study has been done by using the \verb|Meta-knowledge Python| package \cite{mclevey2017introducing}, which accepts raw data from the Web of Science, Scopus, PubMed, ProQuest Dissertations and Theses, and select funding agencies as NSF (United States), or NSERC (Canada), among others. The output of this package is a set of characteristics for quantitative analysis, including Time Series methods, Standard and Multi Reference Publication Year Spectroscopy (RPYS), computational text analysis, and network analysis. 

In particular, this analysis presents a review of works related to SNA using Web of Science as a search engine, covering the highly cited articles over a five years period (from 2014 to 2018) and resulting in a record collection of \textbf{28.805 articles}. It has only been considered in the scienciometric analysis those authors that have used in their publications the keyword (\#hashtag) of \emph{"\#social network analysis (SNA)"}. Although this analysis could discard some relevant authors or publications, which did not use the previous term, this keyword was used to restrict the set of papers to analyze and to obtain some of the highly relevant application domains and research in SNA.

Firstly, to better understand the evolution in the area of SNA in the last sixty years (since the 70s), RPYS has been used. This method was proposed by Marx et al.~\cite{marx2014detecting}, and it is a method for quantifying the impact of historical publications on research fields. Standard RPYS~\cite{comins2015compressing} analyzes the cited references and especially the referenced publication years of a publication collection. This method plots the cumulative distribution of cited references in terms of the referenced publication years. In the first step, all the references from the publications are selected (from a particular research field and period). Then the 5-year median deviations, to the number of cited references from each publication year, are computed to generate a \textit{spectrogram}. The peaks in the spectrogram (deviations from the median) indicate those specific years with highly cited publications within the domain of the sample. Also, Multi RPYS is an extension of the standard method~\cite{comins2015compressing}. It segments the original citing articles based on their publication years and conducts a Standard RPYS analyzes for each one, visualizing the results as a heat map. Therefore, this method is useful for differentiating between historical publications that have a lasting impact, versus those that are influential only within a short time frame.

\begin{figure}[!ht]
\centering
\includegraphics[width=1\linewidth]{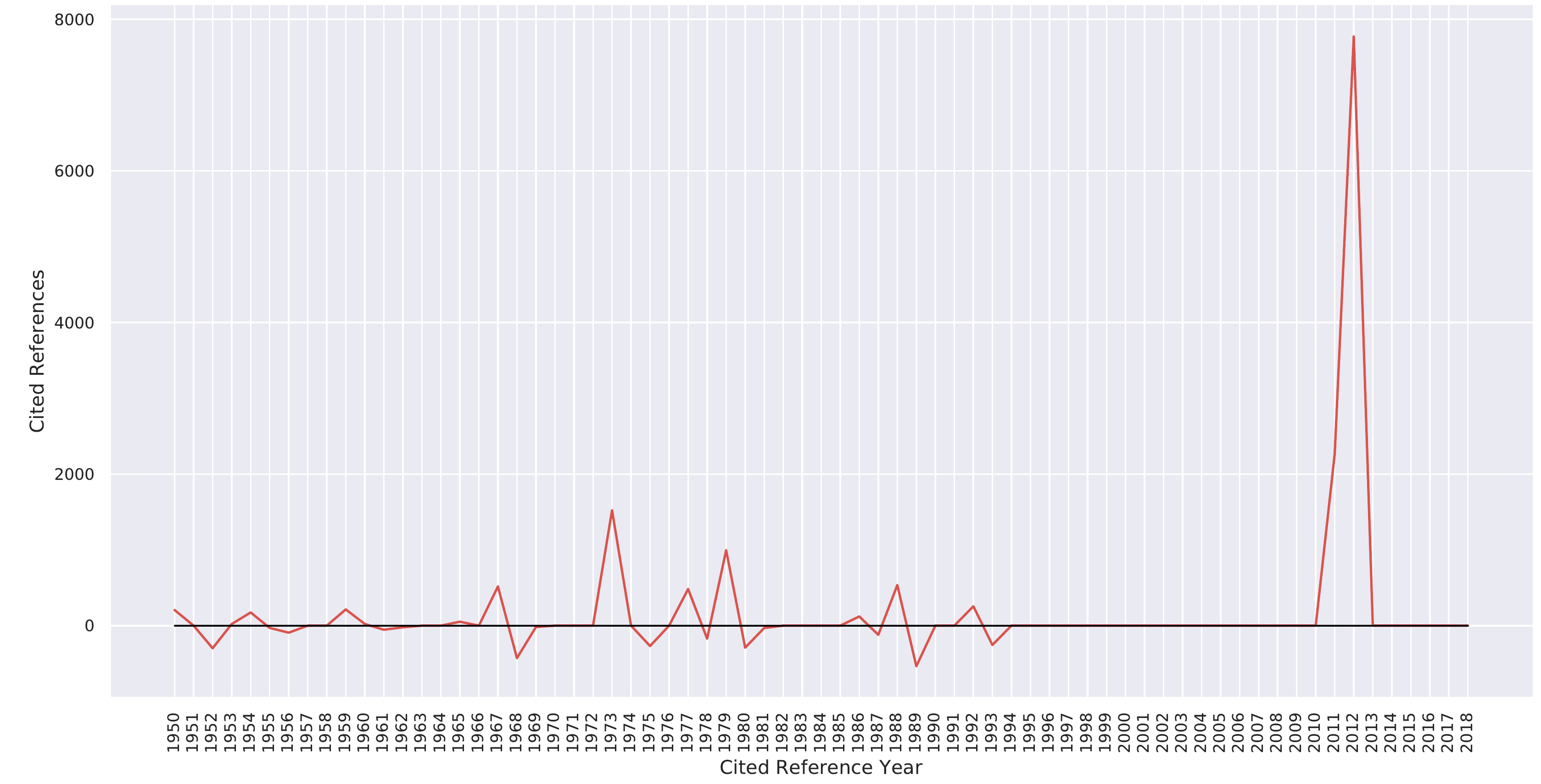} 
\caption{Standard RPYS related to SNA. Pronounced peaks represent years where citations to published books or articles deviate from a 5-year median.}
\label{fig:StandardRPYS}
\end{figure}
\begin{figure}[!ht]
\centering
\includegraphics[width=0.75\linewidth]{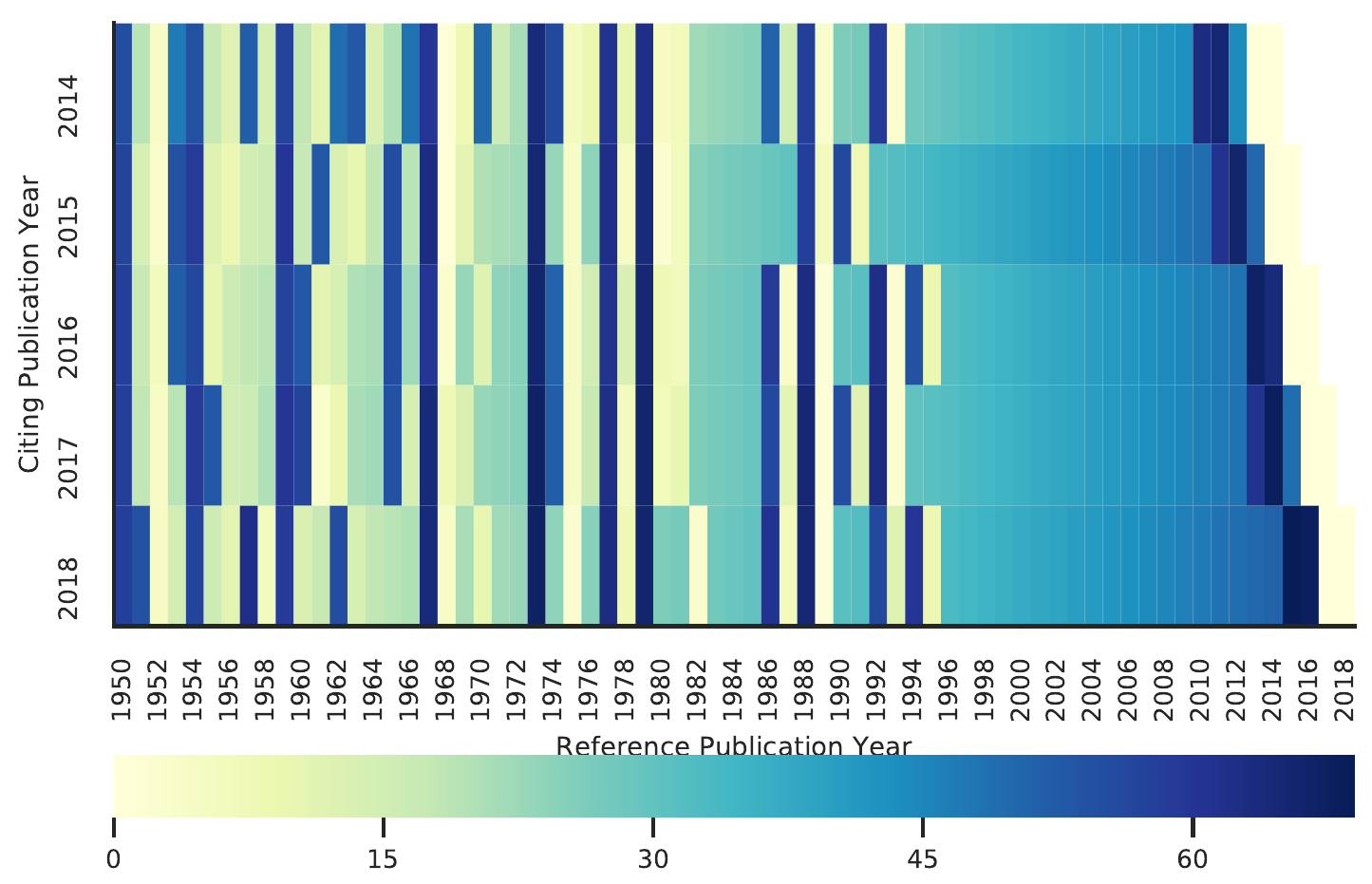} 
\caption{Heatmap showing the results of a Multi RPYS analysis. Darker bands across both periods indicate lasting influence.}
\label{fig:MultiRPYS}
\end{figure}

As shown in Fig.~\ref{fig:StandardRPYS}, the years with pronounced peaks for SNA are: 1967~\citep{milgram1967small}, 1973~\citep{GRANOVETTER1977347}, 1977~\citep{Freeman1977}, 1979~\citep{freeman1978centrality}, 1988~\citep{Coleman1988}, 2011~\citep{BOLLEN20111} and 2012~\citep{Valente49}. Analyzing the most cited papers of this specific years, in the 70s and 80s, there were publications with a high impact on the topic, being published the articles most cited in journals in the area of Social Science~\citep{milgram1967small,GRANOVETTER1977347, Freeman1977,freeman1978centrality,Coleman1988}. While the most recent articles with higher impact are published in journals belonging to Computer Science~\citep{BOLLEN20111, Valente49}. It can also be observed that in the 70s, 80s and 90s there were several publications with a high impact in the area of research, but from 1994 to 2010 there was a period of stability where no high relevant works emerged. Fig.~\ref{fig:MultiRPYS} shows the Multi RPYS analysis, and the fact mentioned above can be seen. Between 1996 and 2010, there are publications for all the years analyzed that are quite quoted, but where no publication stands out from the rest. Therefore, it can be drawn that this period of years has very high activity in the research area of SNA, existing many publications in the topic which are very cited in general terms. Besides, according to the standard RPYS analysis, the oldest years that showed publications with a high impact, also appear as highlighted (1973, 1977, 1979 and 1988). 

From this long period analysis, it can be concluded that most of the papers published in the 70s and 80s were focused on defining the basic principles of the SNA such as the principle of diffusion of influence and information, mobility opportunity, community organization, or links joining people who know each other~\citep{milgram1967small,GRANOVETTER1977347,Coleman1988}. Also, published works at these years introduced a broad background for measures of structural centrality in OSN~\citep{Freeman1977,freeman1978centrality}. However, since 2000, the main contributions have been focused on the application of these principles to specific domains, using OSN data to extract new knowledge, which improves the performance of the organizations. In this case, it can be shown how the areas of highly cited papers have been moved from Social Sciences (the 70s and 80s)~\citep{boyd2007social, GRANOVETTER1977347, freeman1978centrality, Coleman1988,McPherson2001}, to Physics, Mathematics or Computer Science~\citep{Barabasi509, Watts1998Collective, Girvan2002, Blondel_2008}. These articles are mainly focused on the design of new methods, algorithms and applications for SNA, by taking into account the \textbf{structure} of the networks~\citep{Barabasi509, Watts1998Collective} and detecting important structures, as communities, inside them~\citep{Girvan2002, Blondel_2008}. As it has been already seen in the RPYS analysis, the oldest publications that appear are related to the definition of principles and measures to model the dynamics of the OSNs. Whereas, the most recent publications are focused on the development of methods and techniques based on the concepts previously published. For example, it should be noted that community detection algorithms are those that prevail in the last two decades, such as Girvan and Newman~\citep{Girvan2002,newman2004fast}, edge betweenness centrality~\citep{newman2004finding}, fast Greedy~\citep{clauset2004finding,clauset2005finding}, Cfinder~\citep{palla2005uncovering,palla2007quantifying}, Walktrap~\citep{2005-Pons,pons2006computing}, structural algorithm by Rosvall et al.~\citep{rosvall2007information,rosvall2008maps}, clique percolation method~\citep{kumpula2008sequential}, fast modularity optimization by Blondel et al.~\citep{Blondel_2008}, Louvain algorighm~\citep{rotta2011multilevel}, community embeddings~\cite{cavemb}, etc.



\begin{figure}[h!]
\centering
\includegraphics[width=0.65\linewidth]{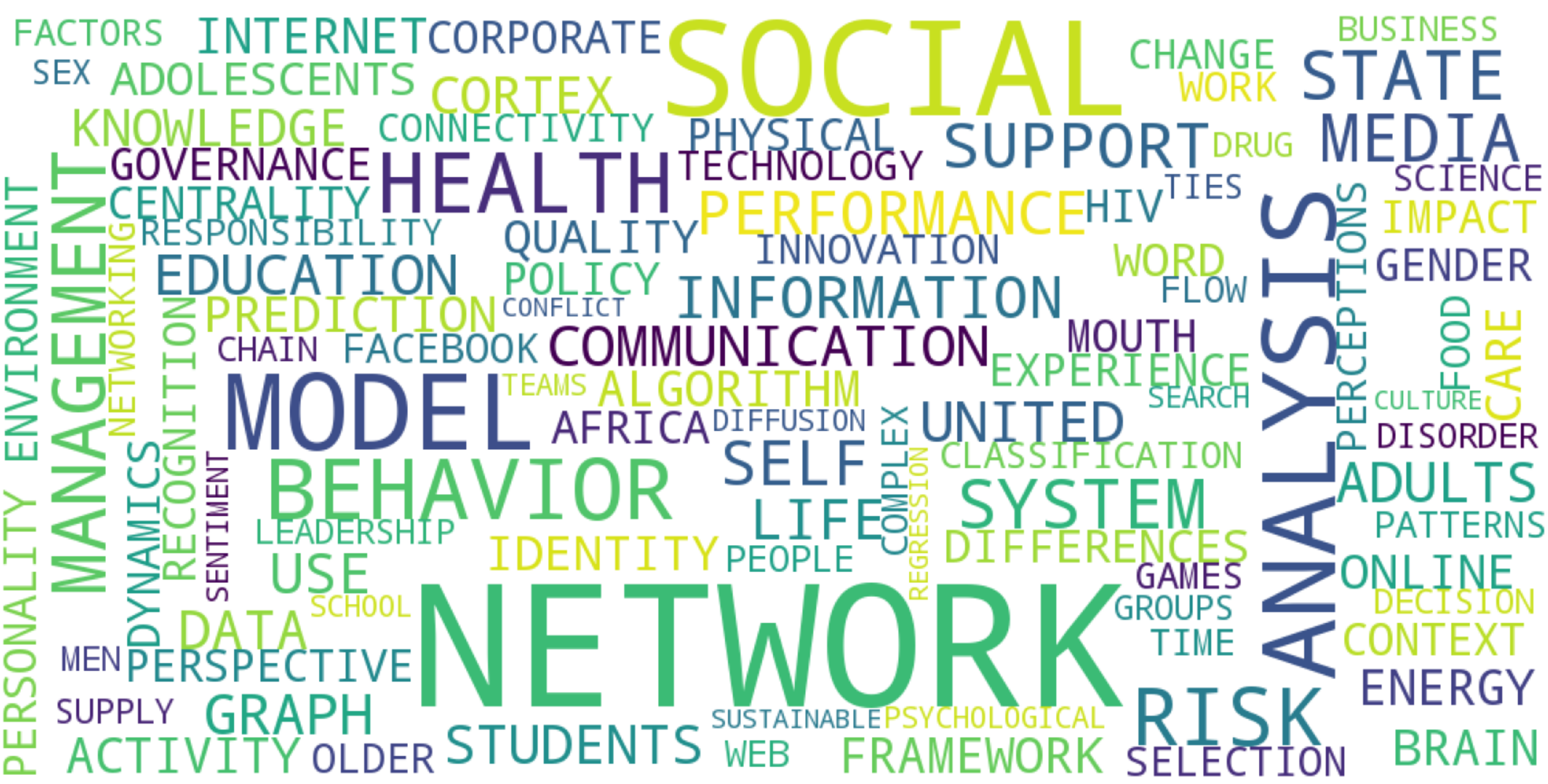}
\caption{World Cloud visualization of the most frequent terms that appear in the top $20$ of extracted topics processing the Keywords of the papers.}
\label{fig:worldCloudKeywords}
\end{figure}

To identify the most relevant topics related to SNA, textual analysis has been performed using the collection of articles gathered from the period of the last 5 years studied. In this case, the latent Dirichlet allocation (LDA) model has been applied to detect the top 20 topics, by processing the ``keywords" used in the article collection, to later visualize the most frequent terms (i.e., keywords) as a world cloud (see Fig.~\ref{fig:worldCloudKeywords}).

Analyzing the most relevant keywords related to SNA, it can be seen at first sight that many of the relevant keywords are related to the key topic of \textit{Pattern \& Knowledge Discovery} are: Model, behavior, Patterns, Classification, Prediction, Knowledge, Regression, Recognition, Dynamics, Groups, Diffusion, or Sentiment, among others. Also, there are several relevant application domains where these methods of knowledge extraction are applied as Healthcare (HIV, Disorder, Brain, Cortex, Drug, Sex, Physical and Psychological), Education, Governance, Policy, Mouth (Mouth-to-Mouth), Business, Food, Culture, or Games. Regarding the topic of \textit{Information Fusion}, there are several keywords related to different data sources and types of information that are used in the area such as Internet, Media, Facebook, Web, Graph, Factors, Activity or Context. The rest of the keywords would be classified taking into account the topic of \textit{Scalability}, where the terms Performance, Complex and Sustainable keywords are directly related to this feature. And finally, related to the topic of \textit{Visualization}, appears some keywords such as Perspective, States, or Flow.

Although the initial study covers the last six decades, a detailed analysis of the last five years has been done to understand which areas and application are more relevant in terms of high impact. From this analysis, it can be concluded that currently some of the hot research areas, from Computer Science, in this topic are: Data Science and Big Data, more specifically Network analysis, Social Media, Sentiment analysis, Text Mining, and Information diffusion. Whereas some of the hot application areas are: Health, Marketing and Business, and Tourism. It is particularly important to point out that a large number of published works are focused on solving problems (through the design of specific algorithms and techniques, such as those related to community finding problems, or information diffusion models), which are related to the problems of pattern mining and knowledge discovery, how to fusion or integrate information, how to visualize adequately the information, or how to handle huge amounts of data.  

Finally, and although the scientometric analysis has not directly shed light on some areas such as fake news and misinformation (e.g., fake news analysis in SN), or cybersecurity (e.g., cyber intelligence, cybercrime, and cyber terrorism), during our analysis of the state of the art we have detected an increasing research activity in these areas (see Section~\ref{sec:emerging} for a further description), so they will be analyzed as emerging areas due to its high potential in the near future. Previous findings related to the research and application domains in SNA have guided the final decision on the selection of what fields will be finally analyzed in this paper.

%% file: sections/sna-research.tex
\section{Techniques and algorithms}
\label{sec:sna-research}
As it has been shown in the previous section, the area of SNA comprises a wide variety of multidisciplinary fields, ranging from Social Sciences to Sciences and Engineering. This section is focused on current findings from the Computer Science point of view. Currently, in Computer Science, and other related areas such as Mathematics or Physics, the design and development of new algorithms to process, extract and visualize (huge amount of) knowledge, are closely associated to \textit{Data Science} and \textit{Big Data} fields. The former can be defined generically, as a multi-disciplinary field that uses scientific methods, techniques and algorithms, to extract useful knowledge from structured and unstructured data. Whereas the latter uses different methods and techniques to analyze, and systematically extract information from data sets that are too large, or complex, to be dealt with traditional data-processing algorithms. From the data engineering perspective, an OSN can be analyzed from two different points of views~\citep{aggarwal2011introduction}: \begin{enumerate*}[label=(\arabic*)] \item \textit{structural data} that represents the connections, interactions (linkage-based), and the topology of the network; or as \item \textit{content data} that is focused on the information contained and shared in the OSN by the users.\end{enumerate*} The isolated analysis of one of these two types of data would provide an incomplete vision of the information stored in the network, so the underlying patterns and knowledge could be lost. Therefore, and taking into account these two types of data, we can distinguish between two different approaches: \textit{Structural-based analysis}~\citep{aggarwal2011introduction, wasserman1994social} and \textit{Content-based analysis}~\citep{aggarwal2011introduction, stemler2015content}, that are described below.

\subsection{Structural-based Analysis}
\label{subsec:structuralSNA}
 This is the research area involved in the analysis of networks using network theory (usually known as graph theory)~\citep{scott1988social, barnes1983graph, borgatti2009network}. Their main methods and techniques have been applied in large network-based problems as OSNs and social media applications (i.e., Facebook, Instagram, Twitter, WhatsApp, LinkedIn, Snapchat, Youtube, Tumblr, Pinterest, Skype, or WeChat to mention just a few)~\citep{lusher2013exponential, carrington2005models}. OSNs are considered complex networks~\citep{vega2007complex} because they present non-trivial topological features. In other words, the connection patterns between nodes will not be random or purely regular~\citep{thai2011handbook}. This section provides a short introduction to some classical, but essential, concepts on graph theory and network analysis in SNA, and their particular application to structural-based problems in OSNs, such as community finding or information diffusion.

\subsubsection{Graph Theory and Network Analytics}
\label{subsec:graphtheory}

SNA examines the structure of relationships between social entities. For example, in mathematics and computer science, graph theory represents social networks as a graph, through nodes connected by edges, where the nodes would be the individuals and the edges would be the relationships that joint them. Nodes and edges compose the graph, a data structure that allows describing the properties of the social network. Graph theory~\citep{west1996introduction, bondy1976graph} is an area of an important contribution to research in data analysis. Through this theory, it is possible to generalize and analyze the existing social interactions between users, and what is their behavior within their contact network~\citep{bello2017detecting, freeman2004development, lara2017measuring}. Currently, the study, analysis and visualization of social networks focus mainly on the association and measurement of interactions between people, groups, organizations, or any set of entities with a relationship between them. From Graph theory, concepts as \textit{betweenness, clique, centrality, cohesion, component, connectivity, degree, diameter, path, transitivity}, and others are used to represent the network structure and topology~\citep{scott1988social, aggarwal2011introduction, wasserman1994social, carrington2005models, barnes1983graph, freeman1978centrality}. Therefore, and from a merely technical point of view, SNA provides both visualization~\citep{freeman2000visualizing, henry2007nodetrix, handcock2008statnet} and mathematical tools~\citep{wasserman1994social, wellman1983network} that allow the analysis and study of non-trivial relationships (patterns, structures, etc.) that can be represented within a network. Following there is a brief description of some basics concepts on Graph Theory and Network analysis~\citep{west1996introduction, bondy1976graph, liu2012survey}:

\begin{itemize}

  \item \textbf{Types of graphs}: a graph $G=(V,E)$ can be defined as a set of vertices, or nodes, denoted by $V=\{v_1, \dots, v_n\}$, and a set of edges $E$, where each edge is denoted by $e_{ij}$, if there is a connection between the vertices $v_i$ and $v_j$, with $v_i,v_j \in V$. Graphs can be \emph{directed} or \emph{undirected}. The graph will be directed if $e_{ij} \ne e_{ji},~\forall i,j$, otherwise the graph is undirected. In some cases, edges have associated one, or more numerical values, called \textit{weights}. In this case, $G$ is a \textit{weighted graph} if exists a function $w: E \to R$ which assigns a real value to each edge ($G(V,E,W)$). In other cases, these attributes can be categorical~\citep{traud2012social, crnovrsanin2014visualization}, or simply signed~\citep{leskovec2010signed, zaslavsky1982signed, 2018-Gmati}. There are a large number of different types of graphs that can be found in the literature, such as: trees and forests (a tree is a graph structure that has not cycles between nodes, a graph that can be represented as a set of disconnected trees is called a forest), spanning trees (for any connected graph is a tree that includes all the nodes of the graph, but not all the edges. The edges included will be those necessary so any two pair of nodes of the tree should be connected. Therefore, for any graph it would be possible to find multiple spanning trees, if a weighted graph is considered, the one with the minimum weight is called the \textit{Minimum Spanning Tree}: MST). Other interesting graphs are Steiner trees, Complete graphs, Planar graphs, Bipartite graphs, Regular graphs, or Bridges among many others~\citep{ west1996introduction, liu2012survey, gross2004handbook}.

  \item \textbf{Graph \& Network models}: graph models are one of the most powerful and useful tools to represent diverse types of data. This kind of models is particularly suitable for network-based domains and problems. Any network (or graph) model has as the main goal to reproduce or mimic the main characteristics from the real network considered. To determine these characteristics, several measures are identified and assessed to ensure that these measurements are consistent with a real-world network. Three network attributes are commonly used to measure this behavior: \textit{degree distribution}, \textit{clustering coefficient}, and \textit{average path length}~\citep{ west1996introduction, gross2004handbook}. Where degree distribution denotes how node degrees are distributed across a network. The clustering coefficient measures the transitivity of a network. And finally, the average path length denotes the average distance (shortest path length) between any pairs of nodes. The fundamental graph or network models which can be found in the literature are the following:
  
  \begin{itemize}
    \item \textit{Real-World networks}~\citep{newman2003social}. One of the basic characteristics shared by this type of networks is the fact that connections (or ties) between nodes depends on a specific probability. In real-world networks, such as social networks, this means that the probability of a tie between two nodes is higher if there exists any kind of relationship (friendship, professional, etc.) between them. 
    The degree distribution values in a real-world network, such as social networks, follow a \textit{power-law distribution}. However, as it was demonstrated in~\citep{newman2003social}, social networks differ from other types of networks, mainly because social networks are more easily divided into communities than non-social networks. Therefore, this characteristic affects the degree distribution, usually with degrees being positively correlated, and a high level of clustering.
    
    \item \textit{Random networks}~\citep{bollobas2001random, newman2002random}. From a theoretical perspective, random networks, most commonly known as random graphs, belong to the intersection between graph theory and probability theory~\citep{bollobas2001random}. Any random graph can be described as a probability distribution, or as a random process, which will be used to generate the connections between any pair of nodes in the graph. Therefore, to generate a random graph, and starting from a set of $n$ isolated nodes, it is only necessary to randomly add (following a predefined probability distribution) successive edges between any pair of nodes. Random graphs are commonly used to compare the structure and properties of any graph. The two most relevant models to generate random graphs are the Erd\"{o}s-R{\'e}nyi model~\citep{erdHos1960evolution}, and the one proposed by Gilbert~\citep{gilbert1959random}. The main difference between both models lies in how the connection probability (i.e., the probability distribution) is managed by each model. In Erd\"{o}s-R\'enyi model, all nodes and edges share the same probability. This means that in this model, two nodes are chosen uniformly random and the link connecting them is created, whereas, when the Gilbert model is used an independent probability for each edge in the graph is defined.
    
    \item \textit{Small-World networks}~\citep{Watts1998Collective}. In many real world networks, there are two basic properties: the first one is related to the distance between two nodes (which it is usually small) whereas the second one is related to the transitivity, or clustering coefficient of these networks (which is usually relatively high). Erd\"{o}s-R{\'e}nyi random networks usually have a low average path length, so it is possible to move from one node to other in the graph with a few numbers of hops. This property, which is shared by a large number of real-world networks, is often called the \textit{small world property}. This concept was popularized by terms like the `six degrees of separation' between two people, meaning that any two persons are distanced by at most six friendship links. The concept of six degrees of separation comes from the well-known Stanley Milgram experiment. This famed social psychologist, conducted an experiment in 1967, which tracked chains of acquaintances in the United States. In this experiment, people were requested to route packages to a fixed recipient (a stockbroker from Boston, Massachusetts), passing them only through their direct acquaintances. In this experiment, Milgram found that the average number of people needed to send the packages between two the original sender and the destination recipient were five (i.e., six degrees of separation)~\citep{milgram1967small}. However, when real-world networks, such as those related to social networks, are analyzed, it is quite common to find one property that does not appear at Erd\"{o}s-R{\'e}nyi random graphs. This property is related to the transitivity degree, which in these real networks have a higher value than in Erd\"{o}s-R{\'e}nyi random graphs. This property appears due to the social behavior of the network (usually your friends are likely to be friends), which means that the number of triangles (i.e., transitivity in the graph, or the clustering coefficient value) will have a larger value in social networks.
    
    \item \textit{Scale-free networks}~\citep{albert2002statistical}. This kind of networks is commonly defined based on their degree distribution. In these networks, the degree distribution of nodes will follow a \textit{power law} (at least asymptotically). That is, the fraction $P(k)$ of nodes in the network having $k$ connections to other nodes goes for large values of $k$ as $P(k) \sim k^{-\gamma}$, where $\gamma$ is a parameter whose value is typically in the range $2 < \gamma < 3$, although occasionally it may lie outside these bounds~\citep{onnela2007structure}. There exists a variety of scale-free network-modeling algorithms, a well established one is the model proposed by Barabasi and Albert~\cite{Barabasi509}, usually called \textit{preferential attachment}~\citep{barabasi1999mean}, or the Barabasi-Albert (BA) model, and it is defined as follows: ``When new nodes are added to networks, they are more likely to connect to existing nodes that many others have connected to". As it was demonstrated by authors, several natural and human-made systems (such as Internet, the World Wide Web, citation networks, and some OSNs) are thought to be approximately scale-free and contain few nodes (named \textit{hubs}) with unusually high degree as compared to the other nodes of the network (e.g., a singer, an athlete, a film actor/actress, or the extremely popular, influencers, youtubers, gamers, \ldots).
    
  \end{itemize}
  
  \item \textbf{Network metrics}: Once, the most general and simple concepts, and models, from graph theory have been introduced, we can proceed with the definition of some basic metrics, or measures, that are used by graph algorithms. There are excellent books and references to these algorithms, which can be consulted to better understand how they have been designed, and applied to the domain of network analysis~\citep{aggarwal2011introduction, west1996introduction, gross2004handbook, agnarsson2007graph, buckley2003friendly, zafarani2014social, cook2006mining}.
  
  \begin{itemize}
    \item \textbf{Centrality}. Centrality is one of the essential metrics in graph and network theory, this metric is used to asses the relevance, or structural importance, of a node in the network. The centrality measure defines how important a node is in a network. In OSNs, this measure can be used to detect or identify, the most influential people in the network. When centrality is assessed, several measures are used: \begin{enumerate*}[label=(\arabic*)] 
    \item \textit{Degree Centrality}, which ranks nodes with more connections higher in terms of centrality; 
    \item \textit{Eigenvector Centrality} tries to generalize the degree centrality by incorporating the importance of the neighbors (in directed graphs it can be used incoming or outgoing neighbors); 
    \item \textit{PageRank} measure takes into account the value of passed centrality by the number of outgoing links (outdegree) from that node, so this measure gets a fraction of the centrality values of the nodes connected to the node from the source node considered;
    \item \textit{Betweenness Centrality} computes the number of shortest paths that traverse any two nodes in the graph, $v_i$ and $v_j$, nodes with higher betweenness values can be seen as ``bridges" between different subgraphs, this measure is used by some specific community finding algorithms; 
    \item \textit{Closeness Centrality}: the idea of this measure is that the more central the nodes are, the easier will be to reach other nodes. Therefore, the smaller the average shortest path length is, the centrality value of the node will be higher; and 
    \item \textit{Group Centralities} such as Group Degree Centrality, Group Betweenness Centrality, Group Closeness Centrality that generalize the previous centrality-based measures to a group of nodes.
    \end{enumerate*}
    
    \item \textbf{Transitivity} and \textbf{reciprocity} are used to represent \textit{linking behavior} in a network. Transitivity analyzes the linking behavior to determine whether it demonstrates a transitive behavior between three nodes, so at least three edges will be needed to create a triangle. Higher transitivity on a graph results in a denser graph, which in turn is closer to a complete graph. Therefore, it is possible to determine how close graphs are to the complete graph by measuring the transitivity. This can be performed by measuring the \textit{[global] clustering coefficient} and \textit{local clustering coefficient}. The former is computed for the network, whereas the latter is computed for a node.
    
    \item \textbf{Balance} and \textbf{status}. A signed graph is a graph in which each edge has a positive or negative sign. This sign is used in OSNs to represent interpersonal relationships (e.g., such as friends or foes, boss or subordinate, social status). This kind of graph is \textit{balanced} if the product of edge signs around every cycle is positive. In real-world social networks, we expect some level of consistency concerning these interactions. For example, it is more plausible for a friend of one's friend to be a friend than to be an enemy. In signed graphs, this consistency translates to observe cycles (triads, triangles) with three positive edges (i.e., all friends) more frequently than the ones with two positive edges and one negative edge (i.e., a friend's friend is an enemy). \textit{Social balance} and \textit{social status} are used to determine the consistency in signed networks. Social balance theory says that friend/foe relationships are consistent when the transitivity between nodes can be propagated, such as ``The friend of my friend is my friend". Therefore, triangles that are consistently based on this theory are denoted as balanced (whereas, inconsistent triangles are denoted as unbalanced), in these balanced triangles, an even number of negative edges will appear. Social status theory measures how consistent individuals are in assigning status to their neighbors. The idea is simple, and it can be summarized that if a person $X$ has a higher status than $Y$, and this last person has a higher status than $Z$, then $X$ should have a higher status than $Z$. In a signed (directed) graph representation, positive and negative signs will be used to show higher or lower status depending on the arrow direction (if the edge has a positive value, it will mean than outgoing node has a higher status than incoming node, if the edge has a negative value, it will represent the opposite).

  \end{itemize}
  
  \item \textbf{Graph algorithms}: from the graph theory area, there are a large number of algorithms that have been designed to work with graphs to compute certain aspects of them (paths, flows, bridge, etc.). Some of the most frequently used in the area of SNA are: Graph/Tree Traversal, Shortest Path Algorithms, Minimum Spanning Trees, Network Flow Algorithms, Maximum Bipartite Matching, and Bridge Detection, among many others~\citep{even2011graph}.

\end{itemize}

\subsubsection{Community Detection}
\label{subsec:cda}




Community detection problem (CDP) can be defined as the division of the graph into clusters of nodes based on the network structure~\cite{xie2013overlapping}. The main idea behind CDP is that nodes belonging to the same cluster are strongly interconnected whereas they maintain sparse connections to the nodes of other clusters. This problem is similar to the idea of graph partitioning into groups of nodes according to the network topology, where a partition is a division of the graph and it can be easily mapped into a cluster~\cite{clauset2005finding, Fortunato201075, bello2012adaptive, 2019-GonzalezPardo}.

To detect the communities, or clusters, on a graph, there is a wide range of techniques such as random walks, spectral clustering, modularity maximization, or statistical approaches~\cite{Fortunato201075}. This kind of algorithms uses the topology of the graph to create the partitions that are validated by taking into account the density of the resulting sub-graph (i.e., a sub-graph is highly connected), and connections from these nodes to the rest. A good community is the one whose nodes are highly connected and it has few connections to the nodes of other communities~\cite{2004-Kannan}. Considering both connectivity and density, a possible definition of a graph cluster could be a connected component or a maximal clique~\cite{Bomze99themaximum}. This is a sub-graph into which no vertex can be added without losing the clique property (a clique, in an undirected graph, is a subset of the vertices, such that every two distinct vertices are adjacent). 

However, it is not always clear that a vertex should be assigned only to a unique cluster, or to several. For this reason, there are several taxonomies to distinguish the different algorithms and approaches developed to solve CDP, but the two most extended taxonomies are: \begin{enumerate*}[label =(\arabic*)] \item the one that pay attention to the number of communities that a single node can belong (at the same time) and \item the second one, that pays attention to the type of analysis (static vs. dynamic) performed in the graph. \end{enumerate*}

The first taxonomy refers to \textit{non-overlapping}, or \textit{overlapping} algorithms depending on the number of clusters that any node can belong at the same time. In this sense, \textit{non-overlapping} methods are those whose nodes belong only to one community, whereas in \textit{overlapping} approaches the nodes may belong to several communities at the same time. 


\begin{itemize}
  \item \textit{Non-overlapping community finding algorithms}. One of the most well-known algorithms for community detection was proposed by Girvan and Newman~\cite{Girvan2002}. This method is based on the ``edge betweenness'' metric that measures the number of shortest paths that go through a given edge. Using this metric it is possible to isolate the different communities by removing those edges with higher betweenness (these edges can be understood as ``bridges" connecting two different communities). The main drawback of this algorithm is its high complexity, and its computational costs if it is applied to large networks.
  
    Another very popular metric to compute communities is called \textit{Modularity}. Despite the computation of this metric is an NP-complete problem, there is a high number of algorithms able to use it to detect communities in a reasonable computational time. One of these algorithms is the one proposed by Newman~\cite{newman2004fast}. This greedy algorithm is based on an agglomerative hierarchical clustering where the groups of nodes are successively joined together to create larger communities with higher modularity. Due to the sparseness of the adjacency matrix, this algorithm involves a high number of operations when this matrix is updated. In order to reduce these operations, Clauset et al.~\cite{clauset2004finding} proposed an algorithm that uses the matrix of modularity variations to improve the performance.
    
    Newman reformulated the modularity measure in terms of eigenvectors by replacing the Laplacian matrix with the modularity matrix~\cite{newman2006modularity}, called the spectral optimization of modularity. This improvement has also been used to improve the results of other optimization techniques~\cite{richardson2009spectral, wang2008vector}.
  
    As has been stated, a good community is the one with high density between the nodes belonging to the community. Therefore, a random walker would spend a lot of time inside a community because there would be a high number of paths that can be followed. This is the idea that relies on behind the \textit{random walk} algorithms~\cite{2005-Pons}. Zhou and Lipowsky~\cite{zhou2004network}, based on the fact that walkers move preferentially towards vertices that share a large number of neighbors, defined a proximity index that indicates how close a pair of vertices is to the rest of vertices. Communities are detected with a procedure called NetWalk, which is an agglomerative hierarchical clustering method by which the similarity between vertices is expressed by their proximity.

  \item \textit{Overlapping community finding algorithms} are those algorithms that rely on the idea that nodes can belong to several clusters at the same time. This idea can be found in real-world networks, for example, people in an OSN can belong to different groups or communities. To represent this feature in the algorithms, fuzzy clustering algorithms applied to graphs~\cite{dong2006hierarchical} and overlapping approaches~\cite{bello2012adaptive} have been proposed.
  
    A good review of the state of the art in overlapping community finding algorithm was carried out by Xie et al.~\cite{xie2013overlapping}. This work noticed that for low overlapping density networks, SLPA, OSLOM, Game, and COPRA offer better performance. For networks with high overlapping density and high overlapping diversity, both SLPA and Game provide relatively stable performance. However, test results also suggested that the detection in such networks is still not yet fully solved. A common feature that is observed by several algorithms in real-world networks is the relatively small fraction of overlapping nodes (typically less than $30\%$), each of which belongs to only 2 or 3 communities. Other approaches based on genetic algorithms, multi-objective GA, and evolutionary strategies can be found at~\citep{bello2012adaptive, bello2014evolutionary, bello2018multi}.
\end{itemize}

The second taxonomy, which can be used to categorize the different algorithms and approaches is the one that takes into account how the variable '\textit{time}' is integrated into the model. In this sense, it is possible to talk about \textit{static} and \textit{dynamic} community finding algorithms.

\begin{itemize}
  \item \textit{Static community finding algorithms} refers to those algorithms and approaches that do not take into account the evolution of the OSN, i.e., the variable '\textit{time}' is not modeled into the system. The algorithms previously described are static, which means that the network is modeled into a single snapshot that contains all the information regarding the OSN. These algorithms present some advantages and drawbacks. On the one hand, static community finding algorithms are quite easy to apply to any problem, because no changes will occur in the network during the algorithm execution. On the other hand, and as one of the drawbacks, this kind of non-temporal models makes that the results may not be very representative, because OSNs are continuously changing due to the high variability in the number of users (nodes) and interactions (edges) between them.
  
  The different algorithms that belong to the static community finding problem can be grouped in four different categories depending on the scope of the corresponding algorithm. In this sense, there are node-centric, group-centric, network-centric and hierarchy-centric algorithms~\cite{2010-Tang}:
  
  \begin{itemize}
    \item Node-centric community finding methods: in this case, each node of the network must satisfy the properties of mutuality, reachability and degrees.
    \begin{itemize}
      \item \textit{Mutuality property} relies on the concept of the clique, which is the maximal complete subgraph of three or more nodes in such a way all of them are adjacent to each other. Finding the maximum clique in a network is an NP-hard problem, for this reason, it is quite popular to develop algorithms able to find approximate solutions. One of these algorithms is the one proposed in~\cite{2002-Abello}, where each time a subset of the network is analyzed. In this algorithm, a greedy-search procedure is executed to find the different cliques in each subnetwork. Then, the maximal clique found is used as a lower bound in a pruning process, i.e., if the maximal clique contains $x$ nodes, those nodes whose degree is lower than $x$ can be removed. This process is repeated until the network is simplified to a reasonable size.
      \item \textit{Reachability property} considers that two nodes may belong to the same community if there is a path connecting them. This property assigns the nodes belonging to the same connected component to the same community. The advantage of this property is that it can be computed in $\mathcal{O}(n+m)$ time, but real-world networks are composed by a big component whereas the majority are singletons and small communities~\cite{2010-Kumar}. For this reason, the identification of communities in the small components is straightforward but some efforts must be done to detect the communities contained in the biggest connected components. One way to do that is by finding the \textit{k-cliques}. A k-clique is a maximal subgraph where the largest geodesic distance between two nodes is less or equal to \textit{k}, i.e., $d(i,j) \leq k~\forall v_i,v_j\in V_S$, where $V_S$ is the set of nodes in the subgraph $S$.
      \item \textit{Nodal degree property} establishes that the nodes of a group must be adjacent to a relatively large number of group members. In this case there are two different structures studied: \textit{k-plex}~\cite{2011-Balasundaram} and \textit{k-core}~\cite{2006-Dorogovtsev}. The former is a subgraph composed of $n_s$ nodes and each node is adjacent to no less than $n_s-k$ nodes in the subgraph. The latter is a subgraph composed of $n_s$ nodes and each node is connected to at least $k$ members.
    \end{itemize}
    \item Group-centric community finding methods. In this category falls all the algorithms and methods, that consider the connections inside the community as a whole. These algorithms are usually known as \textit{density-based} algorithms and are based on the concept of $\gamma$-dense subgraphs, or quasi-clique~\cite{2002-Abello}. A subgraph $G_S=(V_S,E_S)$ is $\gamma$-dense if:
    \begin{center}
      \begin{equation}
        \frac{E_S}{V_S(V_S-1)/2} \geq \gamma
      \end{equation}
    \end{center}
    
    The work performed by Abello et al.~\cite{2002-Abello} uses a GRASP algorithm to find a maximal quasi-clique. This procedure starts the quasi-clique with the vertex with the largest degree in the network. Then, iteratively, the quasi-clique is expanded with those nodes that are likely to contribute to a larger quasi-clique.
    \item Network-centric community finding methods consider all the connections of the network, instead of only the connections of the community as Group-centric methods do. In this type of algorithms, the different nodes are grouped in sets of disjoint communities according to a specific quantitative criterion.
    \begin{itemize}
      \item \textit{Group based on Minimum-Cut}. According to this criteria, a community is defined as a subset of nodes $C\subset V$, such that $\forall v \in C$, $v$ has at least as many edges connecting to nodes in the same community as it does to vertices in $V\setminus C$. In~\cite{2000-Flake} authors showed that the community can be found via $s-t$ minimum cut, where $s$ is the source node in the community and $t$ is the sink node outside the community.
      
      \item \textit{Group based on Modularity}. In this case, the structure of the community is compared against a random graph, more precisely the modularity defines how likely the community structure is created at random. Considering that $A$ is the adjacency matrix, $s_i$ denotes the community that node $v_i$ belongs to, and $d_i$ represents the degree of vertex $i$, the modularity is computed as follows:
      \begin{center}
        \begin{equation}
          Q= \frac{1}{2m}\sum_{ij}\left[ A_{ij} - \frac{d_id_j}{2m} \right] \delta (s_i,s_j)
        \end{equation}
      \end{center}
      \noindent where $\delta(s_i,s_j) = 1$ if $s_i=s_j$. In general, the goal is to find those communities that maximize $Q$, but a negative value for the modularity indicates that vertices are assigned to bad communities.
      
      \item \textit{Group based on Latent Space Model}. The idea is to build a latent space~\cite{2002-Hoff,2007-Handcock} with the nodes of the network in such a way those nodes with dense connections are close to each other. These models assume that the interactions between nodes depend on the position of the nodes in the latent space.
    \end{itemize}
    \item Hierarchy-centric community finding methods try to build a hierarchical structure of communities by taking into account the structure of the network. These methods are quite similar to those from hierarchical clustering research field~\cite{2012-Murtagh,2005-Heller}
    \begin{itemize}
      \item \textit{Divisive hierarchical clustering} starts dividing the whole network into several disjoint sets, and then each set is split into smaller ones until all subsets contain only one node. A popular algorithm in divisive hierarchical clustering is the one based on edge betweenness~\cite{newman2004finding}. In this case, the algorithm removes those edges with higher betweenness in such a way the different communities are isolated.
      
      \item \textit{Agglomerative hierarchical clustering} corresponds to the opposite approach. In this case, the algorithms and methods start assigning each node to one independent cluster. In each iteration, clusters are merged into a larger one according to a specific metric. The most popular metric used in this type of algorithms is modularity~\cite{clauset2004finding}. The idea is to merge two communities if the resulting community improves the modularity. The process is repeated until there are no changes in the modularity. The main problem with this approach is that it generates many imbalanced merges with result in high computational cost~\cite{wakita2007finding}. 
    \end{itemize}
  \end{itemize}

  \item \textit{Dynamic community finding algorithms} refer to those approaches that incorporate the variable \textit{time} into the model. Working with time broadens the range of possible analysis. When time is involved in the model, not only the community structure of the network at any given time can be analyzed, but also their dynamics, i.e., their evolution over time. Regarding community dynamics, one could be interested in analyzing the life cycle of a community, when it appears for the first time, when it grows, when it splits into several communities \ldots etc. Likewise, one could be interested in finding communities that persist over time or dividing a network into periods where the community structure is stable. This variety of analysis, with different outputs and goals, are all encompassed inside the dynamic community finding literature.


Embedding the 'time' variable into a network model is not a trivial task. Rossetti and Cazabet presented in~\cite{rossetti2018community} several models of growing complexity and precision. The first, and simplest model, is the \textit{Aggregate} one. This model consists of generating a weighted graph, where the weights represent the number of times an interaction has occurred. Even though the \textit{Aggregate} model allows doing some types of dynamic analysis, like tie strength estimation, it is very limited and does not capture the network dynamics. Consequently, the \textit{Snapshot} and \textit{Temporal Networks} models were proposed. The former introduces dynamic into a network by generating an ordered sequence of graphs, where each graph represents the state of the network at a given point in time. The latter avoid doing any aggregation at all, and represents the network as a set of timestamped nodes and edges that precisely define when an element appear and disappear from a network. These models allow more expressivity at the expense of requiring more complex analysis. 

Using both the \textit{Snaphsot} and the \textit{Temporal Networks} models entails a series of new challenges~\cite{rossetti2018community, Cazabet2014}. For instance, when using the \textit{Snapshot} model, the granularity of the network (the frequency at which the snapshot are taken) need to be decided. This is not a trivial task and depends on the context of the problem. Selecting an appropriate granularity is of utmost importance as the selection of this threshold can drastically affect the results obtained. Besides, the communities found on adjacent snapshots of the network need to be related to each other to be able to analyze their dynamics (growth, merge, \ldots etc). Finally, another challenge related to community finding that must be addressed is the problem of instability. Usually, community detection algorithms are unstable, many of the efficient community detection algorithms are stochastic, when we observe an event between two snapshots, we cannot be sure if they correspond to real events or an effect due to the stochastic nature of community finding methods. However, when using the \textit{Temporal Networks} model we face similar problems. For instance, community events (growth, merge, \ldots etc) occurring during the evolution of the network still need to be identified. Moreover, it is difficult to ensure long-term coherence when using \textit{Temporal Networks}. Due to the nature of the \textit{Temporal Network} model, communities are continuously being updated over time. In the long run, this could lead to significant differences in the communities found compared to the static state of the network at a given point in time. 

Several taxonomies have been proposed in the literature to categorize dynamic community finding methods. However, in this work, we will use a different one. First, the methods will be categorized according to the model used. Then, these methods will be further categorized according to the analysis followed. Following this taxonomy, we can divide the methods in those that work with \textit{Snapshots} and the ones that work with \textit{Temporal Networks}. As aforementioned, each category will be further divided into more sub-categories depending on their end goal. On the one hand, methods in the snapshot category can be further subdivided into four subcategories (Snapshot Community Tracking, Snapshot Community Detection, Consensus Community Detection, and Change Point Detection). On the other hand, methods in the temporal network category can be further subdivided into three subcategories (Community Structure Update, Temporal Community Tracking, and Persistent Community Finding).

\begin{itemize}
  \item Snapshots-based dynamic community finding methods:
  \begin{itemize}
    \item \textit{Snapshot Community Tracking}: methods in this subcategory try to characterize a community life cycle over the network evolution. They split the community detection method into two steps: \begin{enumerate*}[label=(\arabic*)] \item uncover communities in each snapshot using some other method; \item track communities among consecutive snapshots to define their life cycle\end{enumerate*}. To do so, Palla et al. initially defined the dynamics of a community using six operations (Growth, Contraction, Merging, Splitting, Birth, and Death)~\cite{palla2007quantifying}. Those operations were lately extended by Cazabet et al. with the addition of a `Resurgence' operation~\cite{cazabet2012using}. These methods characterize a community as a sequence of consecutive events that describe what happens between consecutive snapshots. Some relevant works in this category are~\cite{palla2007quantifying, asur2009event, greene2010tracking, brodka2013ged, tajeuna2015tracking, wang2018tracking, tajeuna2018modeling}. 
    \item \textit{Snapshot Community Detection}: these methods are focused on finding suitable communities structure for each snapshot of the network by processing the snapshots in their natural order using previously found results to guide the procedure. While some methods in this subcategory try to stabilize the community structure over time~\cite{chakrabarti2006evolutionary}, others try to reduce the computational effort required using evolutionary strategies~\citep{panizo2019multi, panizo2018genetic} or combining bio-inspired meta-heuristics and novelty search strategies~\citep{osaba2019combining}. Some relevant works in this category are~\cite{chakrabarti2006evolutionary, folino2010multiobjective, lin2009analyzing, lin2011community, tang2008community, gorke2013dynamic, alvari2016identifying, zhou2018multiobjective}. 
    \item \textit{Consensus Community Detection}: methods in this category focus on finding one community structure that fits all the snapshots of the network. These methods process all the snapshots of the network simultaneously in a single process finding communities not only composed of nodes from the same snapshot but also composed of nodes from different ones. Some relevant works in this category are~\cite{tantipathananandh2007framework, yang2011detecting, mucha2010community, mitra2012intrinsically, xu2014dynamic, liu2018global}.
    \item \textit{Change Point Detection}: the methods in this subcategory are related to the change point detection problem. They focus on splitting a network into different homogeneous periods separated by dramatic changes while also finding a suitable community structure for each period. Some relevant works in this category are~\cite{sun2007graphscope, chan2008discovering, duan2009community, aynaud2011multi, peel2015detecting, hulovatyy2016scout}. 
  \end{itemize}
  \item Temporal Networks-based dynamic community finding methods:
  \begin{itemize}
    \item \textit{Community Structure Update}: the methods in this subcategory track community structures in an iterative way. Each change of the network is processed in a streaming way updating the actual community structure when a change demands it. This subcategory includes methods that define a set of rules to update a community given a change or methods that do dynamic optimization of some quality metric. Relevant works in this category are~\cite{cazabet2010detection, nguyen2011adaptive, li2012cdbia, zakrzewska2015dynamic, rossetti2017tiles, boudebza2018olcpm}.
    \item \textit{Temporal Community Tracking}: the methods in this category adapt the community structure to the changes of the network while tracking events on the dynamic communities (birth, growth, split ...etc). Some relevant works in this category are~\cite{falkowski2008studying, bhat2012octracker}.
    \item \textit{Persistence Community Finding}: the methods in this category try to find persistent structures in the whole network evolution. Some methods required a fixed duration to be used as threshold and others can find structures with arbitrary duration. Some relevant works in this category are~\cite{viard2016computing, himmel2016enumerating, li2018persistent, viard2018enumerating, banerjee2019enumeration, qin2019mining}.
  \end{itemize}
\end{itemize}

\end{itemize}




\subsubsection{Information Diffusion Models}
\label{sec:informationDiffusion}

OSNs allow any user ($u_i$) to create new information in the network in such a way any user connected to $u_i$ will receive this information (e.g., in Twitter any user can create a tweet, in Facebook create a new post, or upload a photo in Instagram). Nevertheless, a powerful tool of any OSN is that any user connected to $u_i$ is able to send the received information to their own connections. For example, any user can 'retweet' tweets on Twitter, or 'share' a post on Facebook. 

This property, along with the number of connected users to any OSN, makes that any content created may propagate fast on the network reaching a large number of potential readers. This content that is spread extremely fast, and it reaches a huge number of users in a short period, are commonly known as '\textit{viral}' content of the OSN. Lots of users try to create viral content to gain popularity on the networks, and some others use this propagation procedure for marketing campaigns.

The classic example of marketing campaigns in OSNs is the study of the opinion of customers regarding new products~\cite{2016-Agnihotri,2017-Zhang}, but taking into account the propagation of information in OSNs, the goal is to design marketing campaigns (such as posts, tweets, videos or photos) that reach the larger number of possible users in the shorter period of time~\cite{2008-Thackeray,2015-Ashley}. Another application domain is politics, where different parties use OSN to promote their propaganda~\cite{2016-Howard,2017-Enli,panizo2019describing}. Nevertheless, in the last years, there has been a wide misuse of OSNs. In this sense, ISIS has used OSNs to spread their propaganda and to recruit new members~\cite{lara2017measuring,2015-Klausen,2012-Klausen,lara2017statistical}, or the propagation of fake news~\cite{2017-Allcott,2017-Shu}.

Because of that, many researchers have focused their research on understanding how the information is spread in OSNs. More precisely, researchers have tried to define the different models that describe the diffusion process. In this sense, models can be grouped in two different categories: \textit{explanatory} and \textit{predictive} models~\cite{2013-Guille}. On the one hand, explanatory models are based on the transmission of the epidemic, where there are users infected and users that are susceptible to be infected. On the other hand, predictive models are used to predict how the information will be spread through the network.

Explanatory models consider the diffusion process as an epidemic spread process, where the infection propagates between users in the same way as the information does. These type of models are based on the state of the different users and the models try to extract conclusion from how the users change their states. These states are the following:
\begin{itemize}
  \item \textit{Susceptible} (S): this state represents users that are not infected. In the analogy with the information diffusion, it means any user who has not received the information.
  \item \textit{Infected} (I): it is used to represent those users infected by the virus or the users who have received the information.
  \item \textit{Removed} (R): this state represents an infected user that that has been cured.
\end{itemize}

The classical models that belong to this category are susceptible to infected susceptible (SIS)~\cite{2003-Newman}, susceptible infected removed (SIR)~\cite{2011-Miritello} and susceptible infected removed susceptible (SIRS)~\cite{2014-Liu}. All of them are based on the Susceptible-Infected model~\cite{2001-Pastor-Satorras} that considers two states for the users (susceptible and infected), any susceptible user can be infected and once the user gets infected the state of the user cannot change. Based on this model, SIS, SIR and SIRS models differ in the number of states for each user and the transitions between these states.

\begin{figure}[h!]
\centering
\includegraphics[width=0.5\linewidth]{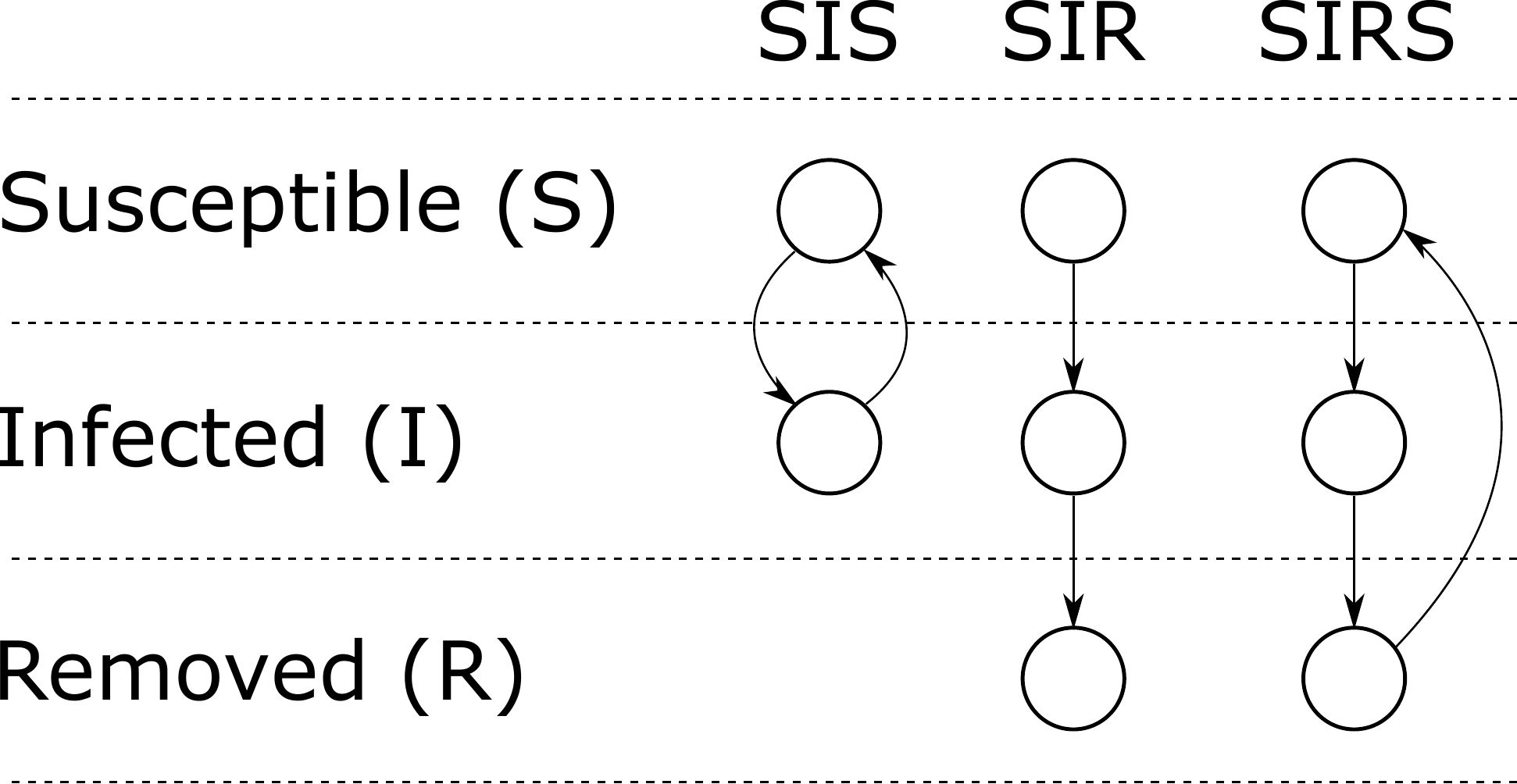}
\caption{Comparison among the different main SI-based models: SIS, SIR, and SIRS.}
\label{fig:si-models}
\end{figure}

The first model is SIS and it considers a daily rate of the cured patients, i.e., an infected user can be cured, and its state will change to Susceptible. The second model was proposed by Kermack and McKendrick in 1927. This model is called SIR, and it introduces a new state named \textit{Removed} that represents those infected users that are cured. The difference between SIS and SIR is that in SIS models any cured infected user is susceptible again (which means that this user can be infected again), but in SIR models an infected user that is cured gets the state of \textit{Removed}, which means that it is immune to the infection. Finally, the third model is called SIRS, and it considers that a cured user can become a susceptible user with a given probability. Fig.~\ref{fig:si-models} shows a graphical comparison of the just explained models.

The second category of diffusion processes are called predictive models, and its most popular, and well-known models are: the independent cascade model (ICM)~\cite{2005-Kempe} and the linear threshold model (LTM). The goal of these models is to predict the future information diffusion process.

On the one hand, ICM was proposed by Goldenberg et al.~\cite{2001-Goldenberg,2001b-Goldenberg} and it is inspired by the theory of interacting particle systems. This model takes into account the probability that an active user infects an inactive one. In this way, given two nodes that are connected in the network: $u$ and $i$ where $u$ is active and $i$ is inactive, the probability of $u$ to infect $i$ will be denoted as $P(u,i)$. There are two hypotheses in this model: the first one is that the probability of a node $u$ to infect node $i$ is independent of the influence of other active nodes connected to $i$. The second hypothesis is that any active node $u$ has only one chance to infect $i$, independently on the result (success or not) node $u$ will never try to infect $i$.

On the other hand, LTM was proposed by Watts~\cite{2002-Watts} and the individuals make a decision based on its neighbors. In this model each individual has a state (\textit{active} or \textit{inactive}) and a threshold ($\phi \in [0,1]$) that will define the activation level of the individual. Initially, there are a small set of users that are active. At each step, if the fraction of active neighbors of a given inactive individual is greater than its threshold, the given individual will change to active. The diffusion process finishes when the number of active individuals becomes stable.

\subsection{Content-based Analysis}
\label{subsec:contentSNA}

 As it has happened since the origin of the Web, the content published by humans is easily understandable by them, however and due to the lack of structure and the multimodality of the information (text, images, video, audio), the automatic analysis of the information has been one of the biggest challenges for those algorithms that must gather and extract knowledge from this data~\citep{poria2016fusing}. Both facts, the lack of structure and multimodality of the information, affect the analysis of information contained in OSNs. This content analysis requires computational methods that transform unstructured content into structured information. One of the most successful areas of research in this context is natural language processing (NLP)~\citep{cambig}. NLP provides a set of methods and algorithms that enable the processing of multimodal information circulating on OSNs, thus allowing unstructured information to be transformed into structured information. Although, other approaches have been used to analyze video and images from OSNs, these methods can be regarded as immature when are compared against NLP~\citep{cha2007tube, weng2009rolenet}. Other successful areas, which have been extensively used to gather knowledge from OSNs, have been Data Science~\citep{carrington2005models} and Big Data~\citep{gandomi2015beyond, wu2014data}. Their main methods and algorithms related to data collection, cleaning, pre-processing or mining, have been used to gather, model and extract patterns from OSNs~\citep{bello2016social, manovich2011trending, tufekci2014big}. The most representative applications of OSN content analysis are user profiling (Section~\ref{subsec:userprofiling}), topic extraction (Section~\ref{subsec:topicextraction}) and sentiment analysis (Section~\ref{subsec:sentimentanalysis}).  

\subsubsection{User Profiling}
\label{subsec:userprofiling}
Content-based SNA mostly focuses on the contents of the interaction between nodes in an OSN to extract topics or opinions (see next sections). Content-based SNA, however, can also be about the information about the nodes themselves, i.e., user profiling.
User profiles are established based on the behavioral patterns, correlations and activities of the user analyzed from the aggregated data using techniques like
clustering, behavioral analysis, content analysis and face detection (the mechanisms used in profiling users vary depending on application and purpose).

Profiling user in OSNs requires data pertaining to the user and the online activities of such user within the OSN. Those activities may depend on one's interest or they can be the effect of some influence on them. 
Pal and McCallum~\cite{palccc} exploited the content of communication messages between users to cluster email recipients into groups. For each user, they built a model that maps keywords and phrases extracted from email messages to the recipients who are likely to receive an email containing those terms. 
Bar-Yossef et al.~\cite{barclu} and Roth et al.~\cite{rotsug} observed how users group their friends when sending email messages. Based on past behavior of user's group communication messages, they developed ranking algorithms that can predict other similar users who can belong to a particular group specified as a seed-set of users. 
De Choudhury et al.~\cite{decinf} proposed an approach to `label' nodes in an OSN as per their roles, e.g., `student', `faculty', or `director'. They applied their approach on an email communication graph, filtering out infrequent email communications below a certain threshold.

Cai et al.~\cite{caimas} proposed a model for mining and identifying the top-k influential bloggers based on parameters such as comments, domain of interest and page link network authority. The developed model can be used for multiple application scenarios. 
Sun et al.~\cite{sunsim} proposed a novel algorithm to recommend influential bloggers based on the observation that the reproduction of blogposts and similar contents is common in blogosphere and this forms implicit links between bloggers. By measuring the text similarity of the blogposts, authors created a link graph between bloggers, and adopted the PageRank algorithm to rank the importance of bloggers.

Akritidis et al.~\cite{akride} proposed a solution for bloggers identification in community blogsites that are both influential and productive. Based on the number of incoming links, two matrices are considered characterizing the bloggers as influential, productive, both, or none. This way, authors can identify blogger activities, temporal patterns and behavioral patterns.
Eunyoung et al.~\cite{mooqua} built a framework for identifying influential and popular bloggers by considering interpersonal similarity, which presents the interaction among bloggers and like-minded readers, and the degree of information propagation, which represents how many readers a blogger has. Their study showed that weighting blog social ties can differentiate influential bloggers from popular bloggers, and what make bloggers influential or popular.
Finally, Cambria et al.~\cite{chaclu} exploited the contents of social interactions to cluster OSNs based on semantics and sentics, that is the conceptual and affective information associated with the interactive behavior of OSN members. In particular, authors created a socio-interaction matrix encoding the concepts discussed by users and applied singular value decomposition (SVD) on it to perform user profiling and topic extraction.

\subsubsection{Topic Extraction}
\label{subsec:topicextraction}
Topic extraction is a technique used for discovering the abstract `topics' that occur in a collection of documents, which is useful for tasks such as text auto-categorization, sentiment analysis but also SNA. Common approaches include mixture of unigrams, latent semantic indexing, LDA, and knowledge-drive methods~\cite{rajcom}. Such methods can be used in combination with SNA to mine user interests in OSNs~\cite{piainf}.

McCallum et al.~\cite{mcctop} described how to determine roles and topics in a text-based OSN by building an author-recipient-topic (ART) model and a role-author-recipient-topic (RART) model. Pathak et al.~\cite{patsoc} proposed a community-based topic model integrating SNA techniques termed community-author-recipient-topic (CART). This model was used to extract communities from an email corpus based on the topics covered by different members of the overall network.  
Later, Wang et al.~\cite{wanmin} proposed an approach to predict the interests of new users or inactive users based on
different social links between them using a random-walk based mutual reinforcement model that incorporates both text and links.
Another work~\cite{waninf} proposed a regularization framework based on a relation bipartite graph, which can be constructed from any type of relationships, and evaluated it on OSNs that were built from retweeting relationships. 

Kang et al.~\cite{kanmod} presented a user modeling framework that maps user-generated content into the associated
category of the news media platform. 
Similarly, Jipmo et al.~\cite{jipfri} proposed a multilingual unsupervised system for the classification of Twitter users'
interests. The system represents tweets and topical interests (e.g., technology, art, sports) as bags of articles and Wikipedia categories respectively, and orders the user interests by relevance, by computing the graph distance between the categories and the articles. 
Faralli et al.~\cite{faraut} proposed a method for modeling Twitter users using a hierarchical representation based on their interests.
This was done by identifying topical friends (a friend represents an interest instead of social relationship) and by associating each of these users with a page on Wikipedia. 
Zarrinkalam et al.~\cite{zarmin} proposed a graph-based link prediction system based on user explicit and implicit contributions to topics, relationships among users, and similarity between topics. 
Finally, Trikha et al.~\cite{tritop} studied the prediction of users' implicit interests based on topic matching using frequent pattern mining. 

\subsubsection{Sentiment Analysis}
\label{subsec:sentimentanalysis}
In recent years, sentiment analysis has become increasingly popular for processing social media data on online communities, blogs, wikis, microblogging platforms, and other online collaborative media~\cite{cambsda}. While most works approach it as a simple categorization problem, sentiment analysis is actually a complex research problem that requires tackling many NLP tasks~\cite{camcls}, including personality detection~\cite{mehrec}, domain adaptation~\cite{banlex,xuuins}, and multitask learning~\cite{yanseg}. It also comes in different flavors depending on granularity of the analysis~\cite{weiasp}, modality adopted~\cite{chafuz} and gender~\cite{mihwha,buktyp}. Applications of sentiment analysis span domains like healthcare~\cite{mahdet,qurmul}, political forecasting~\cite{ebrcha}, tourism~\cite{valsen}, rumors and fake news detection~\cite{akhnoo,reisup} and dialogue systems~\cite{wellea}. 

Sentiment analysis has also been applied to better understand OSN dynamics by looking at the exchange of information between network nodes. This can be useful for trend discovery, user profiling, influencer detection, and study of polarization over a certain topic or political orientation. 
One of the first works in this context was \cite{patkno}, which presented an approach for analyzing knowledge perception in an OSN. Authors provided an evidence-theory based methodology for
constructing and maintaining a knowledge network in an electronic-mail communication environment. 
Later, \cite{bercom} focused on the topic of online radicalization. Authors monitored users and interactions in a specific YouTube group using a combination of sentiment, lexical and SNA techniques. They made a number of interesting observations about the differences in the nature of the discussion and interactions between male and female members of the group.

\cite{grylev} studied the blogosphere's sentiment towards Barack Obama during the 2008 USA presidential elections. Authors used a hybrid machine learning and logic-based framework. The results showed that the classification task in this environment is inherently complex, and learning features that exploit entity-level sentiment and social
network structure can enhance classification. 
\cite{shaapp} proposed an ensemble of sentiment analysis and SNA for extracting classification rules for each customer. These rules represent customer preferences for each cluster of products and can be seen as a user model.
This combination helped the system to classify products based on customer interests. Authors compared the results of their method with a baseline method with no SNA. Experiments on an Amazon meta-data collection showed improvements in the performance of the classification rules
compared to the baseline method. 
More recently, \cite{morcar} proposed an ensemble of sentiment analysis and SNA tools for group decision making. In particular, sentiment analysis was used to model consensus among experts in an OSN and later exploited to automatically generate preference relations, which were used for carrying out the group decision making process.

%% file: sections/sna-applicationDomain.tex
\section{Application domains}
\label{sec:applications}

Due to the number of application domains in OSNs is extremely large, we have selected a subset of them by the scientometric study performed in Section~\ref{sec:scientometric}. This study allowed us to select the following application domains: Healthcare (Section~\ref{sec:app-health}), Marketing (Section~\ref{sec:app-marketing}), Tourism and Hospitality (Section~\ref{sec:app-tourism}) and Cyber Security (Section~\ref{sec:app-cyber}). Also, we have identified two emerging areas of SNA that are described in Section~\ref{sec:emerging}: Politics and Detection of fake news and misinformation. 

\subsection{Healthcare}
\label{sec:app-health}

Over the last years, there has been a huge growth of research papers that study the applicability of OSN on health~\cite{smith2008social,chambers2012social,BERKMAN2000843}. It is widely recognized that social relationships have powerful effects on physical and mental health~\cite{cohen2009can,thoits2011mechanisms,umberson2010social}. With the integration of OSNs into our daily life, new levels of social interaction have emerged, and new possibilities beyond the traditional doctor-to-patient paradigm have arisen. Many patients with different diseases are now using OSNs to share experiences with other patients with similar conditions, providing a new potential source for acquiring knowledge very useful~\cite{Andreu-Perez2015}. In addition, there are numerous health-related behaviors that might easily spread in OSNs, such as obesity, smoking, alcohol consumption, or drug use. Therefore several research studies have been appeared analyzing this effect from the SNA perspective~\citep{smith2008social}. 

The propagation and virality models in OSN have been one of the most used techniques for analyzing health-related behaviors in OSNs. For example, a study carried out by Christakis \& Fowler~\cite{christakis2007spread} used a global network design to analyze how \textbf{obesity} can spread through OSNs in a similar way to a infectious disease. For this purpose, the authors evaluated a densely interconnected OSN of $12,067$ people who underwent repeated measurements over a period of $32$ years (from 1971 to 2003) as part of the Framingham Heart Study. Then, a quantitative analysis of the nature and extent of the person-to-person spread of obesity as a possible factor contributing to the obesity epidemic was performed. This study examined several aspects of the spread of obesity using OSN, including the existence of clusters of obese persons within the network, the association between one person's weight and height gain among his or her social contacts, the dependence of this association on the nature of the social ties (e.g., ties between friends of different kinds, siblings, spouses, and neighbors), and the influence of sex, smoking behavior, and geographic distance between the domiciles of persons in the OSN. The results shown in this study suggests that obesity may spread in OSNs in a quantifiable pattern that depends on the nature of social ties.

Smoking and alcohol consumption among adolescents are prominent risk behaviors for Health in the United States~\cite{us2012preventing}, and there is also substantial evidence that adolescents' use of tobacco and alcohol is highly associated with their friends' use~\cite{trucco2011vulnerability}. OSNs provide a mechanism for adolescents to connect with friends instantaneously, and several research work has been focused on toward uncovering the risks associated with the usage of OSNs, such as the display of inappropriate content like sexual references and substance use~\cite{livingstone2008taking} or tobacco advertisements~\cite{freeman2012new}. Following this research line, Huang et al.~\cite{HUANG2014508} carried out a study to investigate peer offline and online friendships, for determining how online activities with friends might broker the peer influence processes, by either encouraging or hindering the influence of peer risk behaviors on adolescent smoking and alcohol use. For this purpose, a friendship network data about adolescent social media use and risk behaviors, were collected from $1,563$ 10th-grade students across five Southern California high schools, and measures of online and offline peer influences were calculated and assessed using linear regression models~\cite{HOFFMAN2015451}, which were fitted to test the effects of online activity with friends on smoking and alcohol use indicators. The results of the analysis showed, that the frequency of adolescent using OSNs and the number of their closest friends on the same OSN were not significantly associated with risk behaviors. However, exposure to online pictures of their friends of partying or drinking, was significantly associated with both smoking and alcohol use. Whereas adolescents with drinking friends had higher risk levels for drinking, adolescents without drinking friends were more likely to be affected by higher exposure to risky online pictures. Other study on the extent of the person-to-person spread of smoking behavior was introduced by Christakis and Fowler~\cite{christakis2008collective}. The authors analyzed a densely interconnected OSN of $12,067$ people assessed repeatedly from 1971 to 2003 as part of the Framingham Heart Study, applying network analytic methods and longitudinal statistical models to detect clusters of smokers and nonsmokers in the network. The findings of this analysis show that smoking behavior spreads through close and distant social ties, groups of interconnected people stop smoking in concert, and smokers are increasingly marginalized socially.

There are other works that use data from the OSNs to extract valuable information. For example, the text messages posted in OSNs offer new opportunities for the real-time analysis of expressed mood and social behavior patterns~\cite{larsen2015we}. The analysis of these additional types of information can be very useful to the early identification, assessment, and verification of potential public health risks~\cite{paquet2005epidemic}, and the timely dissemination of the appropriate alerts (public health surveillance)~\citep{bello2015survey}. For instance, Twitter users may post about an illness, and their relationships in the network can give us information about whom they could be in contact with. Furthermore, user posts retrieved from the public Twitter API can come with GPS-based location tags, which can be used to locate the potential centre of disease outbreaks. Therefore, a high number of research works have already appeared, showing the potential of Twitter messages to track and predict outbreaks. Based on this idea, a document classifier to identify relevant messages was presented in Culotta et al.~\cite{culotta2010towards}. In this work, Twitter messages related to the flu were gathered, and then a number of classification systems based on different regression models to correlate these messages with CDC statistics were compared; the study found that the best model had a correlation of $0.78$ (simple model regression). Other similar approach was proposed by Aramaki et al.~\cite{aramaki2011twitter} where a comparative study of various machine-learning methods to classify tweets related to influenza into two categories (positive and negative) was carried out. Their experimental results showed that the support vector machine (SVM) model that used polynomial kernels achieved the highest accuracy (F-Measure of $0.756$) and the lowest training time. Bello et al.~\citep{bello2017detecting} focused the research on the detection and tracking of discussion communities on vaccination arising from OSNs as Twitter. Finally, well-known \textit{regression models} were evaluated on their ability to assess disease outbreaks from tweets in Bodnar et al.~\cite{bodnar2013validating}. Regression methods such as linear, multivariable, and SVM were applied to the raw count of tweets that contained at least one of the keywords related to a specific disease, in this case ``flu". The models also validated that even using irrelevant tweets and randomly generated datasets, regression methods were able to assess disease levels comparatively well.

\subsection{Marketing}
\label{sec:app-marketing}

In social media, brands and customers co-exist in such a way that both can interact with each other. On the one hand, companies can use social media to promote new products or to predict what the customers think about the brand or the product. On the other hand, customer can express their opinions and doubts with other customers. For all of that, social media has become the favorite and most popular platform to perform promotional activities to communicate with the targeted customers~\cite{2017-Harrigan,2015-Kohli}.
 
 There is a large number of works that have studied the different aspects related to promotional activities conducted in social media platforms~\cite{2009-Mangold,2012-DeVries,2013-Whitelock,2017-Swani}. For example, it has been studied the relation between how customers perceive and formulate their attitudes towards social media advertising activities, and how these perceptions affects to the efficiency and effectiveness of such activities~\cite{2015-Duffett}. Other work~\cite{2014-Carrillat}, stated that if brands want to generate positive customer attitudes, they have to carefully address hedonic aspects to provide customer pleasure experiences.

 What is clear is the contribution of social media in the generation of the new Word-of-Mouth, \textit{e-WOM} that stands for \textit{electronic word of mouth}. Using social media, customers are able to express their opinions about specific products or brands to many other customers~\cite{2016-Hudson,2016-Viglia,2014-Kim,2011-Chu}. One of the most famous works regarding e-WOM is the one performed by Jansen et al~\cite{jansen2009twitter}. This work studied whether Twitter could become used as an e-WOM advertising mechanism. The conclusion of this work, revealed that $19\%$ of the tweets analyzed mentions a brand, and $20\%$ of these tweets contained expressions of brand sentiments. The conclusions of this work are that \begin{enumerate*}[label=(\arabic*)] \item Twitter reports what customers feel about the brand and its competitors, and \item customers' brand perceptions and purchasing decisions are influenced by social media services.\end{enumerate*} Other study is the one published by Asur and Huberman~\cite{asur2010predicting}. In this case authors used Twitter to forecast box-office revenues for movies. In order to do that, authors built a model based on the rate at which the tweets were created. The results reveals that this model outperforms the market-based predictors.

 The rate at which the tweets, or post, are created also foster the level of interactivity and association with their customers~\cite{2012-Pronschinske,2014-McCarthy} and thus, firms also use social media to contribute to both customer experience and customer relationship management~\cite{2012-Coulter}. In this sense, several authors have studied the usage of social media for the just mentioned purpose~\cite{2013-Malthouse,2014-Trainor,2016-Hudson,2016-Agnihotri} and the general conclusion is that social media is a good tool to help organizations to sustain their relationship with their targeted customers. Nevertheless, research community has also identified that not all OSNs contribute in the same way. Moore et al.~\cite{2013-Moore} concluded that this role of social media in forming customers' relationship with brands could be different according to the kind of platform used: Facebook or Twitter. In this sense, Pereira et al.~\cite{2014-Pereira} noticed that though customers on Facebook are enthused to follow brand accounts, they are less interested in keeping contact with them or re-sharing their content in their own page. Similar research were conducted on Twitter by Kim et al.~\cite{2014-Kim}, where based on the data collected by the authors, the conclusion is that brand re-tweeters show an extent level of brand trust, brand identification and community commitment.

\subsection{Tourism \& Hospitality}
\label{sec:app-tourism}

Tourism and Hospitality is other area that can take advantage of the usage of social media to extract valuable information~\cite{2014-Munar,2018-Li}. On the one hand, hotels and tourism destination use social media with Marketing purposes~\citep{sigala2012social,minazzi2015social}. On the other hand, it has been probed that half of tourists change their travel plans after studying their trip on social media~\cite{2012-Bennett}.

Tourism research has paid attention to user generated contents on social media to extract some valuable information. In this sense, there area two different data sources taken into account: \begin{enumerate*}[label=(\arabic*)] \item online textual data, published by customers on social media, and \item online photo data.\end{enumerate*} The research procedure will depend on the different source of data analyzed (text, images or both), which directly will affects to the algorithms that can be applied, and to the knowledge and patterns that can be extracted~\citep{leung2013social,goh2013social,wilson2012hospitality}.

As in the Marketing area, tourists use OSNs to express their opinions about the different places, their experiences, and their satisfaction and dissatisfaction about tourism products. Analyzing these data, researchers have discovered relevant attributes of tourist satisfaction~\cite{2016-Xu,2017-Guo}, the relations between tourist satisfaction and other related factors, such as guest experience and competitive position~\cite{2015-Xiang}, or how the users use the OSNs for evaluating and improving the e-WOM hotels~\cite{2009-Doh,2015-Xiang}. Nevertheless, the majority of the research works published are focused on investigating the customers opinion and customers experience. In this sense, Guo et al.~\cite{2017-Guo} used LDA to discover the aspects that influence the customers satisfaction. 
Similarly, Poria et al.~\cite{porlda} proposed Sentic LDA, an affective version of LDA based on SenticNet~\cite{camnt5}, which leverages the semantics associated with words and multi-word expressions to extract the polarity associated to tourism-specific aspects such as accommodation, entertainment, food, and transportation~\cite{gueund}.
Other authors used sentiment analysis to extract tourists attitude and opinion toward tourism products such as hotel services~\cite{2017-Hu}. The work published by Bordona et al.~\cite{2016-Bordogna} used clustering to group together those trips with similar \verb|geoslotID| based on the geo-tagged messages in Twitter.

The second data type, which has been used to analyze tourism in social media, is the photos published by tourists. In this sense, the most extended approach is to analyze the metadata associated with the photo, instead of working with the photo itself. Using, for example, the geo-tag information researchers have been able to explore the tourists behavior in Hon Kong~\cite{2015-Vu}, to create a recommendation system that provides travel paths~\cite{2010-Lu}, or to select photo elements from the viewers' perspective and assist marketing organizations~\cite{2018-Deng}. There are other metadata associated with the photos, for example, the user id and the photo id, the date and time when the photo was taken and uploaded, the geographical information expressed by the latitude and longitude, and other information such as the title, descriptions or tags of the photo. On the one hand, there is a wide variety of papers that uses these metadata and clustering techniques to group the different photos analyzed. In this domain it is quite popular to use a density-based spatial clustering to solve the problems related to centroid-based approaches~\cite{2017-Hu,2013-Majid}. On the other hand, other researchers focus their work on travel trajectories, i.e., the sequence of tourism spots and time intervals between them to recommend travel plans for tourists. For example,~\cite{2010-Lu} suggests different routes taking into account the quality and popularity of the routes. Other approach is the one followed by Vu et al. in~\cite{2015-Vu}, where authors applied Markov models to predict the next tourism spots based on the current location of the tourists.



\subsection{Cyber Security}
\label{sec:app-cyber}

SNA can be used to detect and apply different strategies to support law enforcement agencies in the fight against cyber-crime and cyber-terrorism. Criminals, and terrorists, use OSNs, such as Twitter or Facebook, due to the huge number of users that connects everyday to these networks. Moreover, the internal structure of these networks, make that any published message propagates very fast, reaching a high number of potential readers. In this domain, the purpose of SNA techniques and algorithms is to extract the different patterns of criminals and terrorist in OSNs. More precisely, the main goal is to detect and discover crimes and their relations with criminals.

Regarding crime analysis, huge efforts have been made to facilitate the communication between citizens and government agencies. Initially, this communication was roughly performed through telephones, or face-to-face meetings. Then, the information provided by the citizens was saved, or transformed into written text and then archived in a digital format. In order to automate and facilitate crime analysis, Chih-Hao and Gondy~\cite{ku2014decision} designed a decision support system that combines NLP techniques, similarity measures, and classification approaches. Filtering reports and identifying those that are related to the same or similar crimes can provide useful information to analyze crime trends, which allows for apprehending suspects and improving crime prevention.

Other works have been focused on the prediction of the different crimes by trying to discover hidden patterns of criminal behavior. In this sense, it is quite popular geographic knowledge discovery techniques that can be used to discover these patterns that may help in detecting where, when, and why particular crimes are likely to happen. In order to do that, Phillips et al.~\cite{phillips2012mining} presented a crime data analysis technique that allows for discovering co-distribution patterns between large, aggregated, heterogeneous datasets. Authors modeled the aggregated dataset as a graph that stores the geospatial distribution of crimes within given regions. Then, these graphs were used to discover those regions with similar geospatial distribution characteristics.

A different approach is the one followed by Chainey et al.~\cite{chainey2008utility}. This approach is based on the visual identification of regions where the crimes can be produced by using the \textit{hotspot mapping}. This technique is used to predict where crime may happen, using data from the past to inform future actions. Each crime event is represented as a point, allowing for the geographic distribution analysis of these points. A number of mapping techniques can be used to identify crime hotspots, such as: point mapping, thematic mapping of geographic areas, spatial ellipses, grid thematic mapping, and kernel density estimation (KDE), among others. In the just mentioned work, authors performed a comparative study of these mapping techniques and the results revealed that KDE was the most successful technique. This conclusion was also extracted from the work performed by Gerber~\cite{gerber2014predicting}, where KDE was used to automatically identify discussion topics across a city in the United States by applying linguistic analysis and statistical topic modeling over a spatio-temporally tagged tweets extracted from Twitter.

Different approaches are the ones applied to analyze the terrorists' behavior in OSNs. From the last years, it is quite simple to realize the ability of terrorist groups in the usage of OSNs for their own benefit~\cite{UNODC2012,Thompson2011}. Terrorist organizations use OSNs to promote their ideology, and to recruit individuals to their cause. Usually, the first conversations start in the most famous OSNs such as Twitter, Facebook, or Instagram, and then they continue using private message with the target individuals. In order to disconnect these radicalization channels, governments, organizations and social media platforms are continuously searching social media accounts that can be associated with such terrorist groups to block them.

From the research community, a wide range of works has been done in order to help law enforcement agencies. In this paper we are going to talk the four main application domains related to counter-terrorism in OSNs. The first one is focused on the identification and understanding of the language used, or the definition of the linguistic markers. The key idea is the analysis of the text published in the OSNs in order to determine whether the corresponding account belongs to a terrorist or, at least, a supporter. Cohen et al.~\cite{2014-CohenEtAl} discussed about the possibility of detecting some of the linguistic markers defined by Meloy et al.~in~\cite{2012-MeloyEtAl}. More precisely, Cohen et al. discussed about the detection of \textit{leakage}, \textit{fixation} and \textit{identification} warning behaviors because these are the ones have the greatest potential to be discovered with text analysis methods. In a more practical way, Torregrosa et al.~\cite{2019-TorregrosaEtAl} compared the tweets published by pro-ISIS Twitter accounts against the text published by random Twitter users. To analyze the terminology of the tweets they have used Linguistic Inquiry Word Count (LIWC) software. Experimental results reveals that pro-ISIS accounts publish tweets using the third person plural pronouns, they use more words related with death, certainty, and anger, and more negative language than random accounts. Some differences in the language was also evinced by Lara-Cabrera et al.~\cite{lara2017measuring}. They also compared the tweets published by pro-ISIS accounts and random Twitter accounts, using a set of keywords that was extended with different synonyms, and taking into account the stem of the words. Experimental results revealed that metrics defined to measure the indicators performs well in the tasks of identifying those accounts that uses a radical vocabulary. Once the user has been identified as terrorist, or supporter, it is possible to study how this user influence the other users in the OSN \cite{2019-Fernandez}.

All these works, help to highlight those social media accounts that shows a radical behavior based on the content of the tweets. Other approaches model the OSN in a graph and use connections to detect the critical node~\cite{2018-Lalou}. For example, Gunasekara et al.~\cite{2015-Gunasekara} uses the betweenness centrality of the nodes to detect the critical nodes of the graph. The main problem with the betweenness centrality metric is that the metric is really expensive from the computational point of view, it runs in $O(nm)$. In order to alleviate this problem, other works uses heuristic or bio-inspired approaches such as Lozano et al.~\cite{2017-Lozano} that integrates an updated procedure of betweenness centrality metric with an artificial bee colony algorithm.

\subsection{Emerging Areas}
\label{sec:emerging}
Finally, and out of the scope of our scientometric analysis carried out in Section~\ref{sec:scientometric}, we have decided to make a brief analysis of the state of the art in some emerging areas, that are currently experiencing an increasing interest in the area of OSNs. These emerging areas are directly related to recent, but highly societal demanding topics, such as politics and detection of fake news and misinformation, or the integration of multimedia information. With the irruption of 5G and IoT technologies~\citep{singh2017survey, araniti2016context}, it would be expected that in the next years the concept or social Internet of Thinks~\citep{atzori2012social} will generate a huge interest in the field of SNA allowing to generate new kind of intelligent services to end-users through the combination of Machine Learning, Artificial Intelligence and IoT methods. We strongly think that these areas could be in a near future high activity niches, for the research communities involved in areas as SNA, Data Science or Big Data.

\subsubsection{Politics}

The Arab Spring in 2011, and both Obama's campaigns (in 2008 and 2012), marked the beginning of how social media might affect citizens' participation in political life~\cite{2011-Howard}. Since these dates, politicians, citizens and researchers have expressed their interest in how they can take advantage of using social media to participate in political life. In this sense, politicians use social media to attract supporters, and people have been using it to express their political views and opinions about various leaders and issues. In fact, Boulianne~\cite{2015-Boulianne} studied whether social media use, and participation in elections, were correlated. Paying attention to the metadata, more than $80\%$ of the coefficients showed a positive relation.

The research community has used social media to study a wide variety of problems. The most remarkable are the following: \begin{enumerate*}[label=(\arabic*)] \item hate speech detection~\citep{davidson2017automated,schmidt2017survey}; \item topic opinion, or political polarization~\citep{panizo2019describing,conover2011political}; \item community finding problems~\citep{bello2016social,papadopoulos2012community}; and \item information exchange and information diffusion~\citep{Guille-SIGMOD-13,bakshy2012role}.\end{enumerate*} The majority of the research works start with the same idea of analyzing the text of the posts, or comments, and using the results of this analysis to perform a second analysis more focused on the problem to be solved.

Hate speech is used in those posts or comments that defames, belittles, or dehumanizes a class of people on the basis of certain inherent properties such as race, ethnicity, gender, or religion. Due to its goal, it is critical to design systems that automatically classify the posts or comments based on its content and determine whether the corresponding post contains hate speech or not~\citep{davidson2017automated,schmidt2017survey}. In M\"{u}ller and Schwarz demonstrated that there is a significant correlation between increased German hate speech on social media, and physical violence towards refugees in Germany~\cite{2018-Muller}. One recent work that tries to identify hate speech is the one published by Jaki and De Smedt~\cite{2018-Jaki}, where authors tries to understand what disparaging verbal behavior from extremist right wing users looks like, who is targeted and how. In order to do that, they analyzed $55.000$ right-wing German hate tweets from August 2017 to April 2018. The proposed method is able to detect right-wing hate speech with $84\%$ accuracy. Other relevant work in this domain is~\cite{2015-Burnap}, where authors used several classifiers to detect hateful and antagonist content in Twitter. Although the individual results for each classifier reduce false positives and produced promising results regarding false negatives, the combination of classifiers into an ensemble classification approach seems to be the most suitable method.
 
There are other works that are focused on the analysis of the content of the tweets, or posts. In this case, we can talk about topic opinion or political polarization. Although these concepts have different goals, the procedure followed is quite similar. In both cases, works try to extract some valuable information from the content of the tweets. On the one hand, if the goal is to understand what the citizens think about specific topics, we talk about \textit{topic opinion}. 
On the other hand, the goal of \textit{political polarization} works is to discover the alignment of the citizens with the corresponding parties. In this sense,~\cite{2013-Boutet} belongs to topic opinion family. Authors analyzed $1,150,000$ messages from about $220,000$ users to define the characteristics of the three main parties in the 2010 UK General Election and highlight the main differences between parties. In a first analysis they realized that: \begin{enumerate*}[label=(\arabic*)] \item the retweet structure is highly clustered according to political parties, \item users are more likely to refer to their preferred party and use more positive affect words for the party compared with other parties, and \item the self-description of the users can reflect the political orientation of users.\end{enumerate*} Based on these evidences, authors developed a classification method that uses the number of tweets referring to a particular political party, and its semantic content to estimate the overall political leaning of the user. The experimental results achieved an accuracy of $86\%$ for classifying the users' political leanings. Other study~\cite{2014-Gruzd} tries to determine whether the Twitter users can be grouped around the different parties during the elections. To do that, they analyzed around $6,000$ tweets published by $1,500$ Twitter users during the 2011 Canadian Federal Election. Authors concluded that Twitter usage is likely to further embed partisan loyalties during electoral periods rather than loosen them. Therefore, it seems that partisan are closer to their corresponding parties during electoral periods. Other works try to predict the vote intention of the citizens based on their public posts. This is the case of~\cite{2015-Fang}, where they tried to classify people's voting intentions based on the content of their tweets during the Scottish Independence Referendum in 2014. They built a topic-based na\"ive bayesian model (TBNBM) that takes into account the dependencies between topics and user voting intentions. This TBNBM detects the topics using the LDA, and for each topic they built a probability table where each feature has two associated conditional probabilities related to both voting intentions (i.e., 'Yes' or 'No'). In the experimental phased authors realized that this TBNBM improves the classical bayesian classifications. Finally, other relevant work is the one published by Borge-Holthoefer et al.~\cite{2015-Borge}. In this case, authors used social structured and the content of the tweets to understand the opinion evolution in Egypt during the summer of 2013. They observed that the military takeover caused major quantitative (volume of polarized tweets), but not ideological (polarity swaps) shifts among Twitter users. They also observed how the pro-military Twitter users that were very loud before the take-over, became increasingly silent afterwards, and how anti-military intervention Twitter users become significantly louder after the takeover.
 
 Some of the just mentioned works can be also categorized into community finding problems because they detect some clusters, or communities, of users based on different aspects. In this domain, it is important to highlight the work published by Ozer et al.~\cite{2016-Ozer} because they developed three Non-negative Matrix Factorization frameworks to investigate the contributions of different types of user connectivity and content information in community detection. They revealed that user content and endorsement filtered connectivity information are complementary to each other in clustering politically users into pure political communities. Other work~\cite{2011-Conover}, predicted the political alignment of Twitter users based on the content and structure of their political posts after the 2010 U.S.~midterm elections. They used manually annotated data and the different communities are created by taking into account the TF-IDF and the hastags obtained by content analysis. They found a highly segregated partisan structure with few retweets between left and right-wing Twitter users.
 
Finally, some other works are focused on how the information is propagated through the network. In this regard, a well-known work is the one published by Colleoni et al.~\cite{2014-Colleoni}. In this work, authors used a combination of machine learning and SNA to classify users as Democrats or Republicans based on the content shared on social media. Then, they investigated the political homophily in both: the network of reciprocated and non-reciprocated ties. They found that the structures of political homophily differ significantly between Democrats and Republicans. Other works try to study whether the sentiment of a tweet affects to its propagation~\cite{2012-Stieglitz}, i.e., how the affective dimensions of tweets, including positive and negative emotions associated with certain political parties or politicians, affect the quantity of retweets. Experimental results over around $64.431$ political tweets reveals a positive relationship between the quantity of words indicating affective dimensions (positive and negative) and its retweet rate. The last remarkable work~\cite{2015-Barbera}, authors studied whether the online communication of political and nonpolitical issues resembles an ``\textit{echo chamber}'' (i.e., communication between individuals with same ideological segregation) or a ``national conversation''. In order to do that, authors analyzed around $150$ million tweets regarding $12$ political and nonpolitical issues, extracted from $3.8$ million Twitter users. The findings suggest that in terms of political issues, the information was exchanged primarily among individuals with similar ideological preferences; but this effect did not happen with non-political issues.



\subsubsection{Detection of Fake News \& Misinformation}

In 2018, $66\%$ of American adults consume news on social media\furl{journalism.org/2018/09/10/news-use-across-social-media-platforms-2018}, whereas in 2012 only $49\%$ of adults used OSNs for this purpose. This increment is due to two main reasons. The first one is related to the fast propagation of messages on OSNs. The second reason is related to the high number of users connected to any OSN which makes quite easy to comment and discuss any message. Due to the easiness of creating messages and disseminating them in any OSN, it has been significantly increased the number of \textit{fake news}~\citep{2017-Shu,tandoc2018defining,vosoughi2018spread}, i.e., those news with intentionally false information produced online for a wide range of purposes, such as financial and political gain~\citep{2017-Allcott,bakir2018fake}.

In spite of its popularity, the research community is not able to agree in a common,and unique, definition of the term \textit{fake news}. The most extended definition is the one that states that fake news are those news articles that are intentionally false, its truthfulness can be verified, and its intention is to mislead readers~\citep{2017-Allcott,2015-conroy,lazer2018science}. This definition is based on two key concepts: \textit{authenticity} and \textit{intent}. The former means that fake news contain false information that can be verified. And the latter refers to the fact that fake news are created to confuse and mislead customers.

The main risk about fake news is related to the concept of \textit{intent}, just described. The goal of any fake news is to persuade consumers to accept biased, or false, information usually with political messages or influence. This goal and the fast propagation of messages in OSNs made that, for example, the most popular fake news was even more widely spread on Facebook than the most popular authentic mainstream news during the U.S. 2016 president election\furl{buzzfeednews.com/article/craigsilverman/viral-fake-election-news-outperformed-real-news-on-facebook}. For all of that, it is really important not only to understand how fake news propagates through the network (see Section~\ref{sec:informationDiffusion}) but also to develop systems that detect whether a specific new is a fake news or a real one~\cite{2017-Shu}.

Fake news detection is a promising area that tries to define whether a specific new is fake or not~\cite{2017-Shu,2015-conroy}. There are different approaches that can be followed to do that, but all of them can be grouped in two big categories: the \textit{linguistic approaches} and the \textit{network approaches}.

The goal of \textit{linguistic approaches} is to detect the fake news by analyzing its context. The idea is to obtain the different writing styles to detect fake news. In this sense, research works focus on two different linguistic features: \begin{enumerate*}[label=(\arabic*)]
\item \textit{lexical features}, such as total words, characters per word, frequency of large words, etc.~\cite{1998-furnkranz}, and 
\item \textit{syntactic features}, such as ``n-grams'' and bag-of-words (BOW) or parts-of-speech (POS) tagging~\cite{2017-Potthast}.
\end{enumerate*}

Second category, \textit{network-based approaches}, are based on the network that can be built by taking into account the users that publish related social media posts. In this case, researchers create a network based on some specific interaction on the OSN, and then, they apply network metrics to extract the valuable information. There are different types of network that can be built. For example, the \textit{stance network} is a graph where the nodes represent all the tweets relevant to the news, and the edges contains a weight that indicates the similarity of stances~\cite{2016-Jin,2017-Tacchini}. A different approach is the one followed in the \textit{co-occurrence network} where two nodes are connected by a weighted edge that represent how many times both users have written post relevant to the same news articles~\cite{2017-Ruchansky}. And the last example is \textit{friendship network} users who posted related tweets are connected in the network\cite{2013-Kwon}.

Finally, once the model is built, the network metrics can be used to extract valuable information. For example, authors of~\cite{2013-Kwon} used degree and clustering coefficient to characterize the diffusion network. A different approach is the one followed in~\cite{2017-Ruchansky} where SVD is used to learn the latent node embedding.

\subsubsection{Multimedia}


Nowadays, although users are able to generate different types of content in OSN like audio, image or video, text is the most common, and popular, content analyzed. The analysis of audio, image or video, has received less attention compared to text mainly due to the complexity of the analysis tasks, and the initial technological limitations of internet in its early years. The internet connections were slow and encoding algorithms were poor, thus, sharing videos, audios and images were impractical. Nowadays, internet connections have improved and smartphones, all equipped with cameras and microphones, have emerged as a portable alternative to traditional computers with access to the internet. As a result, it is very easy to share videos, photos, or audio using those devices, in fact, some of the most popular modern OSN are completely based on sharing images and videos like Instagram or Pinterest. Consequently, the interest in analyzing multimedia content in OSN has suffered an important increase.

Taking into account the different multimedia types audio, video and images, it is the latter the most common data type analyzed. Some of the works found in the literature have similar goals as the ones analyzing texts, such as Sentiment analysis~\cite{Ji2016}. Furthermore, Image Annotation~\cite{cheng2018survey}, which is generating a set of words that describe the content of a picture, is also commonly applied and it has similar goals to the keyword extraction methods. Besides, Image Clustering~\cite{wazarkar2018survey} can also be found, and it would be the equivalent to topic extraction. Although the just mentioned approaches share the same goals as some of the most popular text-based algorithms, the methods based on images are more complex due to the complexity of processing images. One characteristic of the images, that it is not present in the text analysis, is the fact that images contain information about the geographical location where the image was taken. This fact has allowed SNA researchers to use images as geo-location alternatives.

There is a plethora of works that analyze images for extracting some valuable information. For example, it is possible to predict the popularity of a picture~\cite{khosla2014makes,gelli2015image}, to predict the gender of users~\cite{merler2015you}, to discover events in public places~\cite{jayarajah2016can}, or forecast the ambiance of a place~\cite{redi2015like}, by applying Sentiment analysis and Image Annotation. Other works, like~\cite{2014-hu}, have used Image Clustering to analyze the content of popular images in OSN, such as pets, food, \ldots etc. There is other wide set of works that have developed their own algorithms for analyzing faces and predicting human characteristics~\cite{liu2016analyzing, celli2014automatic} or perceived intelligence~\cite{wei2017smart}. Finally, other works need the supervision of humans to validate the output of the algorithms. Some examples of these works are the one that analyze marijuana-related content~\cite{cavazos2016marijuana} or cyber-bullying content~\cite{hosseinmardi2016prediction}.

It is possible to classify the different algorithms and methods developed to perform some image analysis in three different groups: \begin{enumerate*}[label=(\arabic*)] \item Crowdsourcing, \item Deep Learning, and \item Handmade Features.\end{enumerate*} On the one hand, Crowdsourcing~\cite{hossain2012crowdsourcing} consists of the division of a work package into small pieces, or sets of images, and distributing them between a large number of participants, humans, to achieve cumulative results fast. Usually, this task is performed using some online platform that acts as an intermediary between the owner of the dataset and the workers. Popular platforms for doing image analysis based on Crowdsourcing are Amazon Mechanical Turk\furl{mturk.com} (MTurk), Figure Eight\furl{figure-eight.com} (formerly CrowdFlower) or MicroWorkers\furl{microworkers.com}. On the other hand, Deep Learning and Handmade Features avoid human intervention and use algorithmic approaches to extract significant features from images to be used as representatives. Based on this idea, Handmade Features use custom human-made filters that allow characterizing an image by low-level features, such as, colors~\cite{han2002fuzzy,graps1995introduction}, shapes~\cite{ai2013shape, hong2014shape}, and textures~\cite{fogel1989gabor, liu2016median}, or by high-level ones like the Semantics-Preserving Bag-of-
Words (SPBoW) model~\cite{wu2010semantics} or the Contextual Bag-of-Words (CBoW) one~\cite{li2010contextual}. Deep Learning, specifically convolutional neural networks (CNNs)~\cite{hossain2019comprehensive}, automatically construct these filters on their own. To this end, a supervised learning approach is used to generate a series of convolutions, ordered as layers connected one to another, capable of compressing an image into a representative set of characteristics. It is common to find works that use deep CNN (those with several layers), already trained to identify a huge number of different types of pictures (elephants, cars, plants, houses \ldots etc). For example, in~\cite{deng2009imagenet} authors trained CNN with the ImageNet dataset. Finally, once the characteristics are extracted, common supervised/unsupervised machine learning techniques like neural networks~\cite{haykin1994neural}, SVM~\cite{hearst1998support}, k-means~\cite{lloyd1982least} or DBSCAN~\cite{ester1996density} are used for doing particular analyzes.

Audio analysis in OSN has been largely untapped so far. It is quite common to apply Automatic Speech Recognition~\cite{yu2016automatic} to transcript the audio into a text and then, applying Speech Sentiment analysis extract some useful knowledge~\cite{el2011survey}. In spite of, for the best of the authors' knowledge, there are no works that have used these techniques to SNA, there are some applications of audio analysis in OSN. The first one is related to music. The Internet has been influencing the music scene for the last few decades, making music more accessible to the public and engaging musicians with their audience more easily. Thus, several social media platforms exist focused on music like Last.fm\furl{last.fm} or SoundCloud\furl{soundcloud.com}. With this in mind,~\cite{jacobson2008using} proposed to use audio analysis to categorize music into genres to enhance community finding techniques. The other main application of audio analysis is generating conversational networks from raw audio data~\cite{wyatt2008towards, wyatt2011inferring}. Once the conversational networks are generated, SNA techniques are used to perform speaker role recognition~\cite{vinciarelli2007speakers, garg2008role} and summarizing~\cite{vinciarelli2006sociometry} on broadcast news and podcasts.

Video analysis is, by far, the media type least studied in the context of OSN. It is also the most complex media type to analyze from the three mentioned in this section (image, audio, and video) because it encompasses the other two. However, video media content is prevalent on the Internet and the most popular OSN are completely centered on it, like Youtube or Youku (YouTube's Chinese counterpart). Although, it can be found several papers focused on the video analysis, these techniques have not been applied yet in conjunction with SNA techniques. One example is the work published in~\cite{wang2015video} where authors apply sentiment-analysis to the video once it had been transcribed.




\section{Discussion on Research Methods and Application Domains}
\label{sec:soadiscusion}
\label{subsec:conclusionsSNA}

From the analysis of the state of the art in both research and application areas of SNA, shown in Sections~\ref{subsec:structuralSNA} (OSN Structural-based analysis: graph theory, community detection algorithms, information diffusion), ~\ref{subsec:contentSNA} (OSNs Content-based analysis: topic extraction, opinion mining and sentiment analysis, multimedia), and Sections~\ref{sec:app-health} to ~\ref{sec:emerging} (health, marketing, tourism and hospitality, cyber security, politics, fake news and misinformation, 5G and IoT technologies), some important conclusions can be drawn:

\begin{itemize}
  
  \item On the one hand, OSNs are used as the highest socially trending, and influencing, source of information. Currently these sources have become on the most popular ones in the world, with billions of active users generating huge amount of data (in form of textual information, video, and other multimedia material) per second.
  
  \item Working with only a small part of the available data stored in any OSN, exceed the classical processing algorithms capabilities, and suppose one the biggest current challenges for a wide number of research areas. The complexity of this domain needs by a joint effort from different research areas, so the \textit{multidisciplinarity} between areas, from Social Sciences to Science and Engineering, becomes a necessity to find new methods and technologies to exploit the enormous potential of these sources.
  
  \item Although multidisciplinarity is a necessity, it means that professionals from very different areas, with very different background, have to collaborate together on a wide range of technologies (e.g., sociologists and data scientists). 
  
  \item On the other hand, most of algorithms and methods studied attempt to address the set of \textbf{challenges} (gathering, and processing the data, finding useful patterns, visualize the information, etc.) imposed by the complexity of this domain, in particular:
  
  \begin{enumerate}

    \item Both structural-based and content-based algorithms, are used to \textit{extract or discover} useful knowledge from the network. This challenge is related to the classical Knowledge Discovery problem in Machine Learning, Data mining and Data Science fields. This challenge, or problem, is directly related to our proposed RQ1 (\textit{What can I learn?}).
    
    \item The necessity to manage a huge, and exponentially growing amount of data, needs from new methods, \textit{scalable} algorithms, and technologies. This is one of the hot topics in areas as Big Data, and it is related to RQ2 (\textit{What is the limit?}).
    
    \item Due to the vast number of sources available, and their different data formats (numerical, categorical, textual, metadata, video, images, audio), it's necessary to develop new algorithms capable to \textit{integrate} and \textit{fusion} different sources to allow discovery new and useful knowledge. This problem is usually address by the area of Information fusion, and it's related to RQ3 (\textit{What kind of data can I integrate?}).
    
    \item Finally, one of the essential tools for knowledge discovery in OSNs is related to the \textit{visualization} of the information. This is a challenging, and still open problem, in the area of Information Visualization, so the RQ4 (\textit{What can I show?}) has been defined to cover this aspect.
    
    \end{enumerate}
  
\end{itemize}

Taking into account previous conclusions, this work proposes the definition of four \textbf{dimensions}, which can be used to assess the maturity of technologies currently available in OSN. These four dimensions will later be used to define a set of \textbf{metrics} (which we named \emph{$degrees_*$}) that will be used:

\begin{itemize}
  \item \textit{To quantitative assess the level of maturity of a set of SNA tools and frameworks}. To do that, a set of graphical representations (based on spider graphs), and a new global metric (named \textgoth{C}$_{SNA}$), will be defined to provide a quantitative measure of these tools and frameworks.
  
  \item \textit{To study some possible future trends, challenges, and lines of work in these dimensions} (closely related to research areas such as Data Science, Big Data, Information fusion and Visualization), detecting spaces for improvement, and emerging technologies, where new developments could have a high impact on the scientific community, industry and society.
\end{itemize}


%% file: sections/4-dimensions.tex
\section{The Four Dimensions of Social Network Analysis}
\label{sec:4dimensions}


The Big Data paradigm was characterized by the different V-models that allow any researcher to analyze the capacity of the different Big Data methods. Initially, 3Vs were described in the 3V model~\cite{laney20013d}, but this model has evolved during the last years to the 4V~\cite{data2017four, hashem2015rise}, 5V~\cite{yin2015big}, or 6V model~\cite{ur2016big}. These models allow measuring the maturity of different methods, tools and technologies based on Big Data using simple features such as Volume, Velocity, Variety, Value, Veracity or Variability. There are even some attempts to include new 'V's likes Visualization, in these *V-models. This set of *V-models provides a straightforward and widely accepted definition related to what is (and what is not) a big-data-based problem, application, framework, or technology. We will use this interesting (and successful) approach to describe the challenges, and the current status, of technologies related to OSN. We have mapped our four Research Questions stated in Section~\ref{sec:introduction}, into a set of equivalent dimensions: \begin{enumerate*}[label=\arabic*)]
  \item[D1)] \textit{Pattern \& Knowledge discovery}, 
  \item[D2)] \textit{Scalability},
  \item[D3)] \textit{Information Fusion \& Integration}, and 
  \item[D4)] \textit{Visualization}
\end{enumerate*}. Using these dimensions, we can quantify the different methods, techniques, algorithms and frameworks, allowing us to better understand where we stand and where we could be in the near future in this area.

\subsection{\texorpdfstring{Pattern \& Knowledge Discovery ($D_1$)}{Pattern \& Knowledge Discovery (D1)}}
\label{subsec:dimension1}

The concept of \textit{Value}, as the knowledge or pattern discovery capacity of any method or algorithm, can be easily extrapolated to the area of SNA. This first dimension will be used to define the capacity of Knowledge discovery (mainly from a pattern mining perspective) of SNA technologies. This dimension tries to answer the question: \textit{What can I learn?}, understood as the capacity to discover non-trivial knowledge from OSN.

Knowledge discovery and pattern mining is one of the central topics in different areas as Data Mining~\citep{fayyad1996data, maimon2005data}, Big Data~\citep{wu2013data, lavalle2011big}, and Data Science~\citep{hey2009fourth, provost2013data}, which can be considered as a new paradigm that includes everything from Big Data Analytics, Data Mining, Predictive Modeling, Data Visualization, Mathematics, to Statistics. In the area of Big Data, this topic is related to the concept of Value and usually refers to the process of extracting valuable information from very large data sets (e.g., from social big data~\citep{bello2016social}), and it is usually referred to as Big Data Analytics~\citep{tsai2015big, russom2011big, singh2015survey}. 

Any Data Mining process can generate a large number of patterns from data (from some few ones to hundred or thousands), therefore in Data Mining it is necessary to select or filter those useful patterns from the trivial or redundant. In such case, it is possible to use a set of measures designed for evaluating and ranking the discovered patterns produced by the data mining process. In Data Mining area, this set of measures are usually named as \textit{interestingness measures}~\citep{geng2006interestingness,mcgarry_2005,silberschatz1995subjective}, which are intended for selecting and ranking patterns according to their potential interest to the user. This kind of measures are used to reduce the number of rules discovered by users, so they are used to restrict the final number of rules that are, or could be, interesting to the user. Although there is wide number of interestingness concepts that has been defined, most of them emphasizes conciseness, coverage, reliability, peculiarity, diversity, novelty, surprisingness, utility, and actionability~\citep{geng2006interestingness}. These interestingness measures can be divided into three main categories~\citep{hilderman1999knowledge}:
\begin{enumerate*} [label=(\arabic*)] 
\item \textit{objective measures}~\citep{piatetsky1994interestingness}, which are based only on the raw data, and the statistical strengths, or properties, of the discovered patterns; 
\item \textit{subjective measures}~\citep{silberschatz1995subjective,liu1999finding}, which take into account both the data and the user of these data, and are derived from the user's beliefs or expectations of their particular problem domain (so they are based on the subjectivity of the user who evaluates the patterns on the basis of novelty, actionability, unexpectedness, etc.); 
\item and, \textit{semantic-based measures} that considers the semantics and explanations of the patterns.\end{enumerate*}
The main problem of all these measures is that they cannot be applied directly to quantify the dimension relative to the value provided to the discovery of knowledge by the different techniques and tools used in SNA. These metrics allow us to quantify the value of the resulting patterns that are generated by a specific data mining process after being applied. In other words, a specific technique must be applied to a specific set of data, and the patterns obtained as a result are evaluated. In addition, for each type of mining technique, there are different measures to evaluate its patterns. Therefore, these measures will not able to be directly applied to quantify the value in SNA techniques and tools. 


The objective of this dimension is to evaluate, in a generic way, any type of technique, method, or tool, which is used discover new knowledge in OSN. In the literature, there are several works reviewing the different techniques and algorithms to extract knowledge that are usually applied in OSN. Taking these works into account, we can generate an overall taxonomy of the different functions and methods that SNA tools usually provide. Then, using this taxonomy, it would be possible quantifying the \textbf{degree of value} that each tool provides, according to the functionalities that it covers within the taxonomy generated. 
analyzing the existing research works focused on carrying out a detailed review of the applications and methods to extract knowledge from OSN, the work presented in the 90s by Wasserman and Faust~\cite{wasserman1994social} is one of the most relevant publications on the area over the years. The authors show in this review the following three aspects for extracting knowledge in SNA field:

\begin{itemize}
  \item The application of \textit{Balance theory} is one of the main concepts to emerge from the early days of SNA~\cite{GRANOVETTER1977347}. It is focused on the cognition or awareness of sociometric relations, usually positive and negative affect relations such as friendship, liking, or disliking, from the perspective of an individual. The idea of balance arose in a study introduced by Fritz Heider~\cite{heider1946attitudes} about the individual's cognition or perception of social situations. Heider was focused on a single individual and was concerned about how this individual's attitudes or opinions coincided with the attitudes or opinions of other `entities' or people. The entities could be not only people, but also objects or statements for which one might have opinions. This author considered ties, which were signed, among a pair or a triple of entities. 
  
  \item The use of \textit{Graph theory}~\cite{barnes1983graph} in OSN allow us to identify the most influential, prestigious, or central actors within an OSN. These type of relevant actors are usually located in strategic locations within the network, and there are several centrality measures~\cite{freeman1978centrality,bonacich1987power} in graph theory that measure of how the position of an actor is within the overall structure of the OSN, such as the degree, betweenness, closeness, and eigenvector centrality, amongst many others (see Sections~\ref{sec:scientometric},~\ref{subsec:graphtheory}). In addition, SNA are naturally transitive, which means that friends of a given actor can also be friends. This property of transitivity is quantified in graph theory by the clustering coefficient defined in Watts and Strogatz work~\cite{Watts1998Collective}.
  
  \item The identification of \textit{Cohesive Subgroups or Communities}~\cite{Girvan2002,2003-Newman,Fortunato201075} within an OSN is one of the major concerns of SNA. Cohesive subgroups are subsets of actors among whom there are relatively strong, direct, intense, frequent, or positive ties. The concept of social group as used by social and behavioral scientists is quite general, and there are many specific properties of an OSN that are related to the cohesiveness of subgroups, so there are many possible OSN subgroup definitions (see Sections~\ref{sec:scientometric},~\ref{subsec:cda}).
  
\end{itemize}

Based on these main aspects, different methods and approaches have been designed and developed to discover knowledge, or extract patterns from OSN using data mining techniques. analyzing different publications reviewing the applications and methods to extract knowledge from OSN~\cite{oliveira2012overview,nandi2013survey,bonchi2011social,zuber2014survey,mariam2013survey}, in general terms, it could be considered that the taxonomy of the main functionalities for discovering knowledge which can be embedded in SNA tools are the following:
\begin{enumerate}
  \item \textit{Qualitative and quantitative/statistical analysis ($F_{\textit{Value}(1,i)}$)}:
    \begin{enumerate}
      \item \textit{Computation of measures based on the topology} ($F_{\textit{Value}(1,1)}$) of the network that provides a local and global description of it. These type of measurements are extracted from the graph theory, being some of the most relevant the density, distance, centrality or transitivity, amongst others. A huge number of works~\cite{freeman1978centrality, Watts1998Collective, bonacich1987power}, used previous measures to discover the most influential, prestigious, or central nodes (or actors) within an OSN. The value of this measure, $F_{\textit{Value}(1,1)}$, will be calculated according to the following scale: +1/5 until 1 for each measure employed (diameter, mean degree/distribution, cluster coefficient, connected components, transitivity, triangle count, etc...).
      
      \item \textit{Link analysis} ($F_{\textit{Value}(1,2)}$) algorithms arise with the aim of finding the most valuable, authoritative or influential node (e.g., a webpage in the Web), being the HITS~\cite{kleinberg1999authoritative} and the Google Pagerank~\cite{brin1998anatomy} the most popular ones. This measure will be calculated according to the following scale: +1/4 until 1 for each algorithm provided.
    \end{enumerate}
  
  \item \textit{Pattern Mining} methods ($F_{\textit{Value}(2,i)}$):
    \begin{enumerate}
    
     \item \textit{Community detection} ($F_{\textit{Value}(2,1)}$) algorithms (static and dynamic)~\cite{Girvan2002, Fortunato201075, greene2010tracking} try to find groups of nodes (users) where the set of edges is dense within the group and sparse outside it. One of the main difficulties in this topic is how a community is defined, since it can depend specifically on the domain where it is applied and what the network represents, as well as the type of links that are considered. Naturally, community detection algorithms are based on concepts from graph theory and clustering. Indeed, this problem is very similar to the problem of graph clustering. Recently, the analysis to the dynamics and evolution of OSNs has grown hugely, therefore many of the classic community search algorithms have been extended to study also the behavior of the communities over time. The scale to compute this measure is the following: +1/3 if allows one kind of community detection (overlapping or non-overlapping); +1/3 if allows to do overlapping and non-overlapping community detection; +1/3 if allows to do community detection in temporal networks.
  
     \item \textit{Opinion Mining} ($F_{\textit{Value}(2,2)}$) techniques~\cite{liu2012survey} are focused on the the detection of user opinions, and also feelings or reactions of people about certain beliefs, products, decisions or events. OSN are online sites where people can express their ideas and opinions, exchange knowledge and beliefs or criticize products. Millions of new posts giving opinion on products and services are generated every day in OSN. All this information dumped on OSN mostly in text format is very valuable to discover new knowledge. One of the most famous methods within these techniques is the \textit{Topic Detection} algorithms, that are based on the idea of applying data mining techniques to detect what topics are more popular over the time~\cite{allan2012topic}. In addition, \textit{Sentiment analysis} methods~\cite{pang2008opinion} try to identify how people feel about a specific topic. This issue can be as important as detecting the topic itself on the OSN, and it can be addressed using data mining techniques related to NLP~\cite{manning1999foundations}. This aspect is measured according to the following scale: +1/2 if it provides topic detection; +1/2 if it provides sentiment analysis. 
     
       \item \textit{Homophily Models} ($F_{\textit{Value}(2,3)}$)~\cite{McPherson2001}. Opinions of influential users on OSN often are capable of affecting the decisions of other users. The concept of Homophily refers to how the behavior of users that depict the same opinion (are linked under the same social community), is subject to adjustments in awareness of the behavior of other participants in the same community. The models of this concept are based on the idea that at this stage, the decision of influential participants, who are either efficient in the field or in communications skill, attracts the follower of the minority. So, when cogent information is introduced by these influential users, their opinions are disseminated in cascade affecting the positive or negative decisions of other users. In order to calculate this measure, the following scale is used: +1 if available.

      
      
    \end{enumerate}
  \item \textit{Predictive analysis} ($F_{\textit{Value}(3,i)}$):
    \begin{enumerate}
      \item \textit{Propagation and Virality Modeling} ($F_{\textit{Value}(3,1)}$) consists on the study of the spread of influence through OSN. This issue has a long history in the area of social sciences, where the first studies was emerged on medical and agricultural research areas~\cite{coleman1966medical,valente1996network}. In recent years, these type of models have been applied by marketing researchers, trying to model the `word-of-mouth' diffusion process for viral marketing applications~\cite{2001-Goldenberg,leskovec2007dynamics}. The value of this measure will be calculated according to the following scale: +1 if available.
      
      \item \textit{Link Prediction} ($F_{\textit{Value}(3,2)}$). In dynamic or temporal networks, a typically problem addressed is estimating the probability of two particular nodes are to become connected in the future. This is a classical computational problem underlying in OSN evolution over time, that was introduced by Liben-Nowell and Kleinberg~\cite{liben2007link} as the link prediction problem. It can infer new interactions among members of an OSN that are likely to occur in the near future. This measure will be computed according to the following scale: $+1$ if available.
    \end{enumerate}
\end{enumerate}

As mentioned above, this proposed taxonomy can be used to quantify the \textbf{degree of value} ($ d_{Value}(t)$) that each SNA tool ($t$) provides according to the functionalities that it covers as shown in equation~\ref{eq:dim_value}. Several weights are used ($\alpha$, $\beta$, and $\gamma$), to represent the importance given to each characteristic. In this work, all of the value characteristics will have the same weight (so $\alpha$, $\beta$, and $\gamma$ will be set up to $1/3$). 
This equation has been normalized in the range [0,1] taking into account the different methods, techniques, algorithms and measures that are incorporated by the particular tool ($d_{Value}(t) \in [0,1]$). 

\begin{equation}
  d_{Value}(t) = \alpha \cdot \frac {\sum_{i=1}^{2}{F_{\textit{Value}(1,i)}(t)}} {2} + \beta \cdot \frac{\sum_{j=1}^{3}{F_{\textit{Value}(2,j)}(t)}} {3} + \gamma \cdot  \frac{\sum_{k=1}^{2}{F_{\textit{Value}(3,i)}(t)}} {2}
\label{eq:dim_value}
\end{equation}

\subsection{\texorpdfstring{Scalability ($D_2$)}{Scalability (D2)}}
\label{subsec:dimension2}

Currently, the exponential growth of data has created serious problems for traditional data analysis algorithms and techniques (such as data mining, statistics, machine learning, and so on) to processing the data available in electronic sources. A new generation of algorithms and frameworks is currently being developed in order to manage big data challenges, and they require high scalability in both memory consumption and computational time~\cite{bello2016social}. Therefore, this dimension will be used to define, and quantify, the scalability capacity of a tool or technique (e.g., algorithm) used in an OSN. From this perspective, the amount of information handled will be the key feature considered (mainly using the quantity of nodes and edges that can be processed by the SNA method or tool). A highly scalable software would work correctly on a small dataset as well as working well on a very large dataset (say millions, or billions of nodes and edges). In general terms, scalability refers to those techniques that ensure that some quality of service is maintained as the size of the data set to be managed grows, or the complexity of the addressed problem increases. 

Big Data systems like MapReduce~\cite{Dean:2008}, Hadoop~\cite{shvachko2010hadoop} or Spark~\cite{Zaharia:2010} have been developed as a response for these scalability problems. However, the specific application areas of SNA are usually modeled as graph-theoretical problems, and unfortunately, the direct application of graph algorithms in these big data environments is often not an efficient solution (most of classical graph-based problems are NP-hard). Another approach to tackle this problem is the graph-parallel systems for specialized graph processing problems. These systems perform better than the general tools for Big Data, but their main disadvantage remains that they can only be used for graph specific problems~\cite{soric2017efficient,andersen2016evaluating}. An efficient way to address this problem is to combine the advantages of both approaches: the graph-parallel approach, and the general big data processing tools. For example, GraphX is an Apache Spark's built-in library for graph analytic and graph-parallel computation~\citep{xin2013graphx,gonzalez2014graphx}, which provides an excellent, and highly scalable, solution for graph-based problems. 

In general terms, scalability is a desirable aspect of a network, system, application, or process. This concept can be defined as the capability of a system to handle an increasing number of elements or objects, to process increasing volumes of work adequately, and to be easily enlarged or extended~\cite{bondi2000characteristics}. This means that the application or system should have the ability to continue functioning correctly when the problem it is changed in size or volume, taking full advantage of it resources in terms of performance. The scalability of a system usually depends on the types of data structures and algorithms used to implement it, or if it has different components or modules (and how them have been designed), or on the communication mechanisms used by its components. For example, the data structures of a system affect not only the amount of space required to perform a particular function, but also the time. Through this last observation two of the main aspects of scalability can be appreciated: \textit{space} and \textit{time}. In Bondi~\cite{bondi2000characteristics} work, a more detailed analysis of scalability aspects is presented, where the author considers four main types of scalability on a system: 
 
 \begin{enumerate}
  \item \textit{Load} scalability: the ability to function with agility (without undue delay, without unproductive consumption of resources, and making good use of the available resources). 
    
  \item \textit{Space} scalability: the memory requirements do not grow to intolerable levels, as the number of items that the system supports increases. 
  
  \item \textit{Space-Time} scalability: the ability to continue operating with agility as the number of objects or data to be processed increases by orders of magnitude. 
  
  \item \textit{Structural} scalability: a system is structurally scalable if its implementation does not impede the growth of the number of objects or items it is capable of handling.
 \end{enumerate}

Load scalability may be improved by exploiting parallelism, but the other three characteristics mentioned of scalability are inherent to the architectural design and implementation of the system (such as the length or the choice of data structures), and in many cases are difficult or even impossible to change. In addition, when a taxonomy of characteristics is defined, it is natural to study whether there are dependencies between them. In this particular taxonomy, for example, systems with poor space scalability or space-time scalability, might have poor load scalability, due to the attendant memory management overhead, or search costs. On the other hand, systems with good space-time scalability because their data structures are well engineered, might have poor load scalability due to poor decisions about scheduling, or parallelism, which have nothing to do with memory management. 

Another taxonomy of characteristics related to the scalability was proposed by Hesham and Mostafa~\cite{el2005advanced}, in this particular case, it was defined for those architectures that allow parallel processing. This work presents multiple dimensions to measure the scalability, such as:
  \begin{enumerate}
    \item \textit{Administrative} scalability: The ability for an increasing number of organizations or users to access a system.
    \item \textit{Functional} scalability: The ability to enhance the system by adding new functionality without disrupting existing functions.
    \item \textit{Geographic} scalability: The ability to maintain effectiveness during expansion from a local area to a larger region.
    \item \textit{Load} scalability: The ability to expand and contract to accommodate heavier or lighter loads.
    \item \textit{Generation} scalability: The ability of a system to scale by adopting new generations of components.
    \item \textit{Heterogeneous} scalability: The ability to adopt components from different vendors (or others environments or systems).
 \end{enumerate}

As can be seen, there are several works that try to describe which characteristics should be taken into account when analyzing if a system is scalable. However, any of these mentioned works provide a model to quantify the degree of scalability of a system. The Universal Scalability Law (USL) presented by Gunther~\cite{gunther1993simple}, provided a quantification model for scalability of the systems or applications. This model is defined in terms of three main parameters $\alpha$ (contention), $\beta$ (coherency), and $\gamma$ (concurrency), that can be identified respectively with the three Cs~\cite{gunther2007guerrilla}:

 \begin{itemize}
   \item \textit{Concurrency} ($\gamma$): the maximum throughput (the measure of a number of requests processed over a unit time by the application) attainable at a given level of load.
   \item \textit{Contention} ($\alpha$) queuing time for shared resources.
   \item \textit{Coherency} ($\beta$) delay time for data to become consistent (or coherent), by virtue of point-to-point exchange of data between resources that are distributed.
 \end{itemize}
 
Defining the system throughput $X(N)$ at a given load, $N$, the USL can be expressed as the equation:

\begin{equation}\label{eq:USL}
X(N) = \frac {\gamma \cdot N} {1 + \alpha \cdot (N-1) + \beta \cdot N\cdot (N-1)}
\end{equation}

The independent variable $N$ represents the number of users or data load that is incremented on a fixed hardware configuration. When the scaling is linear-rising (the case for ideal parallelism), the $\alpha = \beta = 0$. In other words, the overall throughput $X(N)$ increases in simple proportion to $N$.

Following the same approach of using throughput to measure scalability, in the work presented by Jogalekar and Woodside~\cite{jogalekar1997scalability}, the scalability is measure based on the `power' metric of Giessler et al.~\cite{giessler1978free} as follows:

\begin{equation}\label{eq:FixedTimeSpeedup}
Power = \frac {\gamma} {T}
\end{equation}

where $\gamma$ is throughput and $T$ is mean delay time. 

As mentioned at the beginning of the section, the technologies developed for big data environments have arose to address the problem of scalability due to the exponential data growth that has occurred in recent years. Nevertheless, how to measure scalability in Big data environments is a question that is still being addressed. In Sanchez et al.~\cite{sanchez2018evaluating} work, the isoefficiency model is introduced as a standard measure of scalability. The isoefficiency function determines the ease with which a parallel system can maintain a constant efficiency and hence achieve speedups increasing in proportion to the number of processing elements~\cite{grama1993isoefficiency}. A small isoefficiency function means that small increments in the problem size are sufficient for the efficient utilization of an increasing number of processing elements, indicating that the parallel system is highly scalable. However, a large isoefficiency function indicates a poorly scalable parallel system.

Taking into account the different taxonomies of scalability characteristics shown by the different studies mentioned above, and adapting them specifically for SNA methods and techniques, the following sets of measures are proposed to quantify the \textbf{degree of volume}, or scalability, for SNA:

\begin{enumerate}
  \item \textit{Space-Time} ($F_{\textit{Volume}1}$): maximum number of elements (nodes and/or edges) it is able to process without degrading its performance. The value of this measure will be calculated according to the following scale: \textit{Low} (1/3) if process networks with less than 10.000 of nodes and/or edges; \textit{Medium} (2/3) if process networks with nodes/edges between $10.000$ to $100.000$; and \textit{Large} (1) if process networks with more than $100.000$ nodes/edges. These values have been set to indicate the maximum size of the elements that can be processed by the tool or algorithm (low, medium, high). Of course, these values would be strongly modified in the coming years as processing and storage capacities increase. 
    
  
  \item \textit{Parallelism} ($F_{\textit{Volume}2}$): capacity of parallel computing. The value of this measure will be calculated according to the following scale: \textit{Low} (1/3) for single or centralized processing; \textit{Medium} (2/3) for distributed processing; and \textit{Large} (1) for parallel processing using Big Data technologies.
  
  
  \item \textit{Functional} ($F_{\textit{Volume}3}$): ability to enlarge the system or application by adding new features or extending existing ones. The following scale allow measuring this characteristic: \textit{Low} (1/3) if none of the existing functionalities can be modified or new ones added; \textit{Medium} (2/3) if some existing functionalities can be added or modified; and \textit{Large} (1) if any of the existing functionalities can be added or modified.

   \item \textit{Heterogeneous-Integration} ($F_{\textit{Volume}4}$): ability to integrate or communicate with components or modules from different environments or systems. The value of this measure will be calculated as follows:
   \begin{itemize}
     \item \textit{Low} (1/3) if cannot be integrated with components or modules from other environments or systems;
     \item \textit{Medium} (2/3) if can communicate with some components, but not all, or modules from other environments or systems;
     \item \textit{Large} (1) if can be fully integrated with components or modules from other environments or systems.
   \end{itemize}

\end{enumerate}

Previous metrics have been combined to generate a scalability degree ($d_{Volume}$) for any tool ($t$) as it is shown in equation~\ref{eq:dscal}, where scalability ($d_{Volumne}(t)$) is rated from $0$ to $1$ depending on its capacity to scale to large data sets ($d_{Volume}(t) \in [0,1]$).

\begin{equation}
\label{eq:dscal}
 d_{Volume}(t) = \frac{\sum_{i=1}^{4}{F_{\textit{Volume}i}}(t)}{4}
\end{equation}






\subsection{\texorpdfstring{Information Fusion \& Integration ($D_3$)}{Information Fusion \& Integration (D3)}}
\label{subsec:dimension3}


OSN popularity is increasing everyday thanks, in part, to the diverse services provided and the target groups that can use them. In this way, it is possible to find OSN for any purpose ranging from personal (such as Facebook, or Twitter), to the professional ones like ResearchGate, or LinkedIn among others. Also these OSN offer a wide variety of services, and it is possible to share photos (Flicker, Instagram, \ldots), videos and music (Facebook, Youtube, \ldots), or micro-blogging (Facebook, Twitter, \ldots), etc. In addition to the service provided by the OSN, all of them allow different kind of interactions between the users (for example in Twitter users can '\textit{follow}' other users, whereas in Facebook they are '\textit{friends}'), and different actions over the content published by the user, i.e., in any OSN anyone can '\textit{like}', '\textit{comment}', and '\textit{share}' the contents published by other users.

With all of this, it is easy to realize the amount, and diversity, of data available in the different OSN that can be used to perform SNA tasks. In order to measure this diversity, we define this dimension called \textit{Information Fusion}, or \textit{Integration}, that tries to answer the question: ``\textit{What kind of data can I integrate?}. This measure would be equivalent to the concept of ``\textit{Variety}" from the Big Data paradigm. In the case of OSN, this dimension will measure different aspects regarding the data used to perform the SNA tasks. In this case, we have defined three different measurements that will take into account: \begin{enumerate*}[label=(\arabic*)]
\item the number of different type of data (\textit{multichannel}),
\item the number of different OSNs used to extract the data (\textit{multimodality}),
\item and the representation of this data into the model (\textit{multi-representation}).
\end{enumerate*}

Following, we provide a detailed description of these three measures and also, a formal definition to measure them:

\begin{enumerate}
  \item \textit{Multichannel} ($F_{\textit{Var}1}$): this indicator measures the diversity of the data taking into account the format of the data. In this regard, SNA algorithms are able to extract knowledge by using data from two different sources of information: the graph resulting from the modeling process, and the content published in the OSN~\cite{2017-GonzalezPardoEtAl-FGCS}. In this sense, the different types of data can be used to perform SNA tasks, are the following:
  \begin{enumerate}
    \item Edges of the graph: this data source is related to the interactions of the different users in the SN.
    \item Text: in this case, the data used to perform SNA tasks is the content published by the users in text format, for example \textit{tweets}, \textit{comments}, etc.
    \item Images: this information is extracted directly from the images, as photos or memes, that users post in any OSN. This data could be the direct analysis of the photo, or the tags that describe the content of the photo, etc~\cite{2014-hu}.
    \item Video: in any OSN is possible to publish videos, and the different SNA algorithms can take advantage of the video content to extract some valuable information.
  \end{enumerate}
  
  To sum up, $F_{\textit{Var}1}$ can be defined as the number of different data formats handled by the algorithms or tools. As it is quite difficult to fix the number of data formats due to the evolution of OSN, in this work we are going to limit the value of this characteristic depending on the different number of data types used. In this sense, we are going to consider three different ranges of values for $F_{\textit{Var}1}$ which are:
  \begin{itemize}
    \item \textit{Low}: this value reflect the case where the algorithm uses one type of data, i.e: only the edges of the graph, or text, or images, etc. When the algorithm, or system, uses only one type of data, it will have low $M_c$ and its value is $1/3$.
    \item \textit{Medium}: this value considers the cases where the algorithms, or systems, integrates $2$ or $3$ different types of data. The value of $M_c$ in this case is $2/3$.
    \item \textit{High}: when the systems, or the algorithms, are able to integrate $4$ or more different types of data, the value for its $M_c$ will be $1$.
  \end{itemize}
  For this reason, $M_c \in [1/3,1]$. It is important to note that this indicator is independent of how data source is modeled, or what kind of information is extracted. Imagine a platform that uses the tweets in text format, the $M_c$ value for this format will be $1/3$ independently of the analysis performed with this data.
  
  \item \textit{Multimodality} ($F_{\textit{Var}2}$): refers to the different number of data sources (independently of the data format) that can be handled by the algorithm. This indicator takes into account the different number of different OSN data formats, which are integrated by the algorithm. In this sense, it is quite difficult to define the limits of this indicator due to the fast evolution of OSN (theoretically this value would be a positive natural value, $\mathbb{N}^+$). Some examples of OSN are: Facebook, Twitter, Instagram~\cite{2014-hu}, Youtube, or LinkedIn, to mention just a few of them.
  
  For this reason, Multimodality is defined in the same way as \textit{Multichannel}, i.e., using three different values that evaluate the different number of OSN taken into account:
  \begin{itemize}
    \item \textit{Low} ($F_{\textit{Var}2}=1/3$): this value represents the systems, or algorithms, that gather data from one OSN.
    \item \textit{Medium} ($F_{\textit{Var}2}=2/3$: it is used when the data used to perform SNA tasks are extracted from $2$ or $3$ different OSN.
    \item \textit{High} ($F_{\textit{Var}2}=1$): this value is used when the system, or algorithm, is able to integrate data from $4$ or more OSN.
  \end{itemize}
  
  With all of that, $M_m \in [1/3,1]$, where $1/3$ means that only one OSN has been used to extract the data, whereas $1$ means that the $4$ or more OSN have been taken into account. Note that the value of this indicator is not ranking the different OSN but only providing a metric about the number of different OSN used by the platform/algorithm. For example, the $M_m$ value for a platform that is able only to extract data from Youtube will be $1/3$; and the value for other platform that integrate data from Twitter and Instagram will be $2/3$.
  
  \item \textit{Multi-representation} ($F_{\textit{Var}3}$) provides a quantitative measure of the data model representation used by the algorithm, and its value will depend on the complexity of the representation model. It is defined as $M_r \in [1/3,1]$, where a value of $1/3$ means the \textit{basic} representation model, whereas a value of $1$, the most \textit{advanced} one. Following, we explain the different representation levels with their corresponding values:
  \begin{enumerate}
    \item \textit{Basic model}: this kind of representation correspond to the case when the OSN is modeled into a simple unweighted graph $G=(V,E)$. In this graph, $V$ is the set of nodes of the graph and represents the set of users of the SN; whereas $E$ corresponds to the set of edges of the graph and represents some sort of connection between the users. Note that depending on the OSN taking into account this graph could be directed or undirected. The value of $F_{\textit{Var}3}$ in this case is the lower one, i.e., $1/3$.
    
    \item \textit{Intermediate model} corresponds to the value of $2/3$. In this case, the problem taken into account is modeled into a simple weighted graph. The resulting graph is quite similar to the previous representation but the edges contain a value providing some kind of information, for example frequency of a specific action, or a meaning of the relation.
    
    \item \textit{Advanced model}: this level represents the most advanced model used to work with the data extracted from the OSN. In this case, we consider a multi-layer SNA representation as the most advance model. \textit{Multi-layer networks} considers multiple channels of connectivity, it is a representation used to describe systems where the different actors are interconnected by different categories of connections. In this kind of networks, each channel (i.e., each type of relation) is represented by a layer, and the same node may have different set of neighbors in different layers. The value of $F_{\textit{Var}3}$ when the data is modeled into a Multi-layer network is $1$.
  \end{enumerate}
  
  Imagine a system that is able to analyze three different types of interactions ($F_{\textit{Var}_i}$) between users in an OSN. Fig.~\ref{fig:Multi-representations} shows the different representations taken into account by the multi-representation measure to model this situation. The first one (Fig.~\ref{fig:Multi-representations}.a) represents all the information into a unweighted graph, i.e., the \textit{Basic model}. Also, it is possible to use the \textit{Intermediate model} with a weighted graph (Fig.~\ref{fig:Multi-representations}.b). In this case, the weighted can be a number to represent, for example, the frequency of an interaction or a label to differentiate the different interactions. Finally, the \textit{Advanced model} is shown in Fig.~\ref{fig:Multi-representations}.c, where the data is modeled into a multi-layer graph and each graph contains the information regarding each kind of interaction.
\end{enumerate}

  \begin{figure}[!h]
	\centering
	\includegraphics[width=0.3\linewidth]{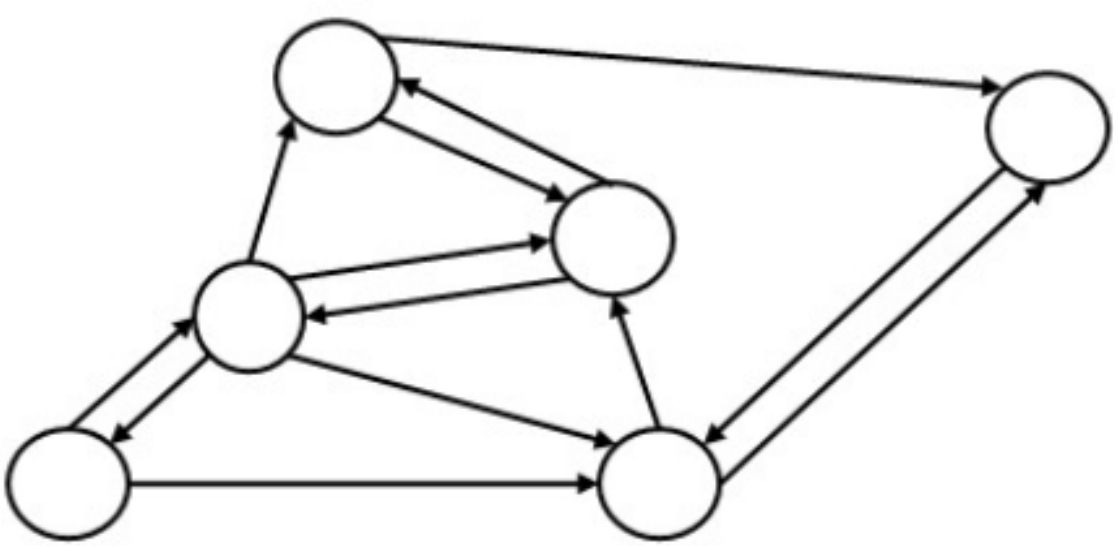}
	\includegraphics[width=0.3\linewidth]{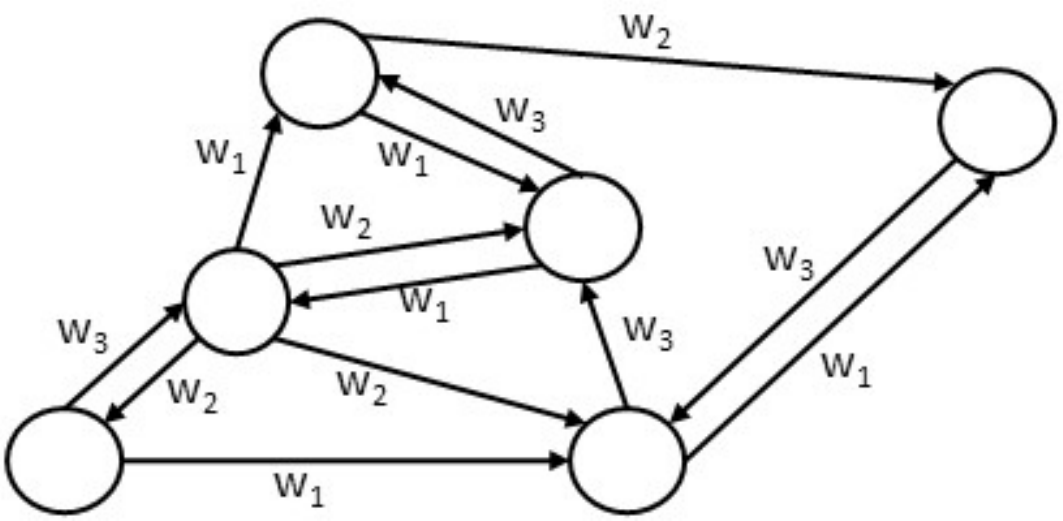}
	\includegraphics[width=0.35\linewidth]{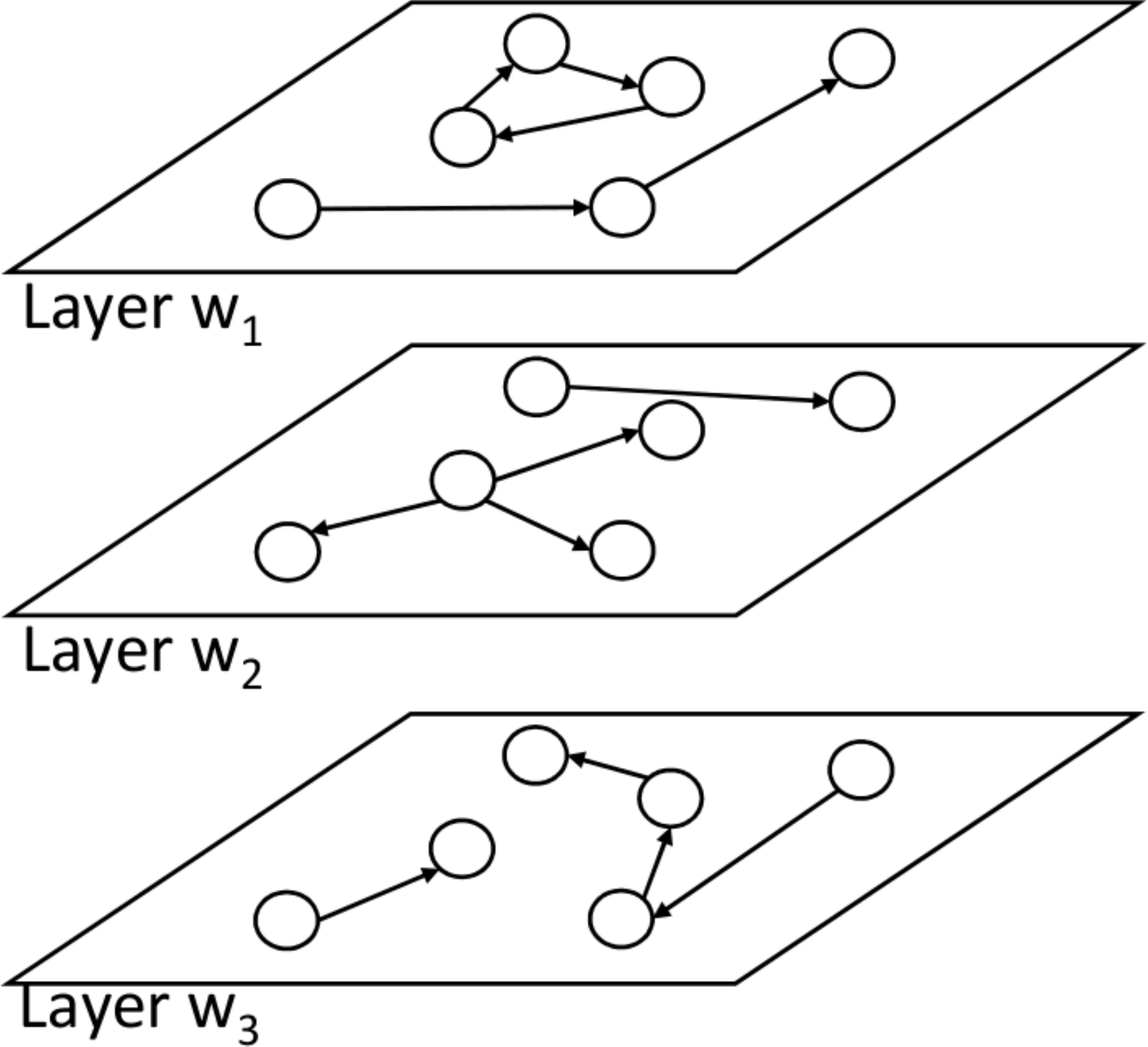}
	\caption{This figure shows the different representations taken into account into the $F_{Var_{3}}$ dimension. The different representations are: a) Basic model, b) Intermediate model, and c) Advanced model}
    \label{fig:Multi-representations}
\end{figure}

The next step is to define the value of the Information Fusion/Integration dimension, i.e., the \textbf{degree of variety}, in terms of the three measures just explained. From these three indicators, we consider that \textit{Multimodality} ($F_{\textit{Var}2}$) and \textit{Multichannel} ($F_{\textit{Var}1}$) play an important role in terms of Information Fusion/Integration because both measures describe, respectively, the number of different OSN analyzed and the number of different data format taken into account. For this reason, this dimension can be defined as:

\begin{equation}
 d_{Variety}(t) = \frac{\sum_{i=1}^{3}{F_{\textit{Var}_i}}(t)}{3}
 \label{eq:variety}
\end{equation}



Following Eq.~\ref{eq:variety}, those algorithms, and tools, using different types of data extracted from different OSN will perform higher than others using different types of data extracted from the same OSN. Finally, the fraction is used to normalize the value of $d_\textit{Variety}$, thus this dimension is assessed using this degree in a range of [0,1] ($d_{Variety}(t) \in [0,1]$).

\subsection{\texorpdfstring{Visualization ($D_4$)}{Visualization (D4)}}
\label{subsec:dimension4}

\begin{figure}[!h]
	\centering
	\includegraphics[width=0.3\linewidth]{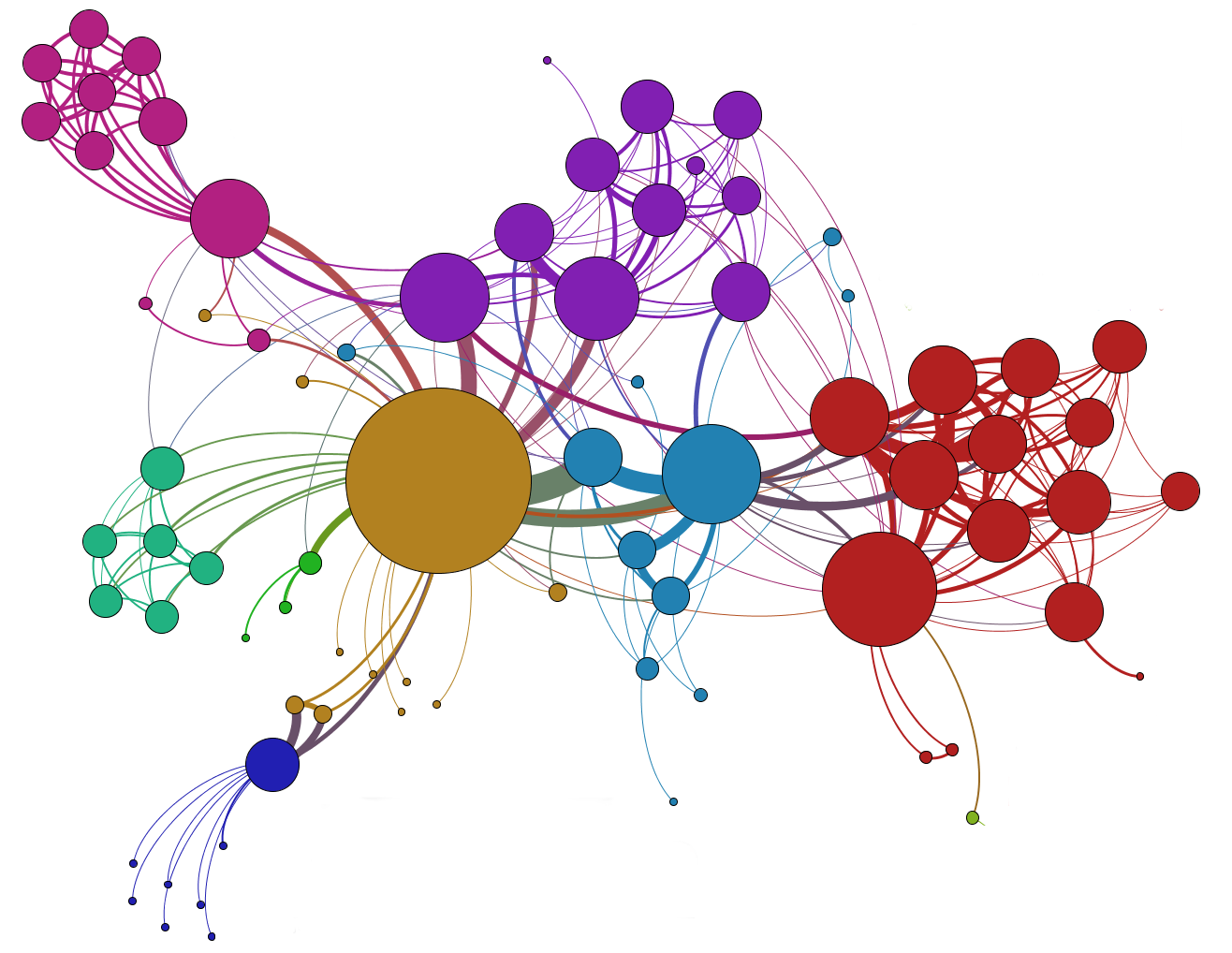}
	\includegraphics[width=0.29\linewidth]{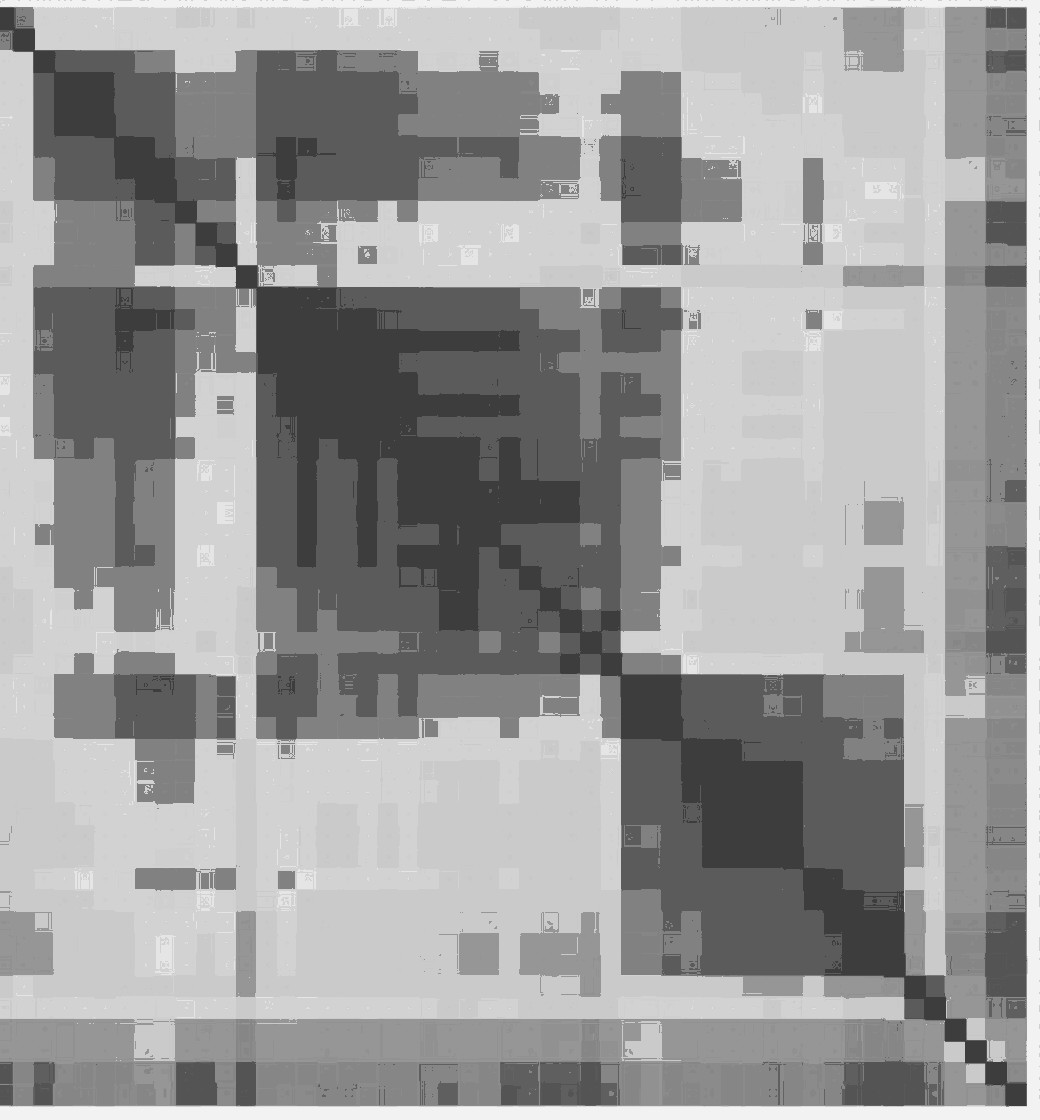}
	\includegraphics[width=0.3\linewidth]{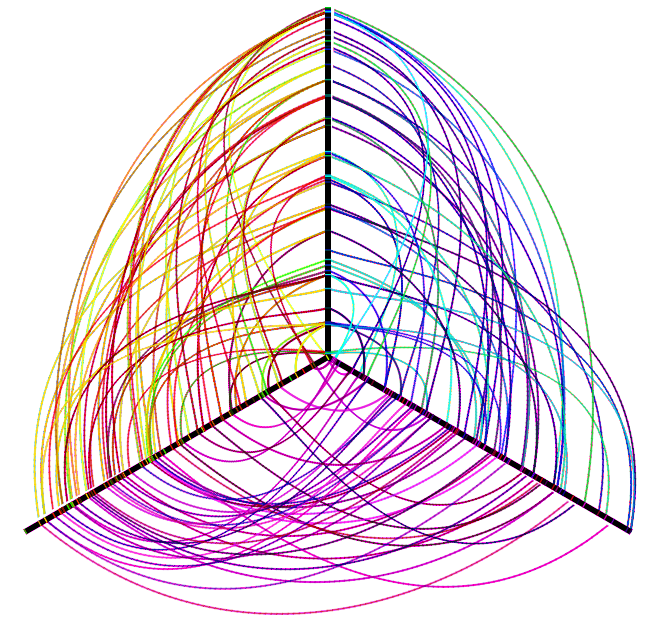}
	\caption{Subfigure a) shows an example of a node-link diagram; Subfigure b) shows an example of a matrix diagram; and Subfigure c) shows an example of an alternative diagram named HivePlot~\cite{krzywinski2011hive}.}
\label{fig:diagrams-types}
\end{figure}

What can the system show from an OSN is one of the key issues for any researcher, analyst or end-user that works in the domain of SNA. The concept of Visualization is used as a dimension to measure the capacity of the tools, frameworks, and methods to visually represent the information stored in the network~\cite{freeman2000visualizing}. Due to the human brain, it is easier for everybody to visualize large amount of data instead of reading tables or reports. For this reason, data visualization is a quick and easy way to convey concepts in a universal manner. In order to create good visualizations, one must first decide which questions want to be answered and select compelling visual encodings that depict the data values as graphical features such as position, size or orientation. Although the amount of possible visualization designs is extremely large, statisticians, psychologists, and computer scientists have studied the most suitable representations for different types of data, easing the difficulty of choosing a proper visual encoding. Regarding visualizing OSNs, the most common diagrams found in the literature are: \emph{node-link} and \emph{matrix} diagrams~\citep{heer2010tour, karampelas2014visual}. Although, lately, new \emph{alternative} diagrams have been proposed that fall outside any of these two categories~\citep{krzywinski2011hive, longabaugh2012combing, wattenberg2006visual}. An example of these diagrams can be found at Fig.~\ref{fig:diagrams-types}, where a node-link diagram (Fig.~\ref{fig:diagrams-types}.a), a matrix diagram (Fig.~\ref{fig:diagrams-types}.b) and the novel HivePlot diagram (Fig.~\ref{fig:diagrams-types}.c) are shown.


The most common visual representation of an OSN is the node-link diagram (see Fig.~\ref{fig:diagrams-types}.a), where individuals of the network are represented as dots, and a relation between two individuals is represented by an arrow connecting them. Depending on the relationship, the edge connecting two nodes can be directed or not. Placing the individuals in such a way that the resulting diagram is legible is not a trivial task and many efforts have been made to ensure this. Since the 80's many publications have addressed this task and a plethora of methods have been proposed~\cite{eades1984heuristic, fruchterman1991graph, kamada1989algorithm, hachul2004drawing, gansner2012maxent, Mathieu2014}. The main approach followed so far considers the network as a physical system where forces are applied to the individuals moving them around a 2D or 3D plane~\cite{hu2015visualizing}, some examples are the Spring-Electrical or the Stress and Strain models. In the former (Spring-Electrical models), individuals are simulated as positive charges and connections between them as springs. First, all users are allocated in a random position. Then, the physical simulation takes place, individuals repels other individuals around them and the springs avoid separating connected individuals too far apart. The simulation continue for a predefined number of steps or until the system stabilizes. The most famous algorithm is this category is the Fruchterman-Reingold~\cite{fruchterman1991graph}. In the latter, the Stress and Strain models, the positive charges are discarded and only the springs are used. These models, first define the desired spring length. Then, the individuals are allocated at random positions. Contrary to the spring-electrical model, there are no simulation and an ``imbalance degree" is defined as the difference between each spring length and the desired spring length. Finally, the ``imbalanced degree" is minimized changing the individual's positions. The most famous algorithm in this category is the Kamada and Kawai~\cite{kamada1989algorithm}. Node-link network visualization is an active research field and a wide variety of methods have been proposed to minimize their computational requirements, or to improve their output on bigger networks~\cite{hachul2004drawing, gansner2012maxent, Mathieu2014}.

Although the node-link diagram with the physical systems simulations works very well for a wide number of different networks, they usually have trouble visualizing small-world networks~\citep{Watts1998Collective}. In a small-world networks, any two nodes are not very far from each other (i.e., there are short path lengths). For that reason, the use of node-link diagrams with the physical system analogy only produces a mess similar to a hairball where it is impossible to see anything. This is a big problem in the area of SNA as, usually, OSN are also small-world networks. Thus, to solve this shortcoming, some alternative representations have been proposed, like the matrix diagrams (see Fig.~\ref{fig:diagrams-types}.b). These diagrams represent the networks by their adjacency matrix. The adjacency matrix of a network has all its users as rows and columns, and the value of each cell represents the strength of the connection between the two corresponding individuals. In a matrix diagram, the adjacency matrix of a network is drawn as a picture, where each cell of it is represented as a colored square depending on its strength. This representation has its advantages and disadvantages. On the one hand, all the connections are always visible, regardless of the network size, in the node-link diagram connections cross one another and, sometimes, it can be impossible to distinguish them. On the other hand, identifying paths in the network is harder than ordering the rows and columns of the adjacency matrix, this process is called \emph{Seriation}, which is essential to unveil the inner structure of a network. \emph{Seriation} is not a trivial task, in fact, it is an active research area where a large number publications are coming out to tackle this problem~\cite{mueller2007comparison}.

There is not a consensus about what is the best strategy for visualizing networks as both of the aforementioned approaches have their advantages and disadvantages. In addition, some hybrid approaches have also been proposed like, MatrixExplorer~\cite{henry2006matrixexplorer} or Matlink~\cite{henry2007matlink}. Finally, within this debate, there are some scientific groups that are proposing alternative diagrams, tailored for specific areas, that deviates from the mainstream. For example, in the Biology area, we can find: HivePlots~\cite{krzywinski2011hive}, where nodes are mapped to and positioned on radially distributed linear axes and edges are drawn as curved links; or BioFabric~\cite{longabaugh2012combing} where nodes are depicted as horizontal lines, not as points. Another example is PivotGraph~\cite{wattenberg2006visual}, that is specialized in multivariate data.

How the network is represented is not the only characteristic used to compare SNA visualization tools, but there are other characteristics that need to be taken into account~\citep{hu2015visualizing, chen2017social, pavlopoulos2017empirical}. These characteristics are: ``volume", ``summary" and ``interaction". The first one, volume, asses the raw processing capacity of a tool, how many users/nodes and connections/edges can be handled by the tool. Although the capacity of visualizing millions of nodes does not imply the generation of outputs in such a way any human can use. Most of the times, visualizing a medium-size network, without any extra help, provides as a result an unreadable hairball. In this situation the second characteristic, name ``summary" becomes useful. ``Summary" evaluates the capacity of a tool to generate simplifications that allows to reduce the complexity of the network while maintaining as much information as possible. This balance, between the summary and the raw data, must not be static and should adapt to the user's requirements. The third and last characteristic, ``interaction", measures the capacity of a tool to adapt its graphical output to the user's needs. These three characteristics together allow us to fulfill the Shneiderman's visualization mantra: ``Overview first, then zoom and filter details on-demand"~\citep{shneiderman2003eyes}. Therefore, these three characteristics have been considered as the essential ones that any SNA visualization tool should have. 

Although these three characteristics are great for describing and comparing SNA tools, two of them, ``volume" and ``summary", are highly correlated with the other three dimensions presented in this work (Scalability, Pattern \& Knowledge discovery, and Information Fusion \& Integration). Therefore, we have decided to move away from the approach used in the literature and remove any aspect not related to graphics from the visualization dimension. Hence, we will describe the visualization dimension in a twofold way. On the one hand, we will use the aforementioned ``interaction" characteristic. On the other hand, the lack of consensus over what is the most suitable diagram to visualize an OSN makes impossible to use it as a comparison method. Thus, we have considered that all the aforementioned diagrams are equally valuable when visualizing an OSN. Therefore, we will introduce an extra characteristic, named ``\textit{Visual Variables}", that will help us to asses the information representation capacity of a tool. \textit{Visual Variables} were proposed in~\citep{bertin1983semiology} by Jacques Bertin based on his experience as a cartographer and geographer. In his work he described the \textit{visual variables} as the fundamental way in which graphic symbols can be distinguished. The author identified the Visual Variables listed below and, according to his work, it can be used for four different purposes: \begin{enumerate*}[label=(\arabic*)] \item Selective (easily distinguish between groups); \item Associative (identify changes among the same group); \item Ordered (allows to identify sequences); and, \item Quantitative (allows to compare numerical values)\end{enumerate*}. The visual variables can be summarized as follows:


\begin{enumerate}
  \item \textbf{Position} ($F_{\textit{VisVar}1}$): refers to the location of an object in the image, position is one of the most versatile variable as it can be used in as a Selective, Associative, Ordered or Quantitative variable. For example, if an OSN has geo-localization data, the position of an user can display his/her country.
  \item \textbf{Size} ($F_{\textit{VisVar}2}$): refers to the size variation of an object. It can be used as: Selective, Ordered or Quantitative variable. For example, changing the size of an user node depending on his/her degree.
  \item \textbf{Shape} ($F_{\textit{VisVar}3}$): refers to the different geometrical shapes an object can have (triangles, rectangles, circles\dots etc.). It can only be used as a Associative variable. For example, showing users from different OSN with a different shape.
  \item \textbf{Orientation} ($F_{\textit{VisVar}4}$): refers to the rotation an object presents. It can be used as a Selective or Associative variable. For example, using an arrow to indicate the direction of a following/follower relation.
  \item \textbf{Color} ($F_{\textit{VisVar}5}$): refers to the color hue of an object. It can be used as a Selective or Associative variable. For example, using the same color for all the users in the same community.
  \item \textbf{Saturation} ($F_{\textit{VisVar}6}$): refers to color saturation, the brighter or lighter a color hue is. It can be used as a Selective, Ordered or Quantitative variable. For example, showing users with a higher centrality colored with a more saturated color.
  \item \textbf{Texture} ($F_{\textit{VisVar}7}$): refers to the fill pattern of an object. It can be used as a Selective or Associative variable. For example, using two different patterns to identify members of different political parties.
\end{enumerate}

In our opinion previous variables are suitable to act as measures of the visualization dimension. Instead of trying to evaluate the quality of a visualization method by the type of representation used, we will evaluate it by the number of visual variables used by a tool to enrich them. The more visual variables a tool is able to handle, the more extra information it can visually represented, and the higher the visualization degree will have the tool. However, as it has been stated by the literature, being able to represent a lot of information does not implies the generation of good quality visualizations. That is why the ``interaction" capabilities of a tool are also being considered when calculating a visualization degree of a tool. The most common interactions analyzed in the literature can be summarized with the next five actions:

\begin{enumerate}
  \item \textbf{Zoom} ($F_{\textit{Inter}1}$): refers to the action of changing the level of details of the elements being shown but maintaining the same number of them. This does not implies that all elements need to be visible in the screen at once. For example, using the mouse wheel to zoom in/out a graph.
  \item \textbf{Filter} ($F_{\textit{Inter}2}$): contrary to zoom, refers to changing the number of elements being displayed but maintaining the same level of detail. For example, dragging the mouse to pan over a diagram.
  \item \textbf{Highlight} ($F_{\textit{Inter}3}$): refers to the action of emphasizing some particular elements of a set. For example, highlighting the neighbors of an user when the mouse is positioned over it.
  \item \textbf{Grouping} ($F_{\textit{Inter}4}$): refers to the action of replacing a group of elements with a simplification that maintains all or some of the properties of the group. For example, joining all the users of the same country into a single node.
  \item \textbf{Multiview} ($F_{\textit{Inter}5}$): refers to the action of switching between multiple representations of the same data. For example, switching between the node-link and matrix representation of an OSN. 
\end{enumerate}

In order to generate a single value capable of evaluating the degree of the visualization dimension, Equation~\ref{eq:dim_visualize1} is proposed. Where $F_{Visual}$ is the number of visual variables a tool can handle, and $F_{Inter}$ is the number of interactions available on a tool, whereas $\alpha$, $\beta$, $\gamma$, $\theta$ are weights that represent the importance given to each characteristic.

\begin{equation}
  d_{Visual}(t) = \alpha \cdot \frac{\sum_{i=1}^{7}\gamma_i \cdot F_{\textit{VisVar}i}(t)}{7} + \beta \cdot \frac{\sum_{j=1}^{5}\theta_j \cdot F_{\textit{Inter}j}(t)}{5}
\label{eq:dim_visualize1}
\end{equation}

 In this work, all of the visual characteristics (visual variables and interactions) will have the same weight (so $\alpha$ and $\beta$ will be set up to $0.5$), and all of the features for each characteristic will have the same importance (therefore $\gamma$ and $\theta$ will be set up to $1.0$).

\subsection{Summary on Social Network Analysis Dimensions}
\label{subsec:C4metrics}

Following the previous subsections, the four basic research questions (RQ) proposed in the introduction, and their related dimensions, have been mapped into a set of measures, or $\textit{SNA}_{\textit{degrees}}$, which can be used to provide a quantitative value for each dimension. These research questions, together with the proposed dimensions, and the metrics, or degrees, defined to assess them are shown in Table~\ref{tab:sumdimensions}. 

\begin{table}[h!]
\caption{Summary on dimensions and the quantitative measures (degrees) defined.}
\begin{center}
\begin{tabular}{|p{2.5cm}|p{2.6cm}|p{7.1cm}|c|}
\hline
 \textbf{Research Quest.} & \textbf{Dimension ($D_i$)} & \textbf{Degree ($d_v*$)} & \textbf{Range} \\
\hline\hline

 \hline What can I learn? & Pattern \& Know. discovery ($D_1$)&
 \centering $d_{Value}(t) = 1/3 \cdot ( \frac{\sum_{i=1}^{2}{F_{\textit{Val}(1,j)}}(t) + {F_{\textit{Val}(3,j)}}(t)}{2} + \frac{\sum_{i=1}^{3}{F_{\textit{Val}(2,i)}(t)}} {3} )$ & $[0,1]$ \\
 
 \hline What is the limit? & Scalability ($D_2$)& \centering $d_{Volume}(t) = \frac{\sum_{i=1}^{4}{F_{\textit{Volume}i}}(t)}{4}$ & $[0,1]$ \\
 
 \hline What kind of data can I integrate? & Information Fusion \& Integration ($D_3$)& \centering $d_{Variety}(t) = \frac{\sum_{i=1}^{3}{F_{\textit{Var}i}}(t)}{3}$ & $[0,1]$ \\
 
 
 
 \hline What can I show? & Visualization ($D_4$)& \centering $d_{Visual}(t) = 1/2 \cdot \frac{\sum_{i=1}^{7} F_{\textit{VisVar}i}(t)}{7} + 1/2 \cdot \frac{\sum_{j=1}^{5} F_{\textit{Inter}j}(t)}{5}$ & $[0,1]$ \\

\hline
\end{tabular}
\label{tab:sumdimensions}
\end{center}
\end{table}	
 
\begin{figure}[!h]
	\centering
		\begin{tabular}{ccc}
		\subfigure[]{
			\includegraphics[width=0.3\linewidth]{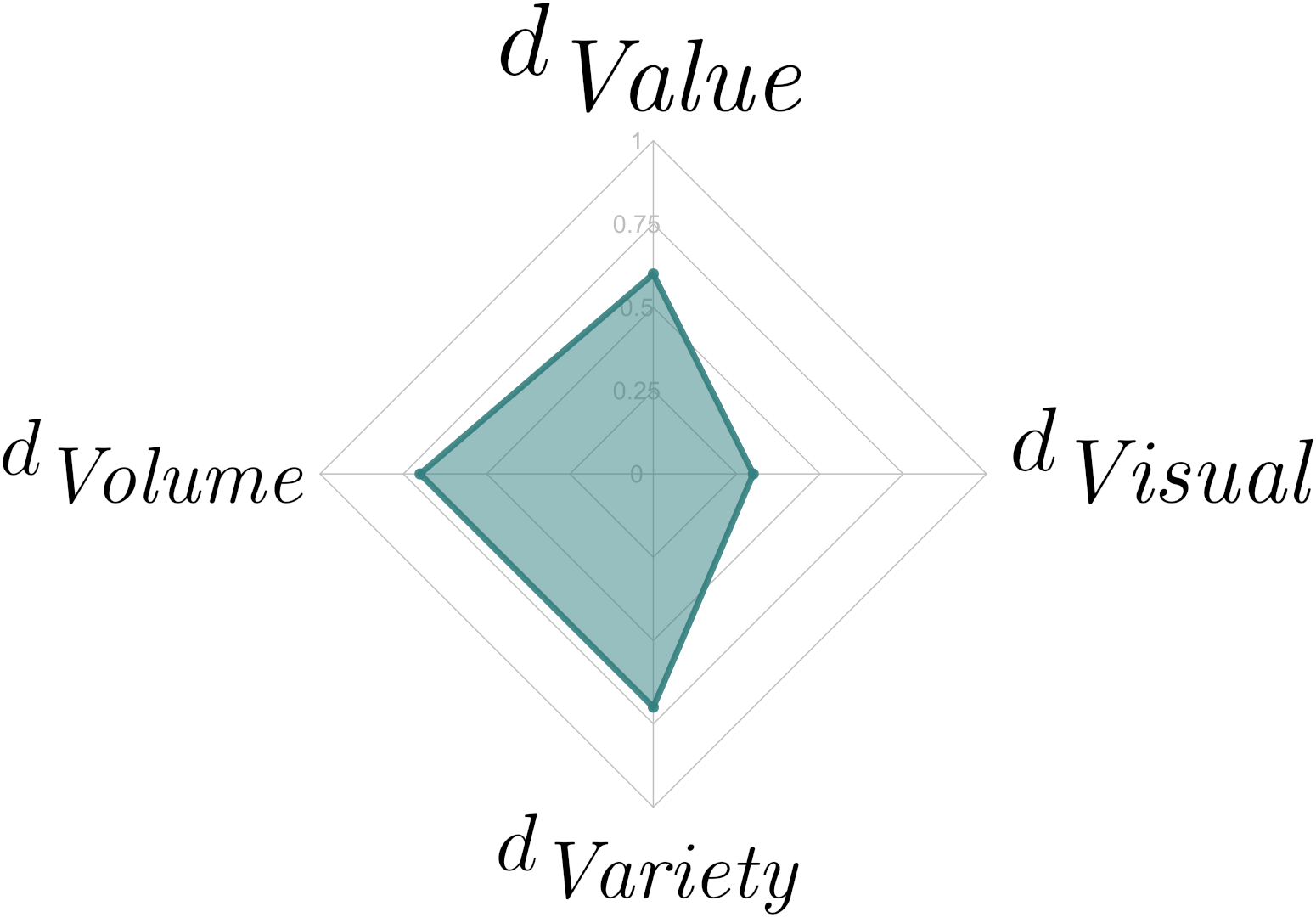}
		}
		\subfigure[]{
			\includegraphics[width=0.3\linewidth]{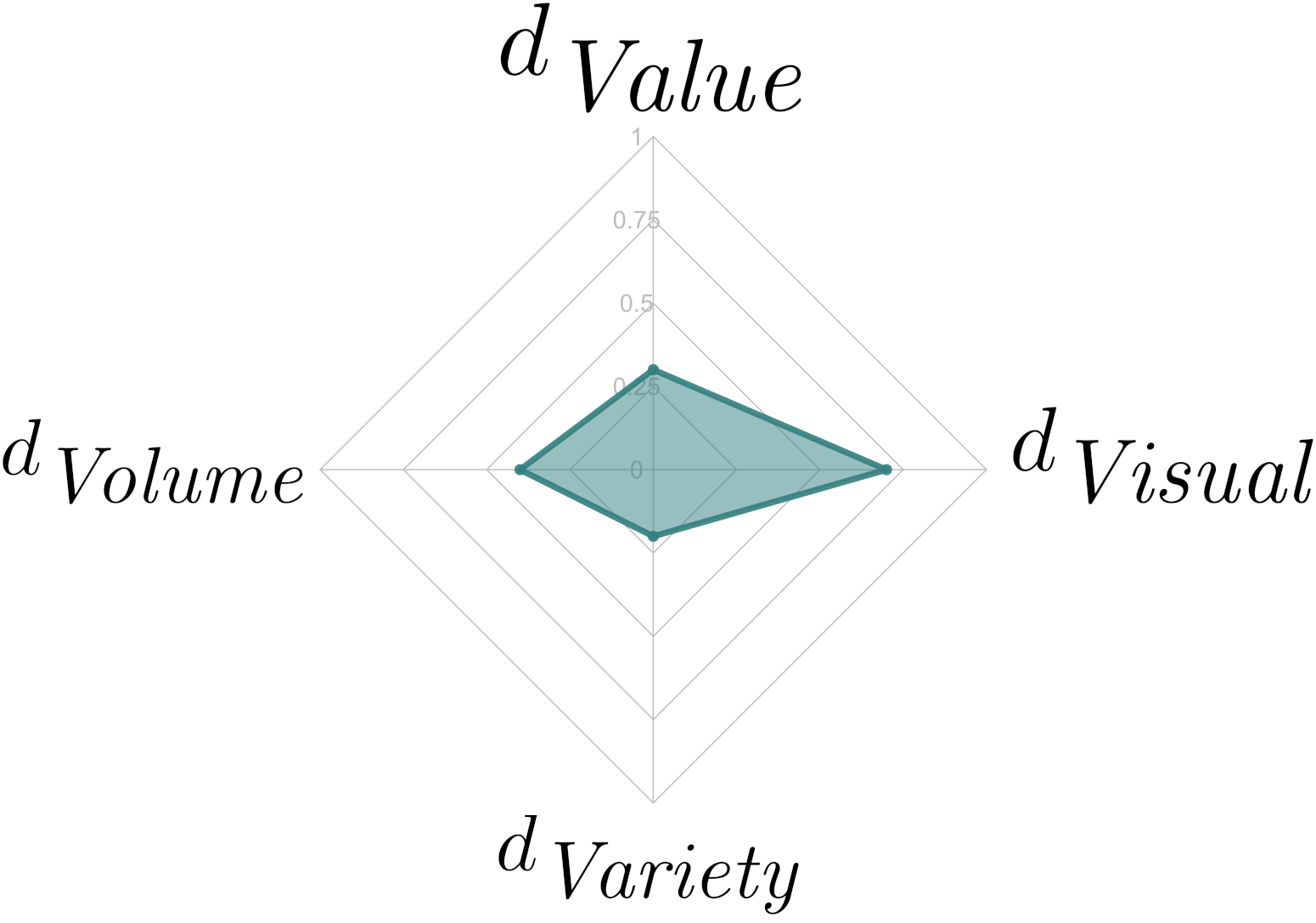}
		}
		\subfigure[]{
		  \includegraphics[width=0.3\linewidth]{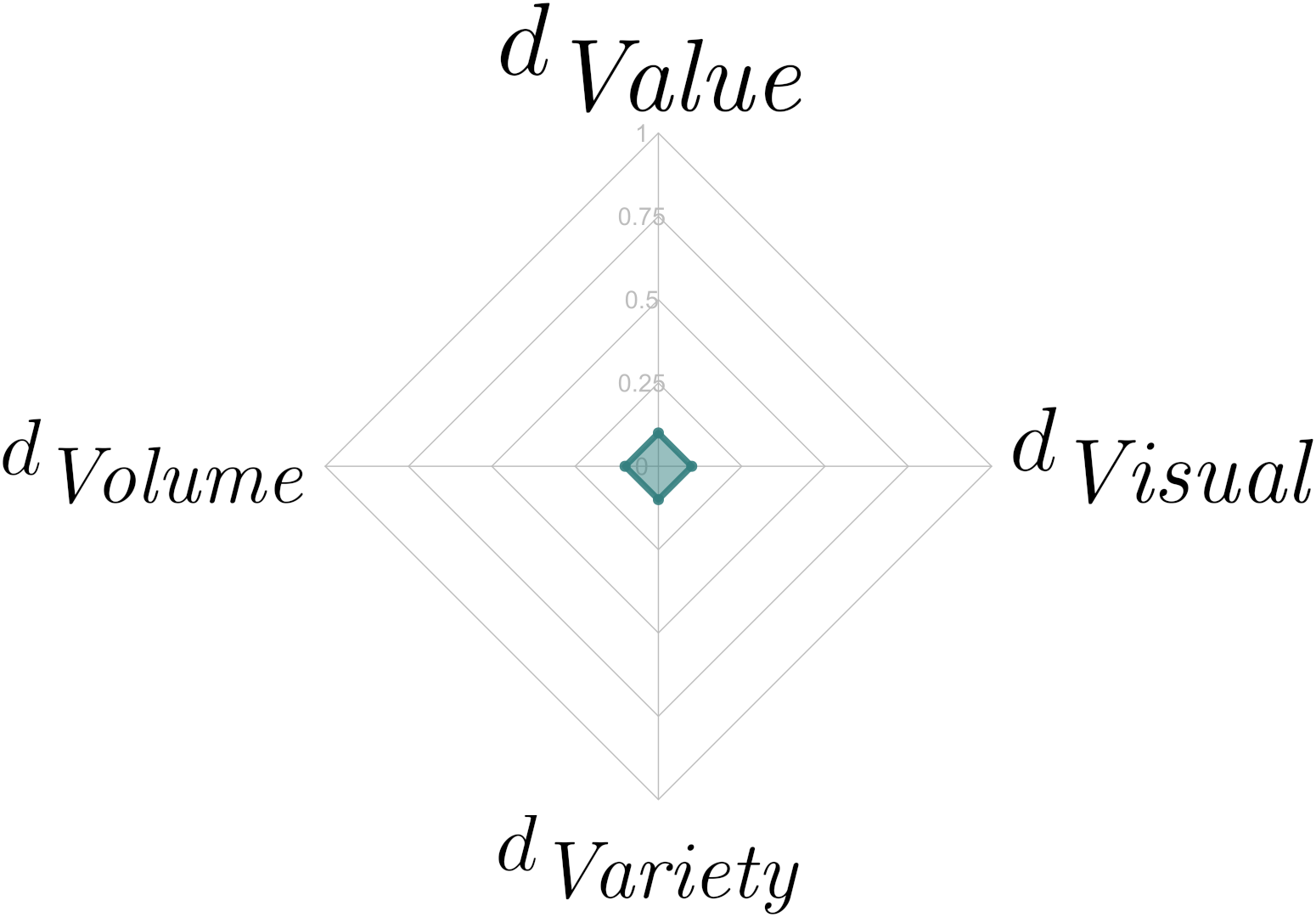}
		}
		\end{tabular}
		\caption{Spider diagrams representing the evaluation of three (hypothetical) SNA platforms by using the $4$ dimensions described in this work. The exact values for each dimensions for these $3$ hypothetical frameworks are shown in Table~\ref{tab:C_SNA metric}. The figures shown correspond to a good platform (a), an intermediate tool (b) and a poor platform (c).}
    \label{fig:spiderchart}
\end{figure}

Fig.~\ref{fig:spiderchart} shows an example of an hypothetical algorithm, framework and a tool, and how the proposed dimensions, can be used to represent the technology maturity to work with OSN data. We have decided to use a spider, or radar, diagram for representing the evaluation of the SNA technology. As can be seen in this figure, the axes will be used to represent each dimension, so the quality, strengths or weaknesses for each method, tool or framework can be easily analyzed.

Finally, and considering these $\textit{SNA}_{\textit{degrees}}$ it is quite straightforward to define a new \textbf{global metric}, that we have called \textgoth{C}$_{\textit{SNA}}$, to represent the ``\textit{Capability}" and power to work with OSN sources, to later use as a metric to rank the technologies analyzed. We calculate the value of \textgoth{C}${i}$, where $i$ represents the number of dimensions to be considered, as the area contained in the irregular polygon defined by the $i$ dimensions used in our previous representation (see Fig.~\ref{fig:spiderchart}). This equation comes from the Shoelace formula, also known as Gauss's area formula and the surveyor's formula~\citep{braden1986surveyor}, and it is a simple formula for finding the area of a polygon given the coordinates of its vertices, see Equation~\ref{eq:areapolygon}, where $A$ is the area of the polygon, $n$ is the number of sides, and $(x_i, y_i)$ are the vertices of the polygon.



\begin{equation}
  A = \frac12 \left| \sum_{i=1}^{n-1} (x_{i} y_{i+1} + x_{n} y_1 ) - \sum_{i=1}^{n-1} (x_{i+1} y_{i} - x_{1} y_{n}) \right|
\label{eq:areapolygon}
\end{equation}




Therefore, the \textgoth{C}$_{i}$ metric is a particular case of previous Equation with $n = 4$, and it can be mapped as Equation~\ref{eq:metricSNA} shows. Taking into account that all of the dimensions defined are represented in an axis, all of the vertices will have at least one of the coordinates ($x$ or $y$) to $0$, due all of the dimensions defined in this work have been normalized to $1$, implies that \textgoth{C}$_{i} \in [0,2]$. Considering: $d_{Value} = (x_1, 0)$, $d_{Volume} = (0, y_2)$, $d_{Variety} = (x_3, 0)$, $d_{Visual} = (0, y_4)$, Equation~\ref{eq:metricSNA} can be easily simplified as Equation~\ref{eq:metricC4} shows.

\begin{equation}
  \textgoth{C}_4(t) = \frac{1}{2} \cdot \left| \sum_{i=1=\textit{Value}}^{3=\textit{Variety}} \Big (d_i^{x_i}(t) \cdot d_i^{y_{i+1}}(t) + d_i^{x_n}(t)\cdot d_i^{y_1}(t) \Big ) - \sum_{i=1=\textit{Value}}^{3=\textit{Variety}} \Big (d_i^{x_{i+1}}(t) \cdot d_i^{y_i}(t) - d_i^{x_1}(t)\cdot d_i^{yx_n}(t) \Big ) \right| 
\label{eq:metricSNA}
\end{equation}

\begin{equation}
  \textgoth{C}_4(t) = \frac12 \cdot \left| \Big (d_{\textit{Value}}^{x_1}(t) + d_{\textit{Volume}}^{y_2}(t) \Big ) \cdot \Big (d_{\textit{Variety}}^{x_3}(t) + d_{\textit{Visual}}^{y_4}(t) \Big )\right|
\label{eq:metricC4}
\end{equation}

Analyzing Equation~\ref{eq:metricC4}, two different (extreme) cases could appear: \begin{enumerate*}[label=(\arabic*)] \item if two of the dimensions are equal to zero (i.e., $d_{\textit{Value}} = d_{\textit{Volume}} = 0$, or $d_{\textit{Variety}} = d_{\textit{Visual}} = 0$), $C_{4}$ will be equal to $0$, in such case the \textgoth{C}$_{4}$ must be calculated as Equation~\ref{eq:metric2dC4} shows (this is equivalent to make a projection of both non-zero dimensions into two different axis). \item If three dimensions are zero, there's no possibility to calculate any area due there's only one dimension with a value, in such case the $\textit{SNA}_{\textit{degree}}$ should be used to evaluate the capability (only in that dimension) of the SNA technology. Fig.~\ref{fig:spiderchart} shows an example of our Capability metric on three hypothetical frameworks).\end{enumerate*}

\begin{equation}
  \textgoth{C}_4(t) = \frac12 \cdot \left| d_{i}^{x_i}(t) \cdot d_{j}^{y_j}(t) \right|, d_{i}, d{i} \neq 0
\label{eq:metric2dC4}
\end{equation}

Following our previous example, the three different SNA technologies previously presented in Fig.~\ref{fig:spiderchart} would be ranked using our metric (see Table~\ref{tab:C_SNA metric}). Although this comparison would not have sense (as we are comparing different kinds of technology), it shows how this general metric can be used to better understand the capability of a particular SNA technology. This new metric will be used in next sections to provide a ranking between the analyzed SNA tools and frameworks.

\begin{table}[h!]
\caption{Example of \textgoth{C}$_{4}$ metric application over three different hypothetical SNA tools and frameworks.}
\begin{center}
\begin{tabular}{| c|c|c c c c |l|}
\hline
\multirow{2}{*}{\textbf{Rank}} & \multirow{2}{*}{\textgoth{C}$_{4}(t)$} & \multicolumn{4}{c|}{\textbf{Dimensions}} & \multirow{2}{*}{\textbf{SNA technology}} \\ \cline{3-6}
& & $d_{Value}$ & $d_{Volume}$ & $d_{Variety}$ & $d_{Visual}$ & \\ 
\hline 
\hline

1 & $0.315$ & 0.6 & $0.7$ & 0.7 & 0.3 & Tool 1 (high value)\\ \hline
2 & 0.170 & 0.3 & 0.4 & 0.2 & 0.7 & Framework 1 (medium value)\\ \hline
3 & 0.010 & 0.1 & 0.1 & 0.1 & 0.1 & Tool 2 (low value)\\
 

\hline
\end{tabular}
\label{tab:C_SNA metric}
\end{center}
\end{table}





%% file: sections/sna-tools.tex
\section{Frameworks \& Tools Analysis}
\label{sec:sna-tools}

This section provides a review on a set of popular SNA frameworks and tools, which are extensively used by the research community and industry. Due, it would be highly difficult to analyze all of the currently available tools, we have selected for analyzing a subset of 20 representatives. In this set, it can be found computing libraries, web applications, distributed applications, or desktop applications among others, we will refer to all of them as \textit{SNA-software}. Initially, a list of 70 SNA-software candidates was generated using Github\furl{github.com/briatte/awesome-network-analysis}, KDnuggets\furl{kdnuggets.com/2015/06/top-30-social-network-analysis-visualization-tools.html/2}$^,$\furl{kdnuggets.com/software/social-network-analysis.html} and the Infovis-Wiki\furl{infovis-wiki.net/wiki/Main_Page} websites. It is important to highlight that we have considered SNA-software with both types of licenses, open source and commercial (proprietary). From the list of 70 SNA-software candidates, a set of 20 tools was selected for analysis. In this selection, it was considered the type of software license, the quantity and \textit{quality} of the software documentation, and its current impact between the community (taking into account the popularity of some tools in published works, websites and other technical material). Following, the list containing the $20$ different SNA-software analyzed in this work is shown. For each software, we provide the name, a bibliographical reference (or a website), a brief description and also, its license type:

\begin{enumerate}[noitemsep,nolistsep]
  \item \textbf{Igraph}~\cite{csardi2006igraph}: a collection of network analysis tools with the emphasis on efficiency, portability and ease of use. License: MIT. 
  
  \item \textbf{AllegroGraph}~\cite{aasman2006allegro}: an ultra scalable, high-performance, and transactional Semantic Graph Database. License: Proprietary.
  
  \item \textbf{LaNet-vi}~\cite{alvarez2006lanet}: a large networks visualization tool. It provides images of large scale networks on a two-dimensional layout. License: AFL. 
  
  \item \textbf{Stanford Network analysis Platform (SNAP)}~\cite{leskovec2016snap}: a general purpose, high performance system for analysis and manipulation of large networks. License: BSD.

  \item \textbf{ORA-LITE/PRO}furl{casos.cs.cmu.edu/index.php}: a dynamic meta-network assessment and analysis tool developed by CASOS at Carnegie Mellon. It contains hundreds of OSNs, dynamic network metrics, trail metrics (path-based metrics), and procedures for grouping nodes. License: Proprietary.
  
  \item \textbf{Network workbench}~\cite{pullen2000network}: a Large-Scale Network analysis, Modeling and Visualization Toolkit for Biomedical, Social Science and Physics Research. License: open-source.
  
  \item \textbf{NetMiner}furl{netminer.com}: a premium software tool for Exploratory analysis and Visualization of Network Data. License: Proprietary.
  
  \item \textbf{Circulo}~\cite{mazzucat}: a ``Community Detection" Evaluation Framework written primarily in Python. License: Apache 2.0.
  
  \item \textbf{Cytoscape}~\cite{shannon2003cytoscape}: a software platform for computational biology and bioinformatics, useful for integrating data, and for visualizing and performing calculations on molecular interaction networks. License: LGPL. 
  
  \item \textbf{JUNG}~\cite{o2003jung}: a software library that provides a common and extendable language for the modeling, analysis, and visualization of data that can be represented as a graph or network. License BSD.
  
  \item \textbf{SparklingGraph}~\cite{sparkling-graph}: a Cross-platform tool to perform large-scale, distributed network computations with Apache Spark's GraphX module. License: BSD 2.
  
  \item \textbf{NetworkX}~\cite{hagberg2008exploring}: a Python language software package for the creation, manipulation, and study of the structure, dynamics, and functions of complex networks. License: BSD.
  
  \item \textbf{Pajek}~\cite{batagelj2004pajek}: is a Windows program for analysis and visualization of large networks. License: Unknown.
  
  \item \textbf{GraphX Apache Spark}~\cite{gonzalez2014graphx}: a module to perform graph-related parallel computation. License: Apache 2.0.
  
  \item \textbf{Gephi}~\cite{bastian2009gephi}: an open-source network analysis and visualization software package written in Java on the NetBeans platform. License GPL3.
  
  \item \textbf{UCINET}~\cite{borgatti2014ucinet}: a software package for the analysis of OSN data. It comes with the NetDraw network visualization tool. License: Proprietary.
  
  \item \textbf{Prefuse}~\cite{heer2005prefuse}: a Java-based toolkit for the interactive creation of information visualization applications (not only for graphs, but also tables and trees). License: Unknown.
  
  \item \textbf{Graphistry}\furl{graphistry.com}: a cloud service that automatically transforms your data into interactive, visual investigation maps built for the needs of analysts. License: Proprietary.
  
  \item \textbf{GraphViz}~\cite{ellson2001graphviz}: a open source graph visualization software. License: CPL.
  
  \item \textbf{Neo4j}~\cite{lal2015neo4j}: an Open source, scalable graph database. License GPL3.\\
  
\end{enumerate}


Once the list of SNA-software has been briefly described, each software has been evaluated using the different metrics proposed in Section~\ref{sec:4dimensions}. The evaluation process carried out in this work follows a ``top-down" approach. First, the global capability metric ($\textgoth{C}_4$) is analyzed for each framework and tool. In a second step, we have analyzed in detail the different dimensions (i.e., the $\textit{SNA}_{\textit{degrees}}$) that compose the global metric. 
The evaluation process for each software was carried out as follows: once the initial set of tools was selected, the authors agreed an evaluation rubric (that is publicly available, see Appendix) to assess the tool. This rubric is based on the analysis of the software documentation, their official websites (or any related site that could store relevant information), and other published works that provide technical details about these tools. From this technical documentation, we assess each of the characteristics that form the different $\textit{SNA}_{\textit{degrees}}$, to finally obtain a quantitative value for each of the proposed dimensions. The features used in this rubric (strictly) follows the characteristics proposed to measure the four dimensions (see Sections from~\ref{subsec:dimension1} to~\ref{subsec:dimension4}), so from these features we can obtain a quantitative value for each degree. Although the authors have previous experience in several of the analyzed tools (such as Igraph, Circulo, JUNG, or Gephi), it is not possible to download, install, and generate experimental datasets and evaluations for each SNA-software. For this reason, we decided to carry out the evaluation of the software following the previous process.



The goal of this analysis is twofold. On the one hand, it allows us to understand the strengths and weaknesses of the different SNA-software. This analysis will help any researcher who is looking for a SNA tool to select the one that best fit to his/her requirements. The type of license used by the SNA-software (public, open-source, BSD, MIT, proprietary ,etc.) can be a determining factor for future research, or the development of new products. Therefore, our analysis will take into account this feature to differentiate between those types of software. On the other hand, the analysis of the disaggregated values allows us to understand the opportunities, and weaknesses, of the tools from the SNA point of view. In this sense, this disaggregated analysis will allow us to understand if the requirements for a specific dimension (e.g., visualization) are currently fulfilled by the SNA-software available, or if there is any specific dimension that requires from some special reinforcement. Therefore, this second analysis will help the reader to detect spaces for improvement in some particular research areas related to SNA.


\begin{figure}
\centering
\includegraphics[width=0.7\linewidth]{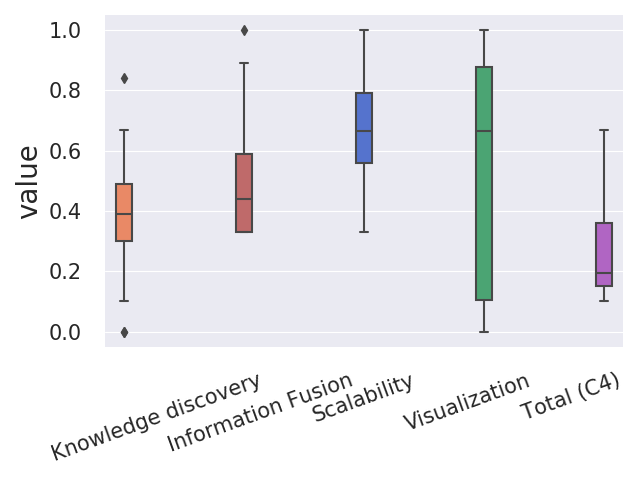}
\vspace{-3pt}
\caption{Distribution of the different dimensions values achieved by the analyzed tools. The X-axis represents the proposed dimensions, whereas the Y-axis shows the values obtained once the quantitative metrics ($\textit{SNA}_{\textit{degree}}$) are calculated.}
\label{fig:dimension-aggregate-boxplot}
\end{figure}







\begin{table*}[!ht]
\centering
\setlength\doublerulesep{0.3cm} 
\caption{Top-5 best SNA tools under Proprietary or Open-Source, and Only Open-source, licenses.}
\label{tab:top5}

\begin{tabular}{| r | c | c c c c |}
\hline
\multicolumn{6}{|c |}{\textbf{Proprietary or Open-Source}} \\
\hline
\textbf{Tool} & \textbf{\textgoth{C}$_{4}(t)$} & $d_{Val}$ & $d_{Var}$& $d_{Vol}$& $d_{Vis}$ \\ 

\hline

\textbf{Graphistry} & 0.67 & 0.33 & 1.0 & 1.0 & 1.0 \\ 
\textbf{Neo4j} & 0.57 & 0.48 & 0.56 & 1.0 & 1.0 \\ 
\textbf{ORA-LITE/PRO} & 0.56 & 0.84 & 0.67 & 0.5 & 1.0 \\ 
\textbf{NetMiner} & 0.52 & 0.65 & 0.89 & 0.67 & 0.69 \\ 
\textbf{Cytoscape} & 0.39 & 0.67 & 0.33 & 0.58 & 0.93 \\ 

\hline
\hline
\multicolumn{6}{|c|}{\textbf{Only Open-Source}}\\ \hline
\textbf{Tool} & \textbf{\textgoth{C}$_{4}(t)$} & $d_{Val}$ & $d_{Var}$& $d_{Vol}$& $d_{Vis}$\\
\hline 
\textbf{Neo4j} & 0.57 & 0.48 & 0.56 & 1.0 & 1.0 \\
\textbf{Cytoscape} & 0.39 & 0.67 & 0.33 & 0.58 & 0.93 \\
\textbf{Gephi} & 0.35 & 0.35 & 0.44 & 0.66 & 0.93 \\
\textbf{Pajek} & 0.31 & 0.48 & 0.56 & 0.50 & 0.73 \\
\textbf{JUNG} & 0.28 & 0.41 & 0.33 & 0.75 & 0.64\\ \hline
\end{tabular}
\end{table*}

\begin{figure}[!ht]
	\centering
		\begin{tabular}{ccc}
		\subfigure[Graphistry]{
			\includegraphics[width=0.3\linewidth]{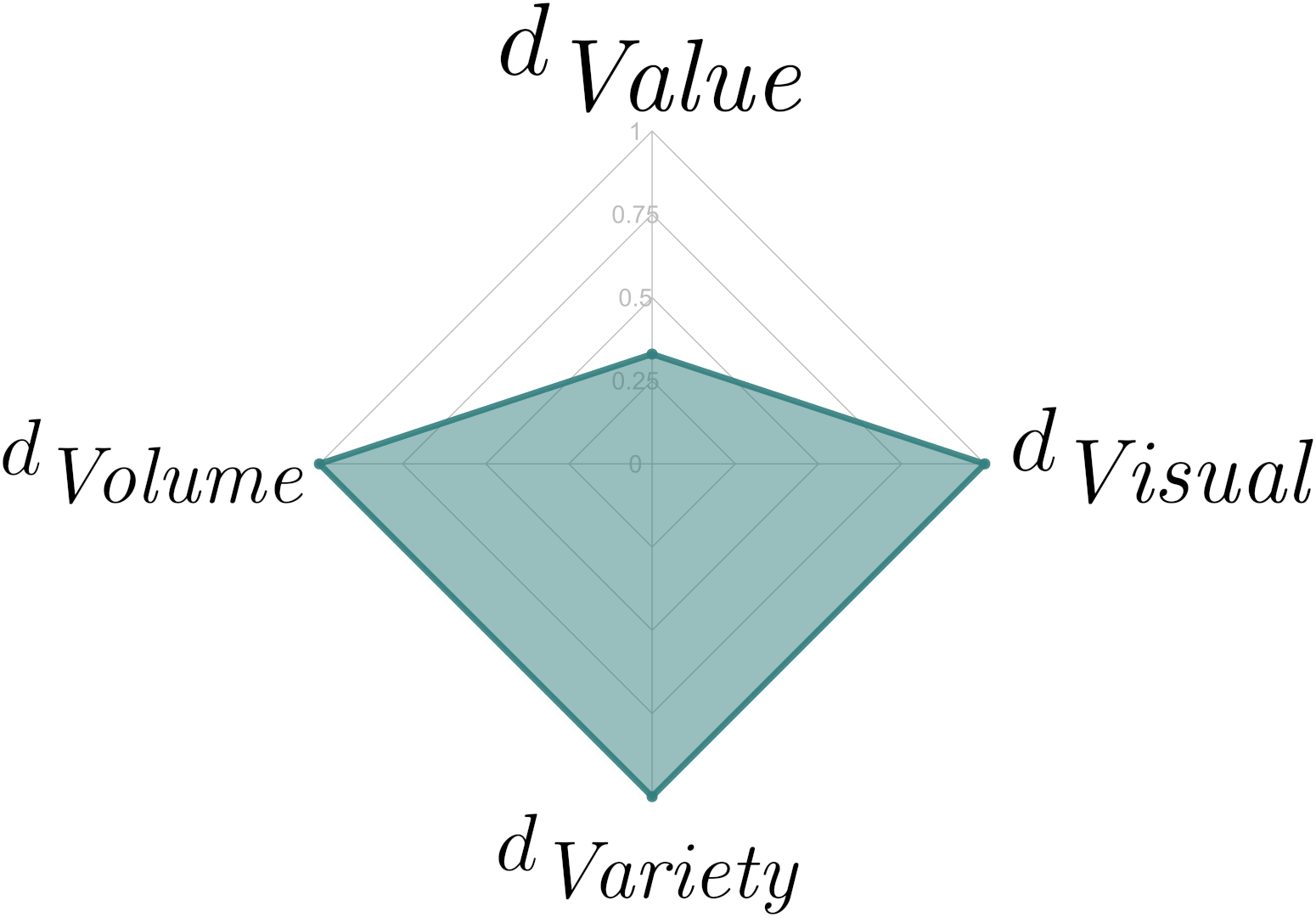}
		}
		\subfigure[Neo4j]{
			\includegraphics[width=0.3\linewidth]{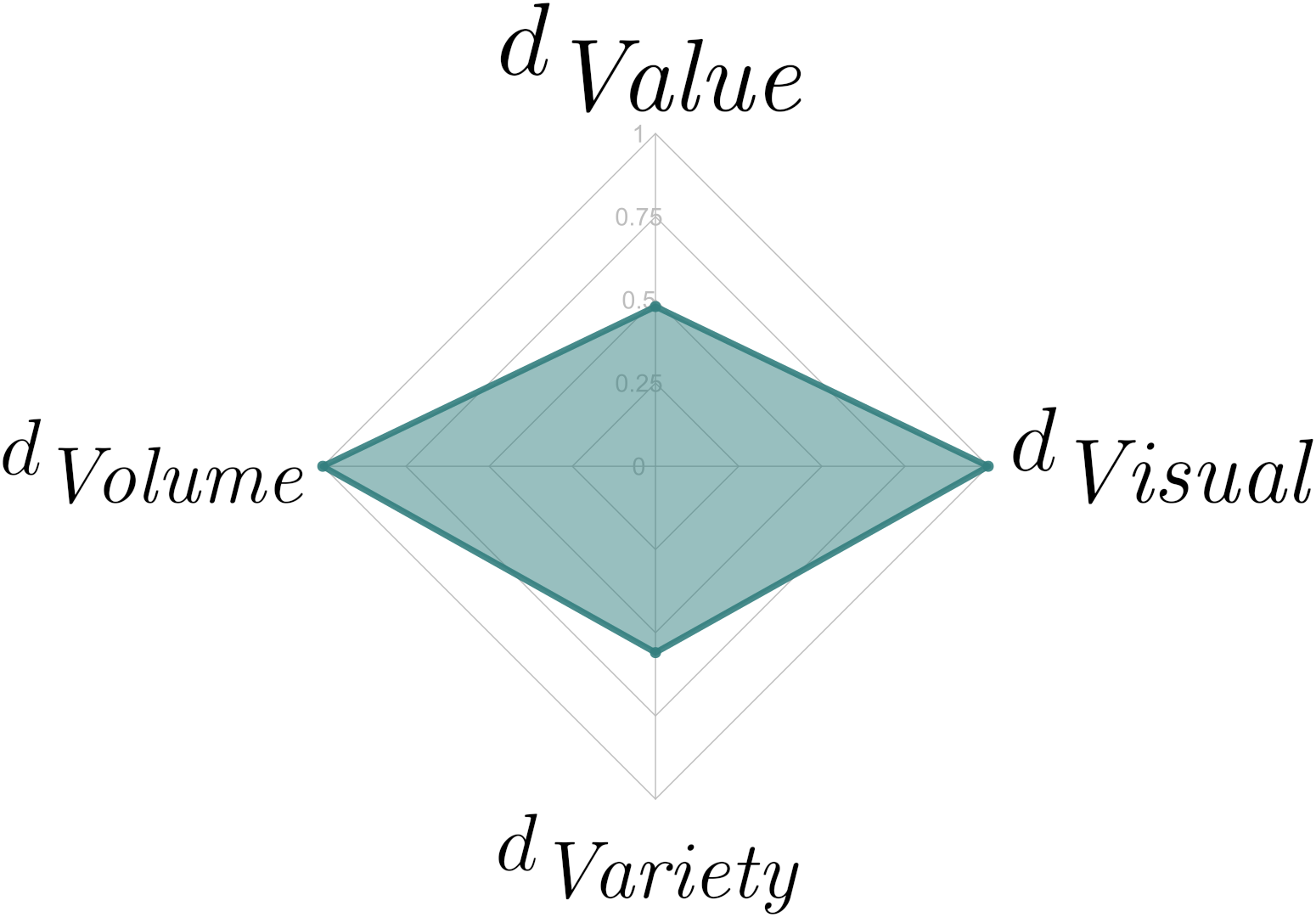}
		}
		\subfigure[ORA-LITE/Pro]{
		  \includegraphics[width=0.3\linewidth]{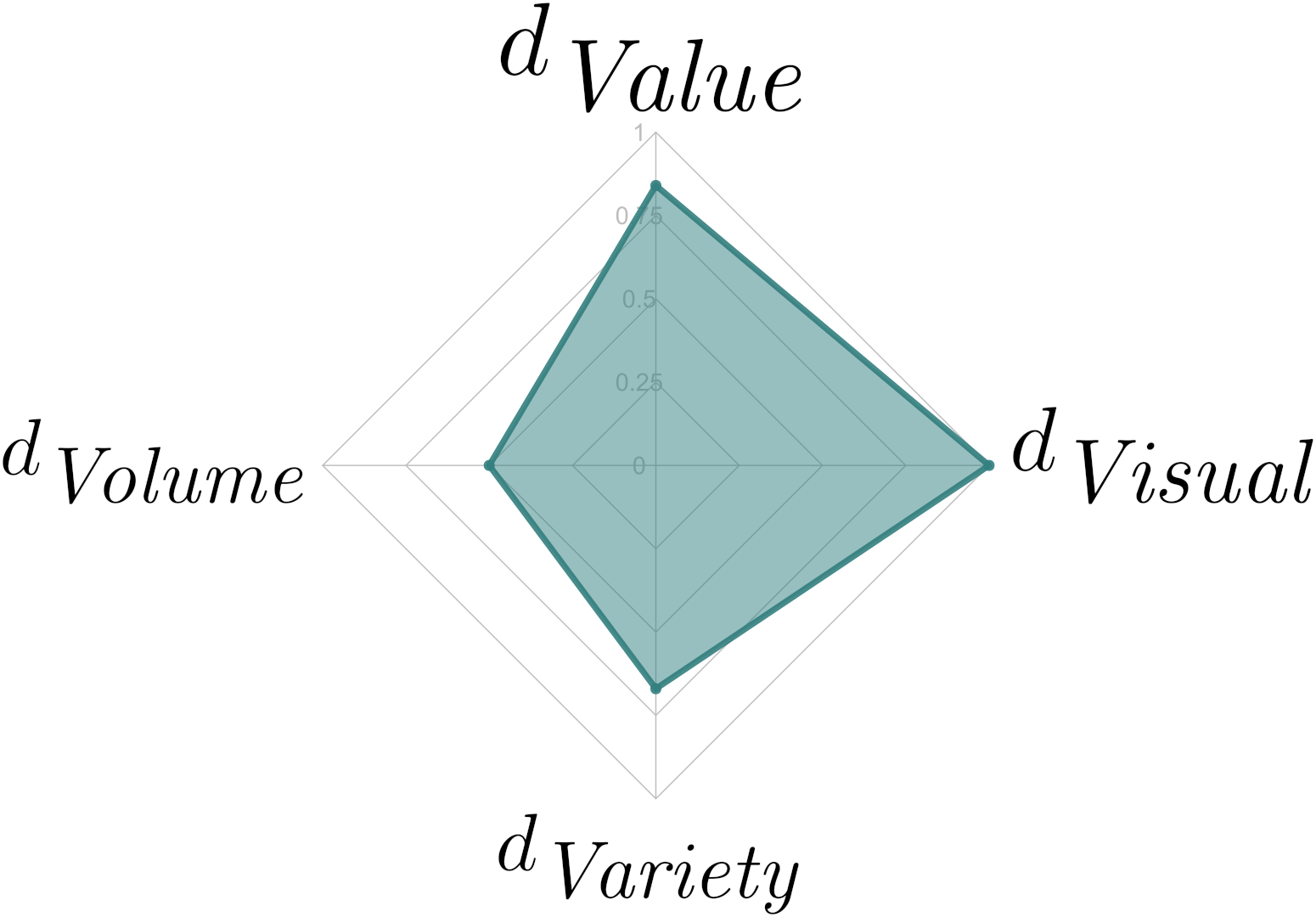}
		}\\
		\subfigure[Neo4j]{
			\includegraphics[width=0.3\linewidth]{images/Neo4j.png}
		}
		\subfigure[Cytoscape]{
			\includegraphics[width=0.3\linewidth]{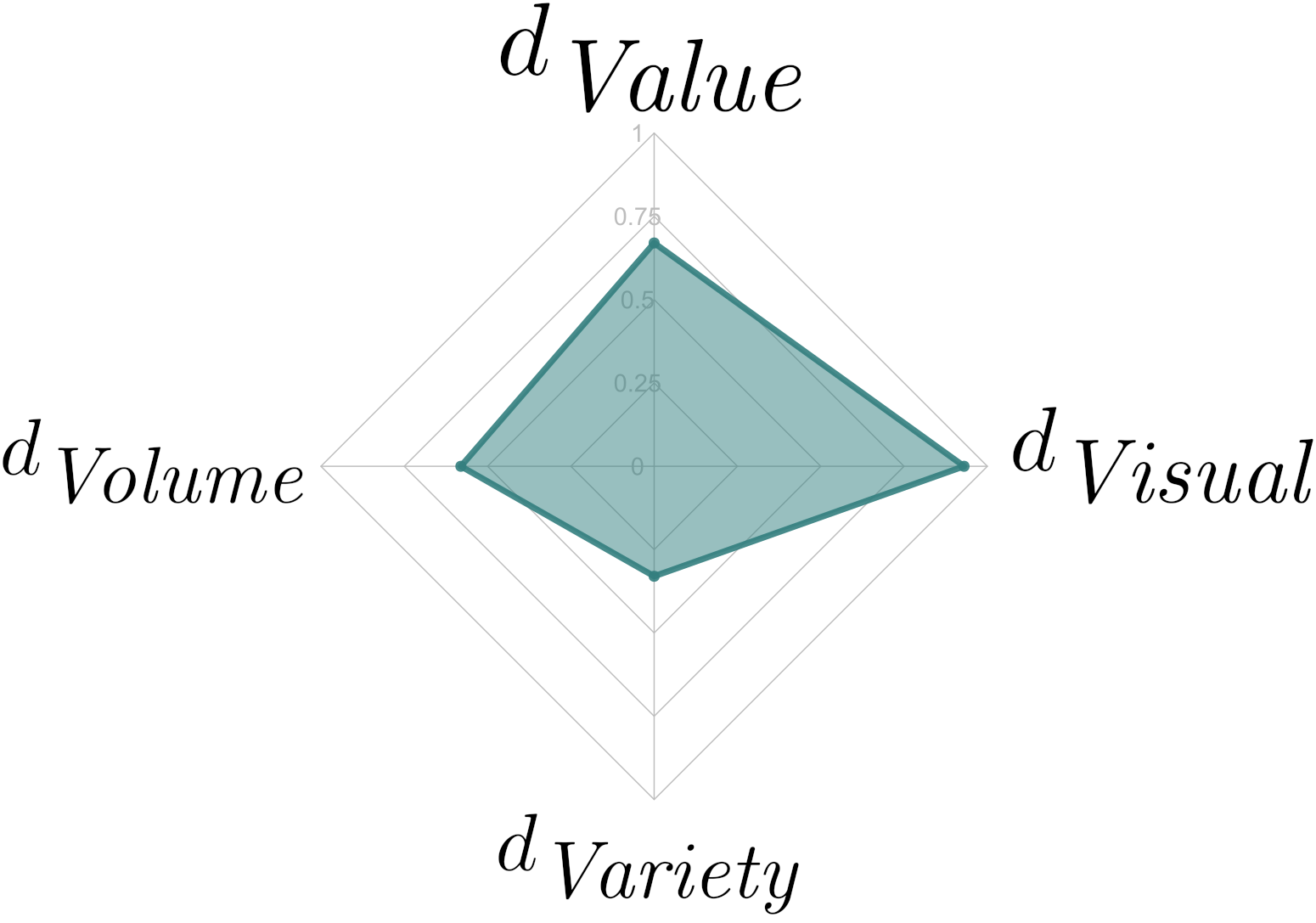}
		}
		\subfigure[Gephi]{
		  \includegraphics[width=0.3\linewidth]{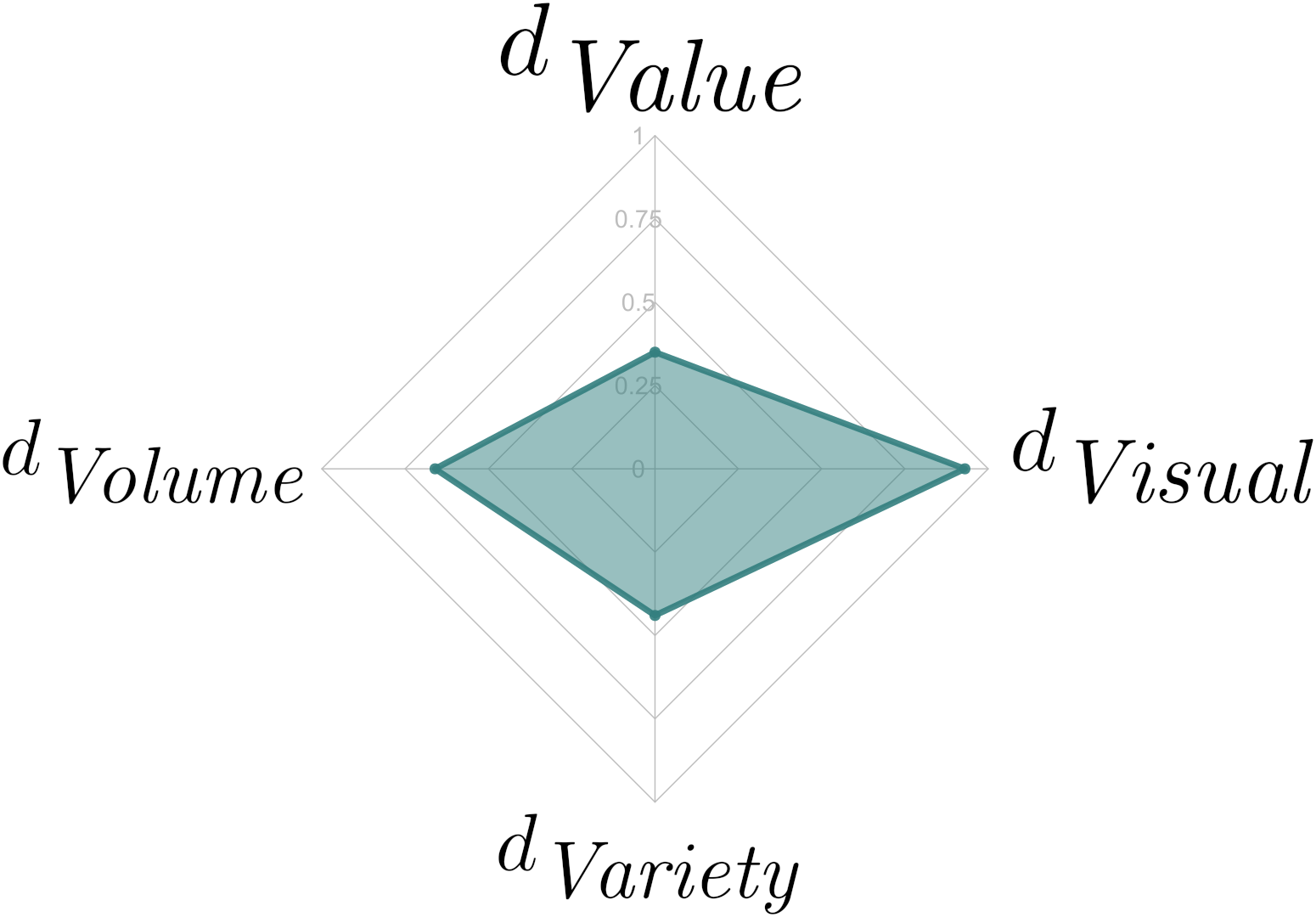}
		}
		\end{tabular}
		\caption{This figure shows the top 3 SNA software with ``Proprietary or Open Source" licenses (first row: Graphistry, Neo4j and ORA-LITE/Pro), and ``Only Open Source" licenses (second row: Neo4j, Cytoscape and Gephi).}
    \label{fig:top3-spider}
\end{figure}

Regarding the \textit{global metric} score (\textgoth{C}$_{4}(t)$), Table~\ref{tab:top5} shows the top-5 ``Proprietary or Open-source", and ``Only Open-Source" SNA-software. In addition to the aforementioned table, Fig.~\ref{fig:dimension-aggregate-boxplot} is presented. This figure contains the distribution of the dimension scores in order to allow the reader to contextualize the values presented. The tool that provides the best global score is \textit{Graphistry}. This tool has the highest possible scores in \textit{Information Fusion}, \textit{Scalability}, and \textit{Visualization}. However, the score on \textit{Knowledge Discovery} is below the average, as a matter of fact, \textit{Graphistry} is the $13^{th}$ tool in that category. Regarding open-source tools, the best one is \textit{Neo4j}. This tool has an above-average \textit{Knowledge Discovery} score, an average \textit{Information Fusion} score, and the maximum possible scores for \textit{Scalability} and \textit{Visualization}. Note that the scores obtained by the tools in the top 5 differ between dimensions. This means that there is not a single tool that dominates all the others for all the dimensions. Furthermore, nearly all the tools analyzed, $15$ out of $20$, have achieved a position in one of the top 5 proposed. 


Fig.~\ref{fig:top3-spider} shows the spider, or radar, diagrams for the top-3 best SNA-software analyzed. The upper row, composed of sub-figures a, b and c, corresponds to those tools under ``Proprietary or Open-Source" licenses, whereas the second row (sub-figures d, e and f) corresponds to those frameworks or tools under the ``Only Open-source" license. The corresponding values for each dimension are shown in Table~\ref{tab:top5}.

\begin{table*}
\centering
\caption{Top 5 SNA-software by dimension.}
\label{tab:topdimensions}

\begin{tabular}{| p{0.12\textwidth} | c | c || p{0.12\textwidth} | c | c |}
\hline
 \textbf{Dimension} & \textbf{SNA-software} & \textbf{Score} & \textbf{Dimension} & \textbf{SNA-software} & \textbf{Score} \\
\hline\hline

\multirow{5}{0.12\textwidth}{\textbf{Knowledge Discovery}} & \textbf{ORA-LITE/PRO} & \textbf{0.84} & \multirow{5}{*}{\textbf{Scalability}} & \textbf{Grasphistry} & \textbf{1.00} \\
& SNAP & 0.67 & & \textbf{AllegroGraph} & \textbf{1.00}\\
& Cytoscape & 0.67 & & \textbf{Neo4j} & \textbf{1.00}\\
& NetMiner & 0.65 & & GraphX Apache Spark & 0.92\\
& NetworkX & 0.52 & & SparklingGraph & 0.92\\
\hline

\multirow{5}{0.12\textwidth}{\textbf{Information Fusion}} & \textbf{Grasphistry} & \textbf{1.00} & \multirow{5}{*}{\textbf{Visualization}} & \textbf{ORA-LITE/PRO} & \textbf{1.00} \\
& Netminer & 0.89 & & \textbf{Grasphistry} & \textbf{1.00} \\
& Network Workbench & 0.67 & & \textbf{Neo4j} & \textbf{1.00} \\
& ORA-LITE/PRO & 0.67 & & Gephi & 0.93 \\
& Pajek & 0.56 & & Cytoscape & 0.93 \\
\hline

\end{tabular}
\end{table*}

In order to further evaluate this phenomenon, the top 5 SNA-software tools for each dimension are shown in Table~\ref{tab:topdimensions}. Starting with the \textit{Knowledge Discovery} dimension. The best performing SNA-software in this dimension is ORA-LITE/PRO and it scores better than the second-best tool, SNAP. This makes ORA-LITE/PRO an \textit{outlier} regarding the \textit{Knowledge Discovery} dimension (see Fig.~\ref{fig:dimension-aggregate-boxplot}). A similar case can be found for the \textit{Information Fusion} dimension, where Graphistry tool provides the highest possible score (see Fig.~\ref{fig:dimension-aggregate-boxplot}). In this case Graphistry also scores better than the second-best tool, Netminer. Contrary, the \textit{Scalability} dimension shows three equally good tools as top-performing (Graphistry, AllegroGraph and Neo4j), followed by GraphX Apache Spark and SparklingGraph. Finally, a similar case can be found on the \textit{Visualization} dimension with the top tools (ORA-LITE/PRO, Graphistry and Neo4j) followed by Gephi and Cytoscape.

\begin{figure*}
\centering
\noindent
\makebox[\textwidth][c]{\includegraphics[width=0.99\linewidth]{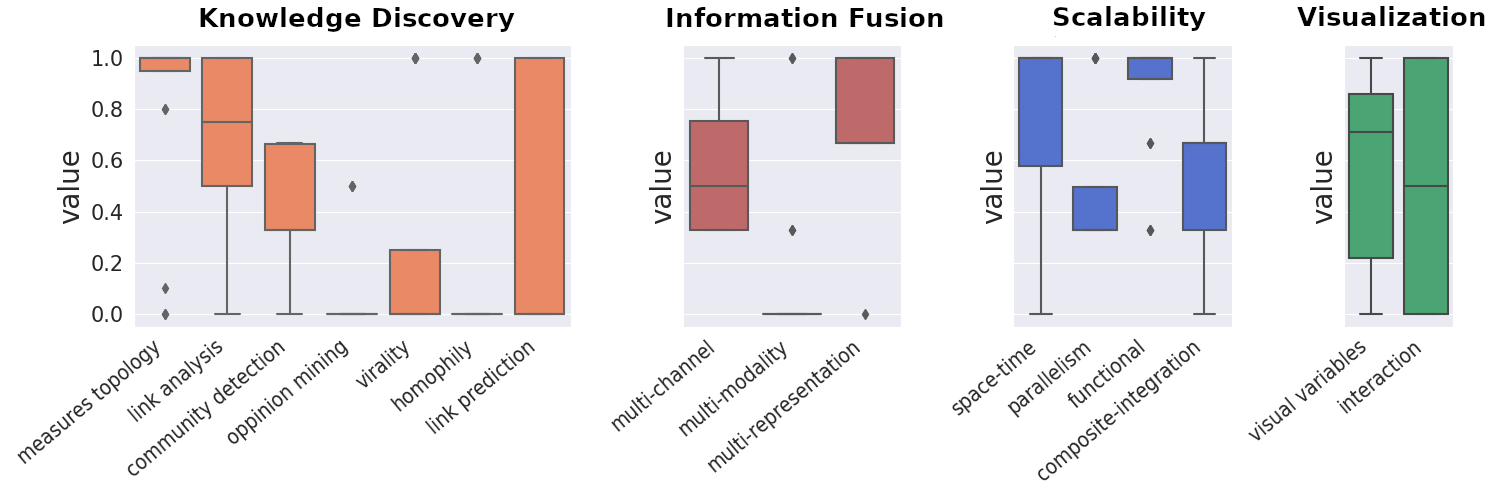}}
\caption{Un-aggregated dimension distributions for the top-5 SNA-software tools, each color represents a different dimension. The X-axis contains each of the features that form a dimension ($\textit{SNA}_{\textit{degree}}$), and the Y-axis their numerical value.}
\label{fig:dimension-unaggregate}
\end{figure*}

The analysis so far has shown that, the ORA-LITE/PRO and Graphistry tools, stand out from the rest in the \textit{Knowledge Discovery} and \textit{Information Fusion} dimensions. Below, the characteristics (or features) that have made these tools to stand out will be analyzed. To do so, each dimension has been split into its basic features (see Table~\ref{tab:sumdimensions}, and sections from~\ref{subsec:dimension1} to~\ref{subsec:dimension4}). Fig.~\ref{fig:dimension-unaggregate} shows the distribution of each of the features that forms a dimension. Notice that in the \textit{Knowledge Discovery} dimension, the \textit{Opinion Mining} and the \textit{Homophily} features are composed mostly by tools that have achieved a score of $0$, and only a few of them have been able to achieve a higher score. Something similar happens with the \textit{Virality} feature but in a less acute way. The tools that work with those features are the ones that appear in the top 5. Moreover, ORA-LITE/PRO is the only tool that has achieved a score (greater than 0) in every feature of the \textit{Knowledge Discovery} dimension. 

A similar case can be found when we analyze the \textit{Information Fusion} dimension. In the \textit{Multi-Modality} feature most of the cases are $0$, and only a few are able of scoring something in this feature. In fact, Graphistry is the only tool that has scored a maximum rating in all of the features of the \textit{Information Fusion} dimension. In general, a similar case can be found on the \textit{Scalability} dimension. However, contrary to the \textit{multi-modality} feature, several tools have achieved the maximum value in this category and not only one. Actually, all the tools that appear in the top 5 in that dimension have achieved a maximum score on that feature. Finally, the \textit{visualization} dimension is the most homogeneous. Nevertheless, the \textit{Visual Variables} feature scores slightly higher than the \textit{Interaction} ones.

To sum up, we have noticed an uneven distribution on the 4 dimensions features. There are some features were nearly all tools score good, while in others only a few are able to obtain some scoring. This makes those tools stand out over the others. For example, the \textit{Measures Topology}, \textit{Link analysis} or \textit{Funcional} features have high values for nearly all the tools analyzed, specially the \textit{Measures Topology} one. Contrary, features like \textit{Opinion Mining}, \textit{Homophily} or \textit{Multi-Modality} are tackled by very few tools. These last features can be used as a foundation of the guidelines that the next iteration of SNA tools must follow.

\begin{figure*}
\centering
\noindent
\makebox[\textwidth][c]{\includegraphics[width=0.80\linewidth]{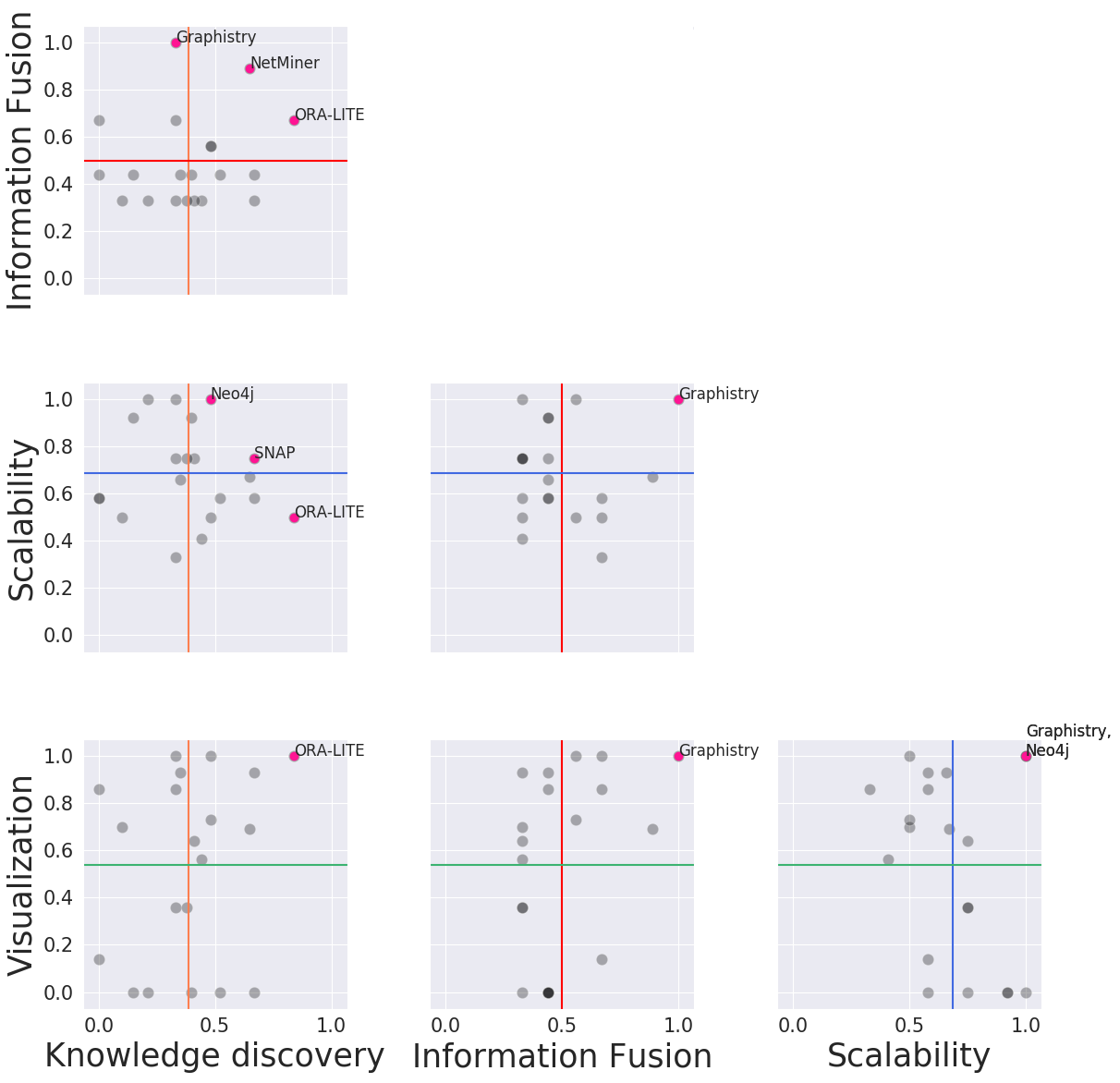}}
\caption{Pairing plot graphics, of the dimensions proposed for SNA-software. Each cell contains a graph where two different dimensions are depicted on the X and Y axis, and each point in the graph represents a tool. The horizontal and vertical lines show the average value at each dimension. The darker a point, the more tools have scored those values. Finally, the pink dots represent the Pareto front of a set, the members of the Pareto front are labeled with their names.}
\label{fig:pointcloud-dimension}
\end{figure*}


Up to now, the 4 dimensions have been analyzed independently. However, it is possible that exist different correlations, or relations, between the dimensions. For example, a proportional relationship between \textit{Visualization} and \textit{Knowledge Discovery} might be observed. It can be hypothesized that the more knowledge a tool can extract, the more visualization capabilities should it have (in order to process it). A similar hypothesis could be drawn between the \textit{Visualization} and the \textit{Information Fusion} dimensions following the same logic. Furthermore, a proportional relation could be also expected between the \textit{Information Fusion} and the \textit{Knowledge Discovery} dimensions. Since, the more types of information a tool is able to handle, the broader the type of possible analysis will be, and therefore the further the knowledge extraction will get. Contrary, an inversely proportional relationship is to be expected between the \textit{Scalability} and the \textit{Knowledge Discovery} dimensions. Fancier knowledge extraction techniques, usually require from a high amount of computational power. Implementing these methods in a scalable way is a daunting task that is currently being tackled by the scientific community. Finally, the same logic could be applied between the \textit{Scalability} and the \textit{Information Fusion} dimensions. 

Therefore, and in order to explore these relations, we have represented the (pair to pair) relations between all of the dimensions for the SNA-software tools analyzed. The results are shown in Fig.~\ref{fig:pointcloud-dimension}, this figure plots two different dimensions in the X and Y axis, and uses dots to represents the analyzed tools. The intensity of the dots indicates the number of tools that have scored a particular tuple of values. A pink dot indicates that a particular tools is a member of the Pareto Front of a graph. The members of the Pareto Front have been labeled with its tool name. Finally, to help the evaluation of the tool distributions, two colored vertical/horizontal lines have been added. These lines indicate the mean obtained for each particular dimension. Two types of graph are present in the matrix. On the one hand, there are the graphs where one tool dominates all the others. You can see this effect in \textit{Scalability vs. Information Fusion}, \textit{Visualization vs. Knowledge Discovery}, \textit{Visualization vs. Information Fusion}, and the \textit{Visualization vs. Scalability} graphs. On the other hand, there are the graphs where that dominance is not present. Notice the \textit{Information Fusion vs. Knowledge Discovery} and the \textit{Scalability vs. Knowledge Discovery} graphs. 

Regarding the hypothesized relations, the data shows that the proportional relation between \textit{Visualization} and \textit{Knowledge Discovery}, and \textit{Visualization} and \textit{Information Fusion} is present in the data. Notice how in their respective graphs there is a higher condensation of points in the top right corner of the plot and how the number of dots grow from left to right and from top to bottom
The points that have a zero value in the \textit{Visualization} dimension have been ignored as those tools do not implement any kind of visualization. However, the hypothesized proportional relation between \textit{Information Fusion} and \textit{Knowledge Discovery} is not present in the data. Notice how the several fronts presented in the graph decreases from left to right. This has surprised the authors and could be interpreted as a future research niche. The data, a priory, shows that collecting extra information types and sources are not being fully exploited by the knowledge extraction algorithms. Concerning the hypothesized inversely proportional relations between the \textit{Scalability} and the \textit{Knowledge Discovery} dimensions , and the \textit{Scalability} and the \textit{Information Fusion} dimensions, the data have shown that both are present. Notice that the dominance fronts of both graphs decrease from left to right. This effect is far more acute between the \textit{Scalability} and the \textit{Knowledge Discovery} dimensions. Nevertheless, the \textit{Scalability/Information Fusion} graph also shows a particular pattern not expected by the authors. The far dominance of \textit{Graphistry} of every other tool in that graph is surprising. In fact, there are only two tools on the top right sector of the graph including \textit{Graphistry}, and the other one is near the mean line. In the authors opinion this huge gap is another research niche, especially taking into account that \textit{Graphistry} is a commercial tool and no other open-source one is near to its capabilities.

%% file: sections/challenges.tex
\section{Conclusion, Future Trends and Challenges}
\label{sec:challenges}

In the last years, we have witnessed the increasing relevance of OSNs in our daily life. This popularity is produced by the high number of users that participate in these social media platforms every day and this participation results in huge amount of data generated by user interactions. OSN popularity and its exponential growth have led to an enormous interest in the analysis of this type of networks. There is a plethora of issues that can be studied from social media data, like the interconnections that originate the network, the structure, the evolution of the network, the identification of user communities, how the information flow and how this information is disseminated, or the patterns that can be extracted from them, just to mention a few of them. As a consequence of becoming a hot research area, the number of papers, conferences, journals, algorithms and tools has risen exponentially. This growth makes almost impossible to analyze in detail the current state of the art related to SNA. In order to limit the state of the art, and to present relevant review of the different research papers published in this topic, we have performed a scientometric study to define the most relevant research areas. From this analysis several fundamental research (graph theory and network analytics, community detection algorithms, information diffusion models, text mining and topic extraction, opinion mining and sentiment analysis), and application domains (health, marketing, tourism and hospitality, cyber-crime and cyber-terrorism, politics, detection of fake news and misinformation, and finally multimedia), have been studied in Sections~\ref{sec:sna-research} and~\ref{sec:applications}.

Any analysis of OSNs rely on the representation of the data as a graph, and then some algorithms are applied to extract some valuable knowledge. In this sense, we have reviewed the basics of SNA techniques and algorithms in Section~\ref{sec:sna-research}. The different algorithms are categorized around what kind of information is used in the analysis. In this sense, any researcher can extract information from the resulting graph, performing a structural-based analysis, or he/she can perform the analysis based on the content published in the OSN (content-based analysis). On the one hand, using the network structure any researcher can apply algorithms to detect the communities that compose the graph, or to study how the information is propagated through the network. On the other hand, using the content of the information published the standard analysis techniques rely on solving the problems of topic extraction, opinion mining and sentiment analysis. Finally, we have also analyzed the current state of the art regarding multimedia content; i.e., how the different media (text, audio and video) is used to perform SNA.

In spite of the previous state of the art analysis, and due to the size of the research community around SNA, it is quite difficult for a novel researcher, or someone (even experienced) who wants to start his/her research on this area, to select the most appropriate tool or algorithm. In order to offer a starting point to the community, we have proposed in this work four research questions that any researcher should answer, or at least keep in mind, before starting his/her research in this area. From these research questions it have been proposed several dimensions, as a tool to obtain a quantitative assessment of the current maturity of SNA technologies, allowing both to better understand what are the main strengths and weaknesses of these technologies, and to look for future trends and possible improvements in the next years in this area. 

To perform this quantitative assessment some specific metrics, or \textit{degrees}, have been defined in Section~\ref{sec:4dimensions}. And finally, a quantitative evaluation of a set of 20 popular SNA-software tools have been carried out, to show how these dimensions (and their related metrics) can be used to evaluate these technologies, in Section~\ref{sec:sna-tools}.\\

The main conclusions from the research questions proposed, their related dimensions and metrics proposed, and the assessment of the SNA- software tools carried out, can be briefly summarized as follows:

\begin{enumerate}
  \item \textbf{{\Large W}hat can I discover?}. This question is related to the different types of knowledge that the tool is able to extract. The goal of this question is to quantify the capacity of the tool to extract valuable knowledge from the data. To answer this question, the dimension \textit{Pattern \& Knowledge discovery} (and its related $d_{Value}(t)$ metric) has been considered. From the current study, it can be concluded that some kind of analyzes, like topology measures, link analysis or static community detection, are fairly common in the tools analyzed. While other analyzes, like dynamic community detection, opinion mining, virality or homophily, are quite rare. In our opinion, this phenomenon is linked to the lack of proportional relation between the \textit{Information Fusion} and \textit{Knowledge Discovery}. This suggests that the content of an OSN is not being fully exploited by actual tools.
  
  
  \item \textbf{{\Large W}hat is the limit?}. Answering this question the researcher will understand the scalability of the tool. This question is quite important due to the amount of data that can be extracted from OSNs. This question has been addressed through the \textit{Scalability} dimension (and its related $d_{Volume}(t)$ metric). As the quantitative analysis of the tools shown in Section~\ref{sec:sna-tools}, it is clear that most of the analyzed tools are capable of handling fairly big graphs (around 100.000 nodes), they are very customizable (their code is publicly available), and allow communication with other tools via an API, even thought just a few achieves fully integration with other applications. However, few tools are capable of doing BigData and the ones that can, have a low/medium average Knowledge Discovery capabilities. Taking into account the fast growth of the OSNs, the size of handled networks (to millions or hundred of millions nodes and vertex) will be swift increased in this tools, jointly to other capabilities, as the fusion and integration from different sources, and with different tools, to improve the knowledge discovery reached by these tools.  
  
  
   \item \textbf{{\Large W}hat kind of data can I integrate?}. This relevant question, related to the capacity to integrate and fusion information from current SNA technologies, has been analyzed through the \textit{Information Fusion \& Integration} dimension (and its related $d_{Variety}(t)$ metric). Again, the quantitative analysis of the tools carried out, shows that most of the analyzed tools used complex graph representation (multilayered graphs or hypergraphs), are capable of processing a medium amount of different data types (two or three different types, and are only capable of extracting data from one unique OSN. Therefore, and related to this dimension, it can be expected a very significant increase in research related to the fusion and integration of information using different types of data formats, and when possible, from different OSNs.
   

  \item \textbf{{\Large W}hat can I show?}. Finally, this last question was explored using the \textit{Visualization} dimension (and its related $d_{Visual}(t)$ metric), and how it has been shown, although there exist a large number of information visualization, and tools that provide flexible methods to visualize the information, this is still an open problem in the area. From the quantitative analysis related to this dimension, it can be concluded that the visualization capabilities of the tools where more evenly distributed compared to the rest of the dimensions. However, we have observed a lack of tools with high Scalability and Visualization capabilities. Taking the absolutely necessity to provide visualization tools to the end users and practitioners, the research and improvements in this area will be a high (and hot) topic in SNA in the next years, for example, in areas as dynamic community finding, data analytics or pattern finding to mention but a few. 
  

\end{enumerate}

Finally, we need to make a reflection on the dimensions and metrics proposed. What is proposed here is an initial work, derived from an intense dedication to the area of SNA in the last ten years. These dimensions, and the defined metrics (or degrees), cannot (and should not) be considered as the only ones that can be defined, even the definition cannot be considered as complete. From the analysis of the state of the art, we have selected those more relevant (from our perspective) features that could be used to better identify and reflect the state of these technologies. It is quite probable, that some highly relevant characteristics have not been considered by authors. On the other hand, the fact that several tools such as Grasphistry have a value of 1.0 in the dimensions of Information Fusion, Scalability, or Visualization, or that other tools such as Neo4J, also reach the value of 1.0 in dimensions such as Scalability or Visualization, do not mean that these features cannot be improved in the future. These values only indicate that given the current state of technology (and using our evaluation rubric), a higher value is achieved than other systems or tools for those dimensions. Obviously, and given the huge and fast growth of these technologies, we are sure that these values will change in the coming years, but the authors think that these dimensions, along with their related metrics, could be an important decision tool for future researchers, and practitioners, in the field of SNA.

\section*{Appendix: Open Access to Social Network Analysis Dimensions}
\label{sec:appendix}

In order to allow researchers from social sciences, science and engineering, SNA analysts and professionals, developers and engineers, etc. not only to access the data used in this article, but to foster for a future collaboration among the community interested in network analysis, a website has been designed to facilitate accessing to:


\begin{itemize}

   \item An open github repository to allow including new software, documents, open papers, and technical data, related to SNA technology.

  \item \textit{Rubric.odt}: The evaluation rubric designed to assess the SNA-software tools.
  
  \item \textit{analized\_tools.ods}: The specific evaluation made for the SNA-software tools carried out in this paper.
  
  \item \textit{README.md}: A collaborative website to evaluate SNA-software (tools, frameworks, algorithms, etc.) by the community, included as part of the github repository.
  
  \item \textit{4Dimensions.pdf}: A brief summary of this paper.

\end{itemize}

\myexample{The SNA 4-Dimensions website:}{ \textcolor{blue}{https://ai-da-sna.github.io/}}



We would like to encourage the community to provide its own evaluations, of both the tools that have been evaluated in this paper as well as others in which you have previous experience. This collaboration will promote the use of SNA technologies and fostering new developments, but it will not be possible without the cooperation of the community, so your contribution will be highly appreciated.

\label{anex:web}